%% file: main.tex
\documentclass[12pt]{article}

\usepackage{fullpage}
\usepackage{epsfig}
\usepackage{latexsym,amsmath,amssymb,amsthm}
\usepackage{subfig}
\newcommand{\ceil}[1]{\left\lceil {#1} \right\rceil}

\def\maps11{\stackrel {1-1}{\longmapsto}}

\begin{document}

\title{A New Aggregation based Scheduling method for rapidly changing IEEE 802.11ac Wireless channels}

\author{%
Oran Sharon
\thanks{Corresponding author: oran@netanya.ac.il, Tel: 972-4-9831406,
Fax: 972-4-9930525} \\
Department of Computer Science \\
Netanya Academic College \\
1 University St. \\
Netanya, 42365 Israel
\and
Yaron Alpert\\
Lantiq\\
13 Zarchin St.\\
Ra'anana, 43662, Israel\\
yaron.alpert@lantiq.com
}


\date{}

\maketitle

\begin{abstract} 
In this paper we suggest a novel idea to improve the Throughput
of a rapidly changing
WiFi channel by exploiting the standard aggregation schemes
in IEEE 802.11ac networks, and by transmitting several copies of the same
MPDU(s) in a single transmission attempt. We test this idea in scenarios where
Link Adaptation is not used and show a significant improvement, in the
order of tens of percents, in the achieved Throughput.
\end{abstract}

\bigskip


\renewcommand{\baselinestretch}{1.3}
\small\normalsize


\input intro
\input model

\input performance

\input summary

\clearpage


\bibliographystyle{abbrv}
\bibliography{main}


\end{document}

%% file: intro.tex
\section{Introduction}

\indent
The IEEE 802.11 Standard (WiFi), created and maintained by 
the IEEE LAN/MAN Standards Committee (IEEE 802)~\cite{IEEEBase1},
is currently the most
important solution within the range of Wireless Local
Area Networks (LAN). Since its first release in 1997
the standard provides the basis
for Wireless network products
using the WiFi brand, and it has been improved
in many ways. One of the main goals of these improvements
is to optimize the Throughput of the MAC layer, and to improve
its Quality-of-Service (QoS) capabilities.

To fulfill the promise of increasing IEEE 802.11 performance and
QoS capabilities, 
and effectively supporting more client devices on a network,
the IEEE 802.11 working group introduced the fifth 
generation in IEEE 802.11 networking standards; namely,
the IEEE 802.11ac amendment, 
also known as Very High Throughput (VHT)~\cite{IEEEac, IEEEBase1}. 
IEEE 802.11ac
is intended to support fast, high-quality data streaming and nearly
instantaneous data syncing and backup to notebooks, tablets
and mobile phones. The IEEE 802.11ac final version, IEEE 11ac-2013, released
in 2013~\cite{IEEEac},
leverages new
technologies to provide improvements over the previous generation, i.e.
IEEE 802.11-2012~\cite{IEEEBase} .

The IEEE 802.11ac amendment improves the achieved Throughput 
coverage and QoS capabilities, compared with
previous generations, by introducing improvements
and new features in the PHY and MAC layers. In the PHY layer,  
IEEE 802.11ac (VHT) 
continues the long-existing trend towards higher 
Modulation and Coding rates ( 256 QAM 5/6 modulation), working
in wider bandwidth channels ( up to 160 MHz ) and using 8 spatial
streams that enable higher spectral efficiency.

In the MAC layer IEEE 802.11ac includes many of the improvements
that were first introduced with IEEE 802.11n~\cite{IEEEBase},
also known as 
High Throughput (HT). 
A key performance feature first introduced in
IEEE 802.11n MAC layer is the ability to aggregate packets 
in order to reduce transmission overheads in the PHY and
MAC layers.

Frame aggregation is a feature of the IEEE 802.11n and IEEE 802.11ac
Wireless LAN standards that increases Throughput by sending
two or more consecutive data frames in a single transmission,
followed by a single
acknowledgment frame, denoted {\it Block Ack}
(BAck). Aggregation schemes benefit from amortizing
the control overhead over multiple packets. The achievable benefit
from data aggregation is often of interest, especially in the 
face of several factors that can impact its performance, e.g., 
link rates, error-recovery schemes, 
inter-frame spacing options, QoS guarantee, etc. 
IEEE 802.11n introduces, as a pivotal part of its MAC enhancements, three
kinds of frame aggregation mechanisms:
The  Aggregate MAC Service Data Unit (A-MSDU) aggregation,
the  Aggregate MAC Protocol Data Unit (A-MPDU)
aggregation and a Two-Level aggregation
that combines both A-MSDU and A-MPDU. 
The last two schemes group several MPDU frames
into one large Physical Service Data Unit (PSDU).
IEEE 802.11ac also uses these three aggregation schemes, but enables
larger MPDU and PSDU sizes. 

The IEEE 802.11 standard also defines an Automatic Repeat-Request (ARQ)
protocol that enables a transmitter to retransmit lost MPDUs
and guarantee in-order reception of MPDUs at the receiver.
This protocol is also used to improve the quality of the wireless
channel.

Given the use of this protocol, e.g. in QoS constrained 
applications such as Voice and Video, in this paper we consider
several methods to further improve the Throughput of the wireless
channel by using aggregation. We consider methods in which
some MPDU(s) are retransmitted several times
in a lossy channel and in a single 
transmission attempt. We do not assume
the model of Transmission opportunities 
(TXOP)~\cite{IEEEBase1}, and a station
transmits only one PSDU in every transmission attempt.
We are not aware of any other research using aggregation
to retransmit several copies of the same MPDU(s) in a single
transmission attempt in order to increase Throughput.

Improving the quality of the wireless channel is also possible
through Link Adaptation (LA) methods, in which a more robust
Modulation/Coding scheme (MCS) is used, at the cost of
reducing the available PHY rate. However, there are
scenarios in which the Signal-to-Noise-Ratio (SNR) is either
rapidly changing, or it is changed in small amounts (dBs).
In these cases LA is not used, either because the channel's
SNR is not stable and it is changing faster
than the LA tracing capability, or the changes are too small to trigger
LA. For these scenarios we suggest methods to improve the 
Throughput of the wireless channel by using the new aggregation schemes.

Notice that by retransmitting MPDUs one
actually reduces the PHY rate. However, we suggest
in this paper to retransmit only {\it few} MPDUs,
i.e. we reduce the PHY rate for only few MPDUs
and not for {\it all} MPDUs as in LA. We show that
such a change increases the Throughput considerably.

Another important point to mention is that we
improve the Throughput of the {\it current} IEEE 802.11ac
standard which has an upper bound of 64 MPDUs on
the Transmission window size. We do not look for solutions
that change the current standard as e.g. increasing
the above upper bound.

Finaly, our proposal is not mandatory in the sense
that it should or should not be used by all
the stations all together in a givan time.
For example, stations that are close to the AP
and have a high SNR and so a low Packet Loss Rate
should not use it. Stations that are located
far from the AP and has a high Packet Loss Rate
should use the proposal. In our later results
we show when the proposal has high benefit and
should be used.

\subsection{Our work}

We consider a single pair of transmitter/receiver, over a 
WiFi  wireless channel. Such a scenario is possible 
when the WiFi channel is used as a Point-to-Point Backhaul
(Usage Models 4a, 4b in~\cite{UM}). The transmitter transmits
MPDUs to the receiver using the above 
mentioned ARQ protocol, and using the
A-MPDU/Two-Level aggregation schemes
defined in the IEEE 802.11ac standard~\cite{IEEEBase1}. We assume a saturated 
scenario where the transmitter has an infinite number
of MPDUs to transmit. 
We also assume UDP like traffic; the receiver
does not transmit above Layer 2 acknowledgments such
as TCP Acks to the
transmitter. It only transmits Layer 2 
Acks. Therefore, the receiver does not contend on the channel
and there are no collisions. As mentioned, we also
do not assume the use in Transmission opportunities (TXOP)
and only one PSDU is transmitted in every transmission event.

We investigate the performance
of several methods to improve the Throughput by retransmitting
several copies of the same MPDU(s) in a single transmission attempt,
using aggregation.

Given the SNR of the wireless channel,
the MCS and the aggregation scheme in use, there
are two factors that influence the Throughput. First, the transmission
success probability of an MPDU.
Second, the {\it Transmission Window (TW)} of the ARQ protocol. Only
MPDUs in the $TW$ are allowed for transmission. Therefore, lost MPDUs
not only need to be retransmitted, thus wasting transmission time,
but if failed MPDUs are located at the beginning of the $TW$, the
$TW$ cannot move forward over the MPDUs' sequence, and new MPDUs
cannot be transmitted. This results in fewer MPDUs transmitted
in a single transmission attempt, again reducing the Throughout.

Blinded retransmission of several copies 
of the same MPDU(s) in a  single
transmission attempt has two advantages: First, the success
probability of an MPDU that is retransmitted several times
is improved. Second, the probability that the {\it Transmission
Window} moves forward, therefore containing new MPDUs, is also
increased. A disadvantage of this approach is increasing
the transmission time of the PSDU frame by the same
data bits. 
Investigated in this paper is whether this increase
is beneficial.

\subsection{Our results}

We consider the A-MPDU and Two-Level aggregation schemes
over several PHY rates: 433.3, 866.7, 1299.9 and 3466.8 Mbps. We consider
four MSDUs' lengths: 128, 512, 1024 and 1500 Bytes. In the Two-Level
aggregation we vary the number of MSDUs per MPDU in the range 1-7 .
We also consider several Packet Error Rates (PER): 0.50, 0.45, .... , 0.05 .
We show that the Throughput is improved using our methods, especially
in low PERs, and the improvement can sometimes be in the order
of tens of percentages ! Another important aspect of our proposed
methods is that they are simple to implement and fully comply
with the IEEE 802.11 standard~\cite{IEEEBase1}.

\subsection{Previous works}

The performance of the IEEE 802.11 protocol has been
investigated in dozens of papers over the years.
We only mention a few of those
that relate to our current research. The first set
of papers deals with the basic access scheme of IEEE 802.11 .
In~\cite{XR,L} the Throughput and Delay
performance of the legacy transmission mode
(no aggregation) are investigated,
with upper and lower limits set on the
Throughput and Delay achievable~\cite{XR}.
In~\cite{B,CBV,CA} an analytical study of the Throughput of the
basic IEEE 802.11, together with collisions, is performed, taking into
account the RTS/CTS control mechanism~\cite{IEEEBase1}. 
In~\cite{WLI,TC,CSV,GPE,C2}
the performance of the legacy transmission mode using Block Ack and
RTS/CTS is investigated.

In~\cite{SC,LW,GK,SNCSKJ,KHS,C1,WW,SS,Z,DAM,SOSH,KKS,KMLPC}
the Throughput and Delay performance 
of the A-MSDU, A-MPDU and Two-Level aggregation schemes is
investigated. Several papers assume an error-free channel
with-no collisions,
several papers
assume an error-prone channel and some papers also
assume collisions. In~\cite{P,OKACHN,CAHNOKR,BBSVO, SA}
the performance of IEEE 802.11ac is investigated. Papers~\cite{OKACHN,SA}
consider the performance of the aggregation schemes in IEEE 802.11ac
and compare the performance of IEEE 802.11ac to that of IEEE 802.11n. 

Another set  of papers, e.g.~\cite{CS,SNMB,TQDDYT,SNZ,MGCR,SA1},
deals with QoS together with the aggregation schemes. 
In particular, in~\cite{SA1} the use of the ARQ protocol
of the IEEE 802.11 standard~\cite{IEEEBase1}, together with the 
aggregation schemes, is investigated in relation to QoS guarantee.
In this paper we also investigate the use of the ARQ protocol
with the aggregation schemes, but this time we investigate
another aspect of the aggregation: Blinded retransmission of several
copies of the same MPDU(s) in the same transmission attempt,
in order to improve the
Throughput when Link Adaptation is not used. As far as we know,
such an aspect of the aggregation schemes has not 
previously been investigated.

The rest of the paper is organized as follows: In Section 2 
we describe the network model used. In Section 3 we show
the performance of our methods. Section 4 is a summary of
the paper.

%% file: model.tex
\section{Network Model}

\subsection{Successful transmissions}

\begin{figure}
\vskip 4cm
\includegraphics{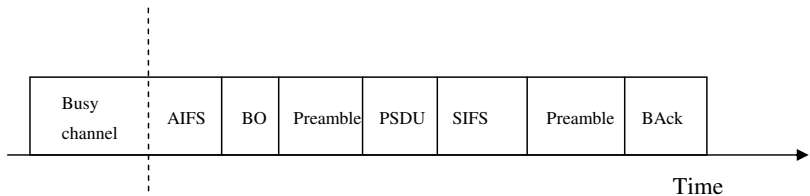}
\caption{The activity on the channel in the case of a successful transmission}
\label{fig:success}
\end{figure}

In Figure~\ref{fig:success} we show the IEEE 802.11 channel access method
in the case of successful transmissions, i.e. without
collisions. This is the model that we consider
in this paper. The channel access scheme in IEEE 802.11 is based
on the Carrier Sense Multiple Access with
Collision Avoidance (CSMA/CA) protocol, which we assume
the reader is familiar with. Therefore,
we only remind the steps that a station performs in a successful
transmission. 

After a station senses
an idle channel for a duration equal to the
Arbitrary Inter Frame Space (AIFS) and BackOff
intervals, it transmits its data frame,
denoted Physical Service Data Unit (PSDU).
The PSDU's transmission is also preceded by a PHY
preamble. 
After an SIFS the receiver
acknowledges the reception.  In the case of the
A-MPDU and Two-Level aggregation schemes, the BAck frame is used. 
The BAck transmission is also
preceded by a PHY Preamble.
The process of generating the PSDU 
frame, in the case of the A-MPDU and Two-Level aggregations,
is shown in the following
Figures~\ref{fig:ampdu} and~\ref{fig:twole} respectively.
As mentioned, 
in this paper we assume 
UDP like traffic, where the receiver does not transmit
above$-$Layer 2 acknowledgments. It only transmits the Layer 2
BAck acknowledgments. Therefore, there are
no collisions. We also assume that the
transmitter has an infinite number of MPDUs to transmit.
Our proposed methods have significant under-saturation 
scenarios only where the loss of MPDUs delays
the transmission of following MPDUs.
Therefore, the transmission scenario in Figure~\ref{fig:success} 
repeats itself.

We assume that the transmitter
is using the Best Effort Access 
Category and the following values
are taken from the WiFi Alliance (WFA)
publications~\cite{WFA}. The WFA is an organization
that performs certification tests. The tests
ensure reliability of the WiFi brand,
and certification programs can be seen from
the certified products. 

In the Best Effort Access Category
the AIFS is $43 \mu s$.
The BackOff is a multiple of the $SlotTime$ size,
which for the OFDM PHY layer is $9 \mu s$.
We assume that there are no collisions and
so, on average, one half of the minimum BackOff interval is
used, i.e. $7.5 \cdot 9=67.5 \mu s$. 
The duration of the PHY preamble, preceding
the PSDU transmission, is changed according
to the number of spatial streams in use~\cite{IEEEBase1}. In this
paper it is $43 \mu s$ for the most part, corresponding
to 3 spatial streams.
The SIFS is $16 \mu s$.
The Block Acknowledgment (BAck) frame is 32 bytes
long. Its transmission
time, denoted BAckTime, is $32 \mu s$, using
the Basic PHY Rate of 24Mbps, and including
the legacy PHY Preamble of $20 \mu s$.
If the PHY rate $R$ used for data frame transmissions
is lower than 24Mbps then $R$ is also 
used for the BAck transmission.
However, in this paper we use $R$s with higher values than 24Mbps. 

\subsection{The Aggregation Schemes}

\subsubsection{The A-MPDU aggregation scheme}

\begin{figure}
\vskip 9cm
\includegraphics{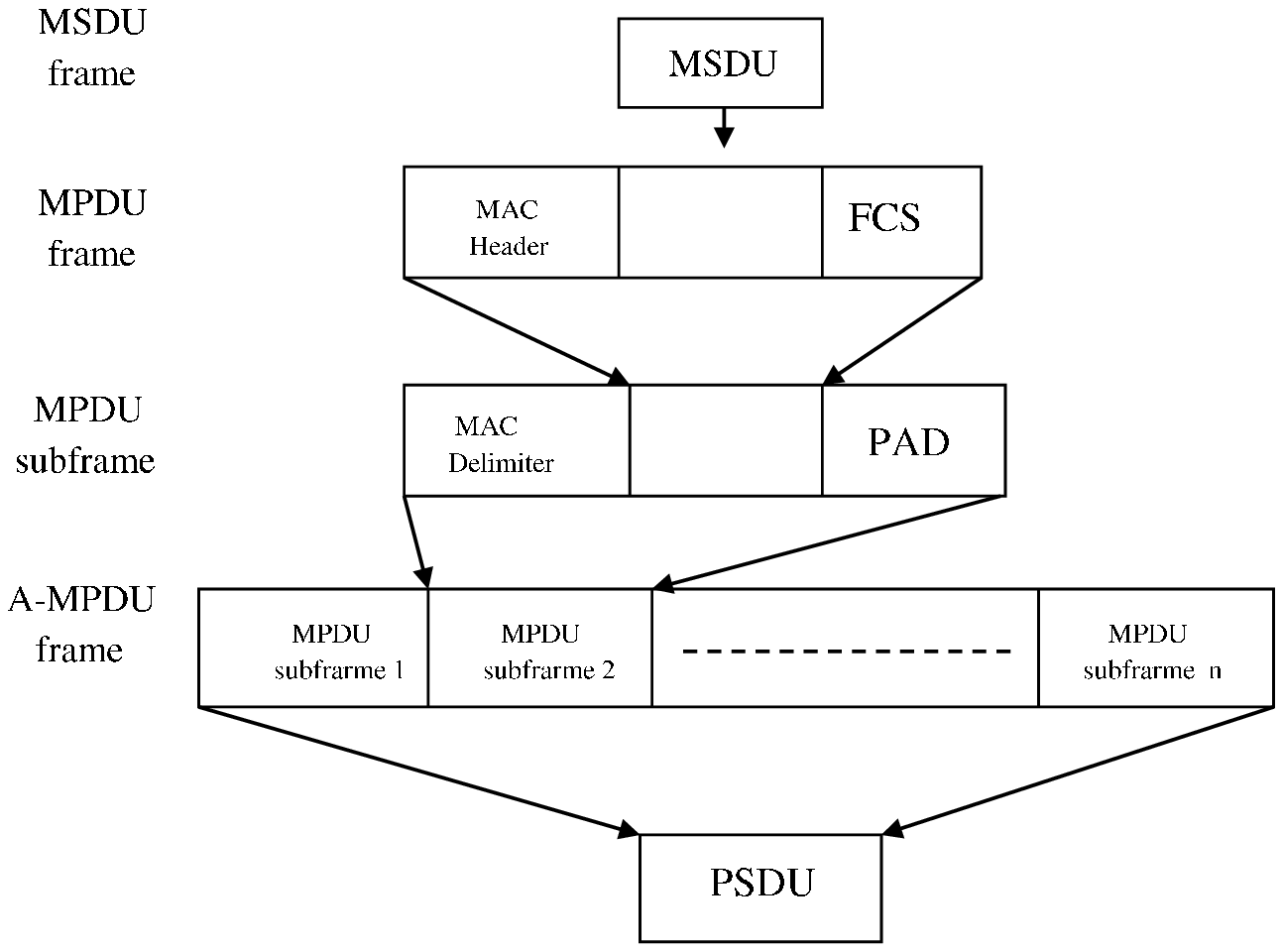}
\caption{The generation of an A-MPDU frame in A-MPDU aggregation}
\label{fig:ampdu}
\end{figure}

The A-MPDU aggregation scheme is shown in Figure~\ref{fig:ampdu}.
Several {\it MAC Protocol Data Units} (MPDUs) 
are inserted for transmission into one
{\it Physical Service Data Unit} (PSDU) where each
MPDU contains only a single {\it  MAC Service Data Unit} (MSDU). 
Up to 64 MPDUs, {\it with different sequence numbers}, are allowed
in one PSDU. Such a PSDU is
denoted {\it A-MPDU frame}. The MPDUs are separated
by a MAC Delimiter of 4 octets, and every MPDU is rounded
with its delimiter by a PAD, to a length that is an integral
multiply of 4 octets.

The advantage of this aggregation scheme is that every MPDU
is protected by its own Frame Control Sequence (FCS), so MPDU
either arrives successfully or un-successfully at the receiver,
independent of the other MPDUs. The acknowledgment frame is
now the Block Ack (BAck) frame, which acknowledges every MPDU separately.
Other advantages are the relatively small size of the MAC Delimiter
and sharing of the AIFS, BackOff, PHY Preambles, SIFS and
transmission of the BAck frame overheads among several MPDUs.
See Figure~\ref{fig:success}.

In IEEE 802.11ac the maximum length of an A-MPDU frame
is 1048575 octets.
The maximum length of an MPDU within an A-MPDU frame is
11454 octets.
However, neither of the
above two size limits is ever reached because of the limit
of 2304 bytes on the MSDU's size~\cite{IEEEBase}.

\subsubsection{The Two-Level aggregation scheme}

\begin{figure}
\vskip 9cm
\includegraphics{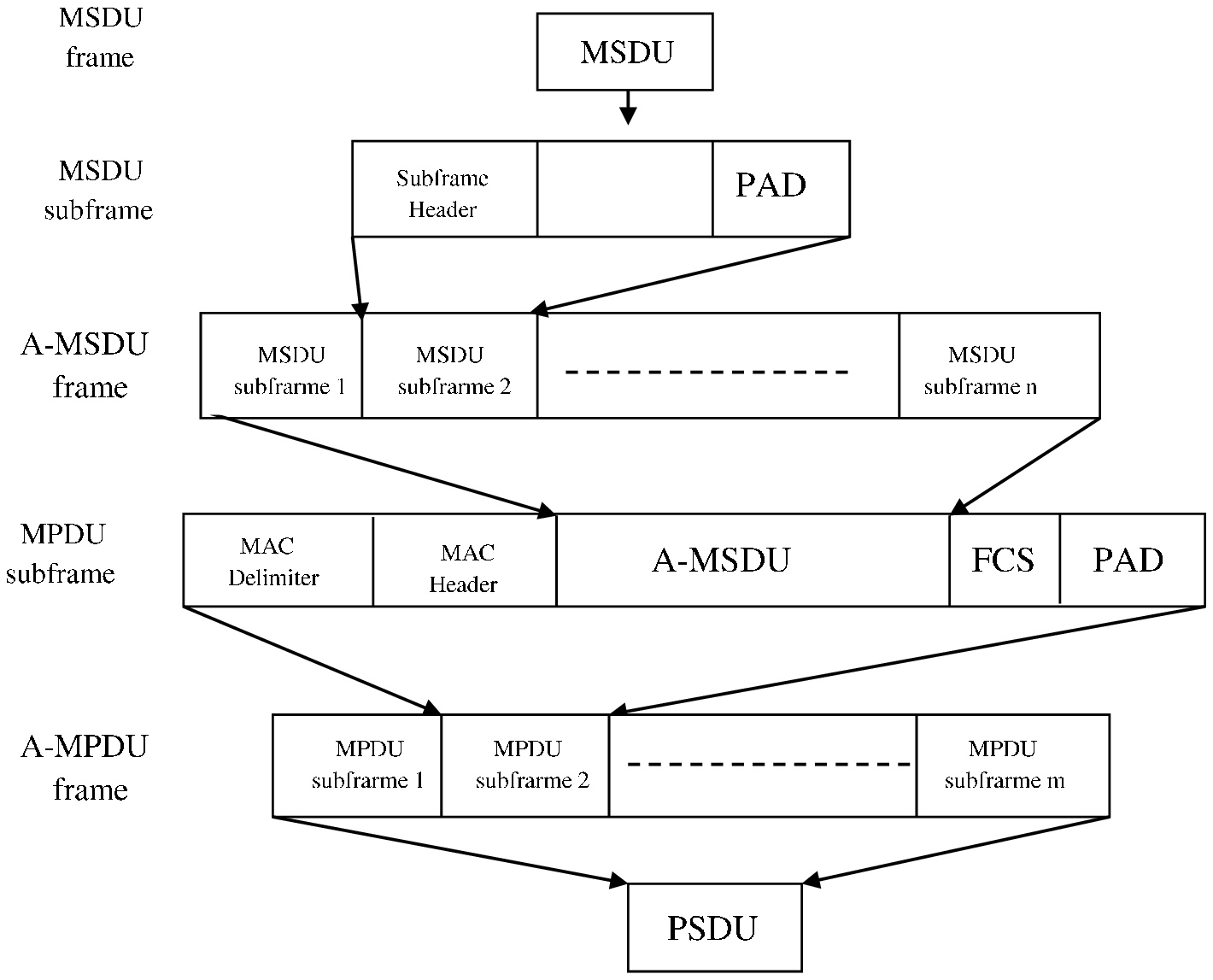}
\caption{The generation of A-MSDU and A-MPDU frames
in Two-Level aggregation}
\label{fig:twole}
\end{figure}

The Two-Level aggregation scheme is shown in Figure~\ref{fig:twole}.
In this aggregation scheme several MPDUs are inserted for
transmission into one PSDU, as in the A-MPDU 
aggregation scheme. However, an MPDU can now contain
several MSDUs. Such an MPDU is denoted Aggregated MSDU ({\it A-MSDU}).
Every MSDU is preceded by a SubFrame Header of
14 bytes, and every MSDU with its SubFrame Header
is rounded, by a PAD, to a size that is an integral
multiple of 4 bytes.
The Two-Level aggregation scheme achieves a better
ratio than A-MPDU, between the amount of Data octets transmitted to the
PHY and MAC layers' overhead. 

In IEEE 802.11ac the maximum PSDU's size is
1048575 octets and the MPDU's maximum
length is 11454 octets.
The upper limit of 64 MPDUs {\it with different
sequence numbers} in the A-MPDU/PSDU also holds 
in this aggregation scheme.

\subsection{The Error model}

We assume that the process
of frame loss in a Wireless fading channel
can be modeled with a good approximation by a low order Markovian 
chain, such as the two state
Gilbert model~\cite{L1,ZRM}.

In this model the state diagram is composed of
two states, "Good" and "Bad", meaning successful or unsuccessful
reception of every bit arriving at the receiver, respectively.
{\it Bit-Error-Rate} (BER) is the 
probability of moving from the Good state to the
Bad state.
($1-BER$) is the probability of remaining at the Good state.
According to the above model, the success probability of a frame
of length $B$ bits is $(1-BER)^B$ and the failure probability
$p$ is given by Eq.~\ref{equ:failureprobbasic}:

\begin{equation}
p=1-(1-BER)^{B}
\label{equ:failureprobbasic}
\end{equation}

By the above model
one can see that as the frame length $B$ increases, so does its
failure probability. 

Notice that errors in the MacDelimiter field(s) in an
A-MPDU frame can make the
receiver unable to detect the starting point(s)
of subsequent MPDUs. In this paper the shortest MPDUs that
we consider are of 168 bytes and the longest are of 1540 bytes.
The MacDelimiter is 4 bytes and in the worst case
about $2-3\%$ of the MPDU's length, when considering
MPDUs of 168 bytes. For MPDUs of 1540 bytes it is about $0.3\%$.
Therefore, and as observed in real systems, the probability
of not detecting the next MacDelimiter after a corrupted
MPDU is very slight, and is therefore not
mentioned in this paper.


\subsection{Proposed transmission methods}

In the A-MPDU and Two-Level aggregation schemes
it is possible to transmit up to 64 MPDUs, {\it with different
sequence numbers}, in an A-MPDU/PSDU
frame~\cite{IEEEBase}. 
In 
the {\it compressed BAck} frame described in section 8.3.1.9.3
in~\cite{IEEEBase1} is a {\it  Block Ack Bitmap } field
containing 64 bits. Every one bit in
this field acknowledges the reception of one MPDU in
increasing order of sequence numbers, starting from
a sequence number that is also included in the frame.
This is the reason for the limit of 64 MPDUs with
different sequence numbers per A-MPDU frame.


Let $K$ be the maximum number of MPDUs, with
different sequence numbers, that
are actually allowed in a PSDU frame. $K$ ranges between
1 to 64. Notice however, that the standard does not
prohibit the transmission of more than 64 MPDUs per A-MPDU frame,
as long as there are at most 64 different sequence numbers, and the
transmission time of the PSDU is not larger than $5.4 ms$. 
This limit is derived from Eq. 9-12 in Section 9.26.4 in~\cite{IEEEBase1}.
We base our proposed methods on this observation.

We are given an infinite sequence of MPDUs
to transmit from transmitter A to receiver B. 
All MPDUs are of the same length. 
Every MPDU has a probability $1-p$ to 
move successfully from
A to B . This probability $1-p$ is the same for 
all MPDUs and all MPDUs' transmissions are independent.

We also have a {\it Transmission Window (TW)} of size $W$. 
This $TW$ is part of the ARQ protocol defined in the IEEE 802.11 
standard~\cite{IEEEBase}. According to this ARQ protocol,
only MPDUs within the $TW$
are permitted for transmission from A to B. 
For example, if the MPDUs are numbered 1, 2, 3, 4, ... 
and $W=20$ to start, MPDUs 1-20
are in the $TW$ and only these MPDUs are allowed for transmission.
The $TW$ slides over the MPDUs' sequence. After MPDU 1 
is transmitted successfully the $TW$ slides
one position and now includes MPDUs 2-21 etc.

In every single transmission from A to B 
it is permissible to transmit up to $K$ MPDUs
with different sequence numbers, $K \le W$. 
$K$ is fixed and given. After the transmission, B 
notifies A which MPDUs arrived successfully. 
MPDUs that did not arrive successfully 
must be retransmitted. 

Let's assume for now that only a single 
copy of an MPDU can be transmitted in a given
transmission. 
Let's also assume that $K < W$. 
At the first transmission $K$ MPDUs are transmitted. Some
arrive successfully and others not. 
However, if MPDU number 1 does not arrive
successfully, the $TW$ does not slide ! 
Consider the extreme case where MPDU 1 does not
arrive successfully over several transmissions. 
In such a scenario, the $TW$ does not slide
and a stage will occur where A will 
not have $K$ different MPDUs to transmit to B !

Assume the A-MPDU aggregation
scheme and that a station transmits $X$ different
MPDUs in a given A-MPDU frame; one copy of each MPDU.
The throughput of this
single transmission, denoted $Thr$, is defined in Eq.~\ref{equ:equ1}:

\begin{equation}
\label{equ:equ1}
Thr = \frac{8\cdot L\cdot X\cdot Psucc}
{C_1+Tsym
\ceil
{
\frac{8\cdot X\cdot 4
\ceil{\frac{MacDelimiter+MacHeader+FCS+L}{4}}+22
}
{BitsPerSymbol\cdot R}
}
}
\end{equation}

From Figure~\ref{fig:success} we define $C_1$ to be
$C_1 = AIFS+BackOff+PHY preamble+SIFS+BAckTime$ 
( BAckTime contains the PHY Preamble preceding the BAck transmission). Assuming
OFDM PHY layer, $T_{sym}$ is $4 \mu s$ and $BitPerSymbol$ equals 4.
$L$ is the MSDU's size in bytes and the additional 22 bits
in the denominator are due to the SERVICE (16 bits) and TAIL
(6 bits) fields that are added to every transmission by the
PHY layer conv. protocol~\cite{IEEEBase}.
Finally, $Psucc$ denotes the probability
that an MPDU arrives successfully at the receiver.

The rationale behind our proposed methods is to blindly retransmit several
copies of MPDUs at the beginning of the $TW$, in order to increase
the probability that the $TW$ will slide forward and will contain
new MPDUs for later transmissions. We examine the cases of retransmitting
only several copies of the 1st, 2nd, 3rd and 4th
MPDUs in the $TW$ respectively. These cases
are denoted later by {\it Set1, Set2, Set3} and {\it Set4} respectively.
As an extreme measure, we also check the possibility of retransmitting
several copies of each of the MPDUs are transmitted in a 
transmission attempt. This is done in {\it Set5} further on.

We now describe the five methods for retransmission
of the MPDUs and compare between the Throughputs
that these schemes achieve.

For example, there are $I$ MPDUs in the $TW$ that were received 
successfully at B. Let $X=min\{K,W-I\}$ and let $X_{min}$
be the set of $X$ MPDUs with the
smallest indexes in $TW$ that have not yet arrived successfully at B.
The MPDU with the smallest index in $X_{min}$ is the one
at the beginning of the $TW$. Let's denote this MPDU
by $MPDU_{min}$.
For example, say $W=10$, $K=9$ and that only MPDUs 2, 4, 5, 6, 7 and
8 arrived successfully at B. Then $I=6$, $X=min\{9,10-6\}=4$,
$X_{min}=\{1,3,9,10\}$ and $MPDU_{min}$ is MPDU number 1.
We use this example in the description of the schemes below.

\begin{itemize}

\item
$Base:$ 
The X$_{min}$ MPDUs are transmitted,
a single copy of each MPDU.
In our example the transmission contains
one copy of MPDUs 1, 3, 9, 10 .

\item
{\it Set1 - 1MPDU2, 1MPDU3, 1MPDU4, 1MPDU5:} 
One copy of each MPDU in $X_{min}$ is transmitted once, except
to $MPDU_{min}$. In 1MPDU2 this MPDU is transmitted twice
in every transmission attempt. In {\it 1MPDU3i} this MPDU
is always transmitted 3 times in every transmission attempt, and so on
until 5 times.



\item
{\it Set2 - 2MPDU2, 2MPDU3, 2MPDU4, 2MPDU5:} 
This set of methods is similar to {\it Set1} but this time the first
two MPDUs in a transmission attempt, i.e. $MPDU_{min}$ and
the next MPDU, are transmitted several times. 




\item
{\it Set3 - 3MPDU2, 3MPDU3, 3MPDU4, 3MPDU5:} 
Same as {\it Set 1} and {\it Set2}, except that now the first 3 MPDUs in a transmission
attempt are retransmitted several times.

\item
{\it Set4 - 4MPDU2, 4MPDU3, 4MPDU4, 4MPDU5:} 
Same as {\it Set1, Set2, Set3} 
except that now the first 4 MPDUs in a transmission
attempt are retransmitted several times.

\item
{\it Set5 - All2, All3, All4, All5:}
All the MPDUs in the transmission attempt are transmitted
several times: In the All2 method every MPDU is transmitted
twice etc. 



\end{itemize}

%% file: performance.tex
\section{Performance results}

Our performance results are based on simulations.
In all the simulations we set $W$, the size
of the {\it Transmission Window}, to be 64, the
maximum possible in the IEEE 802.11ac standard~\cite{IEEEBase1}.
We checked
the Throughput for all possible $K$, the number of
MPDUs in every transmission, $1 \le K \le 64$, and
picked the maximum Throughput that is achieved by any 
of the $Ks$.

As mentioned, we consider 4 MSDUs' sizes, 128, 512, 1024 and 1500
bytes, which, with the MAC Header, MAC Delimiter, FCS and the rounding
to an integral multiply of 4 bytes, become MPDUs of size
168, 552, 1064 and 1540 bytes respectively. We also
consider 4 PHY rates:  433.3, 866.7, 1299.9 and
3466.8 Mbps.
All these PHY rates correspond to Working
Point MCS9. The first assumes 4 spatial streams and a 160 MHz
channel. The other three assume a 80 MHz channel with 3, 2 and
1 spatial stream(s) respectively~\cite{IEEEBase1}.

Our performance results are organized as follows: 
In Figures~\ref{fig:fig01} and~\ref{fig:fig02} we justify
our decision to retransmit only the first 4 MPDUs
of an A-MPDU frame (Section 3.1). 
In Figures~\ref{fig:fig1}-~\ref{fig:fig3}
we show the performance of {\it Set1, Set2} and {\it Set3} for
MPDUs of 1540 bytes, and in Figures~\ref{fig:fig11},~\ref{fig:fig13}
we show the performance of {\it Set1} and {\it Set4} for MPDUs of
168 bytes respectively (Section 3.2). 
In Figures~\ref{fig:fig4},~\ref{fig:fig41}
we show the performance of {\it Set5} for MPDUs of 1540 and 168 bytes
respectively (Section 3.3). 
In Figures~\ref{fig:fig5}-~\ref{fig:fig8} we show
the Throughput improvement over the $Base$ method,
taking into account all the suggested methods. 
In Figure~\ref{fig:fig9} we show that our methods
improve the Throughput over a wide range of BER values (section 3.4)
and 
finally, in Figure~\ref{fig:fig10} we show results for the
Two-Level aggregation (Section 3.5).

\subsection{Why {\it Set1-Set4} only}

\begin{figure}
\vskip 16cm
\includegraphics{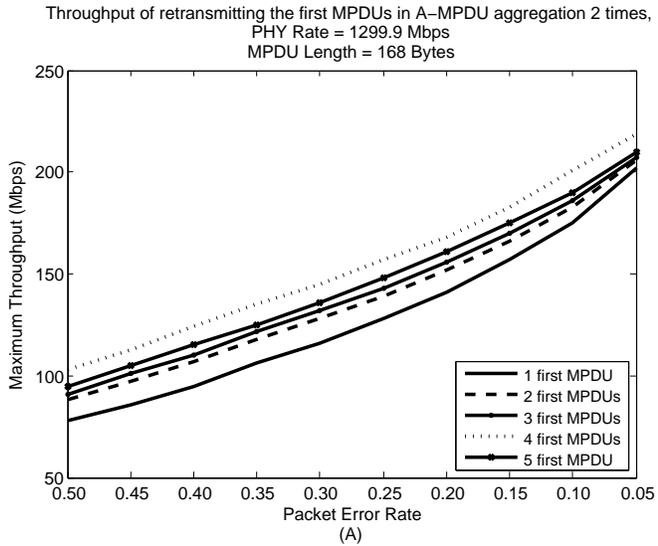}
\includegraphics{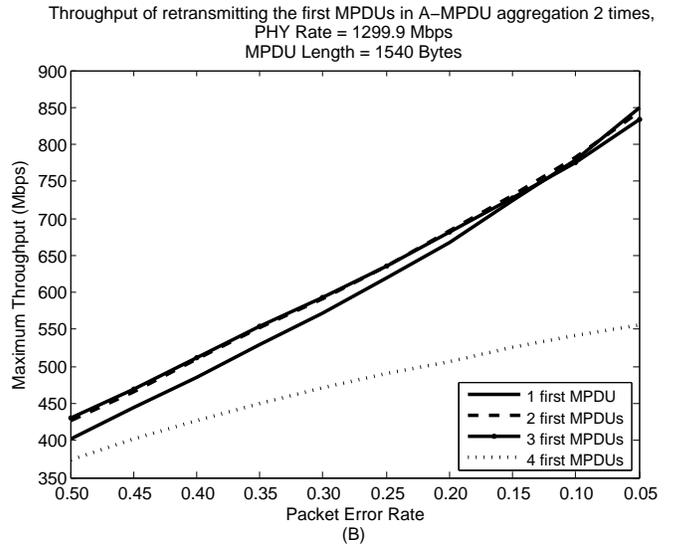}
\includegraphics{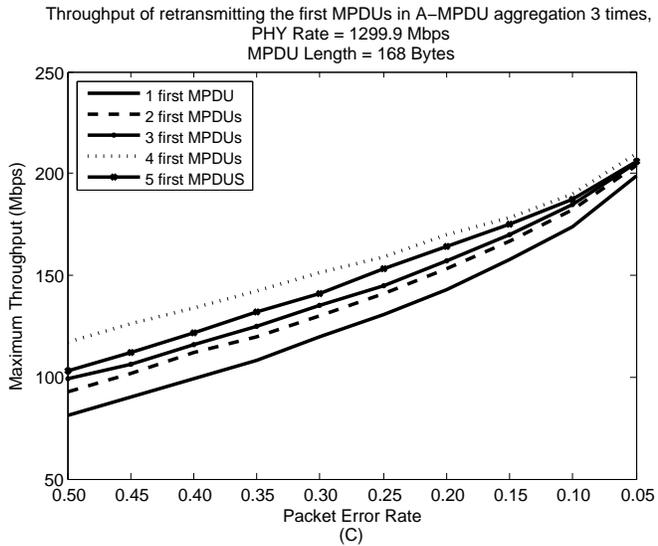}
\includegraphics{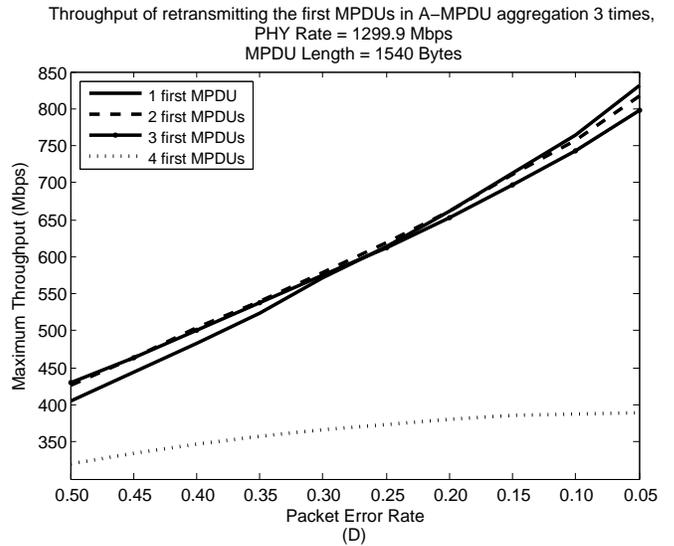}
\caption{Throughput comparison when retransmitting the first 1 to 5 MPDUs in
an A-MPDU 2 and 3 times, A-MPDU aggregation, MPDUs of 168 and 1540 bytes, PHY rate = 1299.9 Mbps }
\label{fig:fig01}
\end{figure}

\begin{figure}
\vskip 16cm
\includegraphics{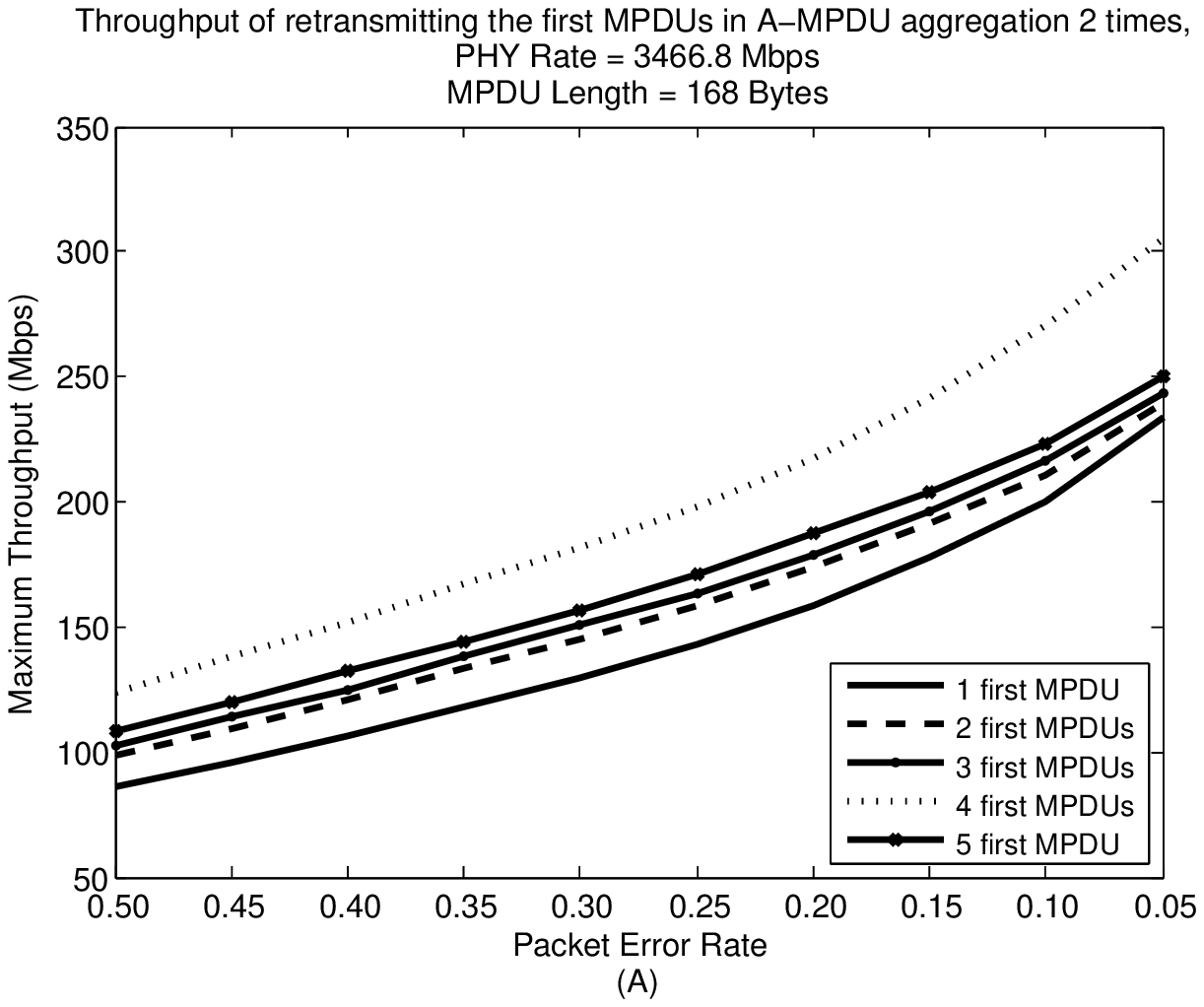}
\includegraphics{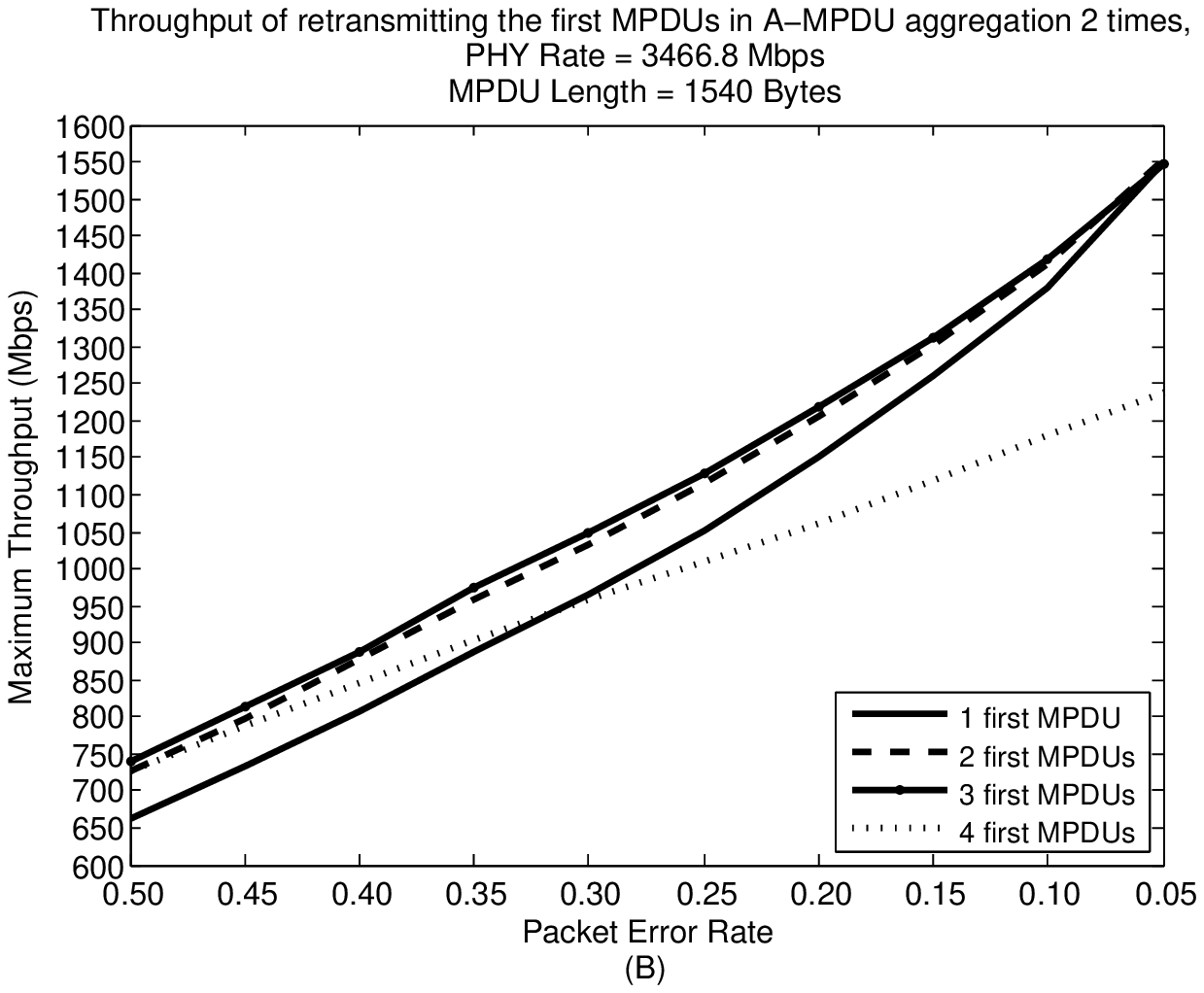}
\includegraphics{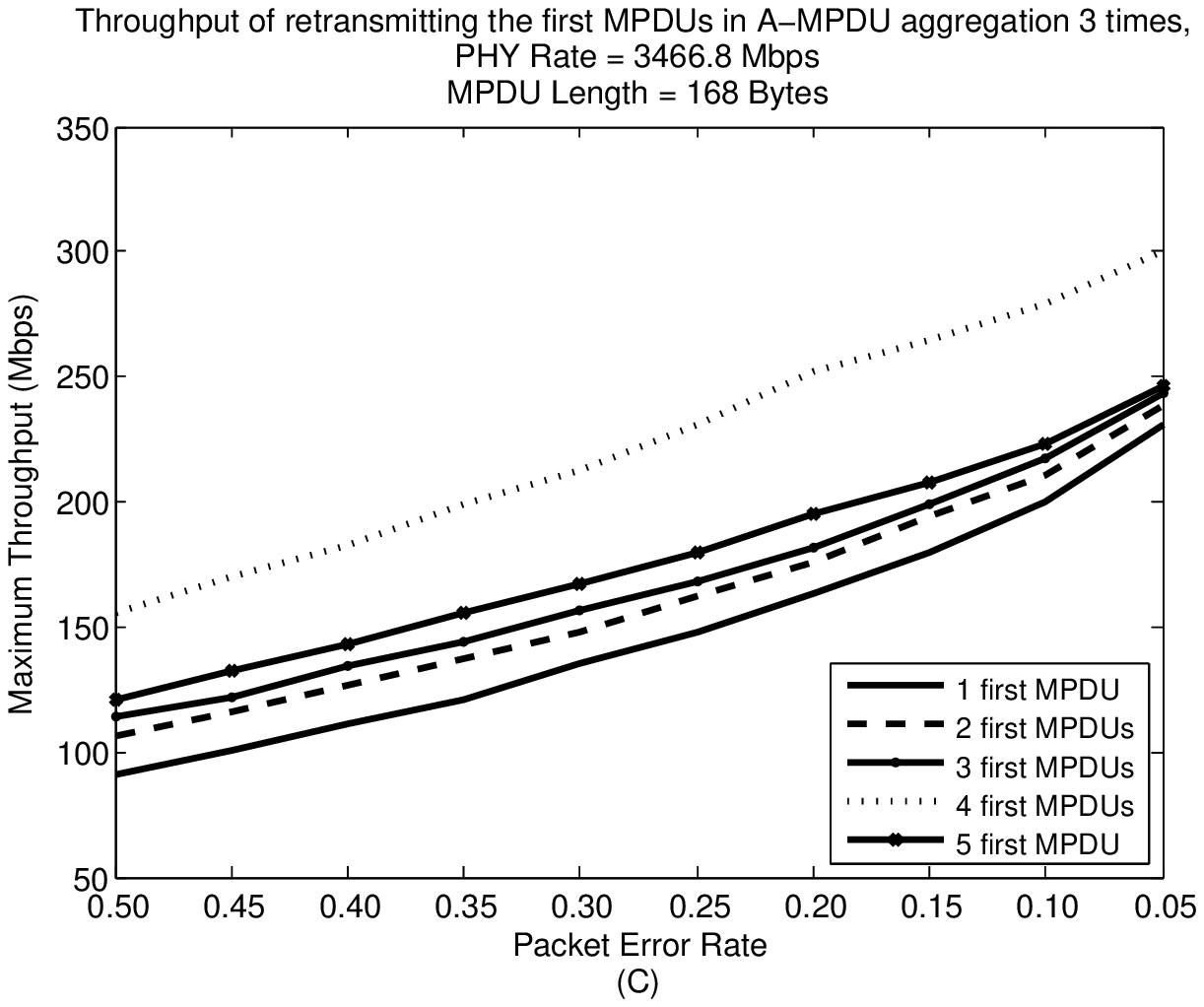}
\includegraphics{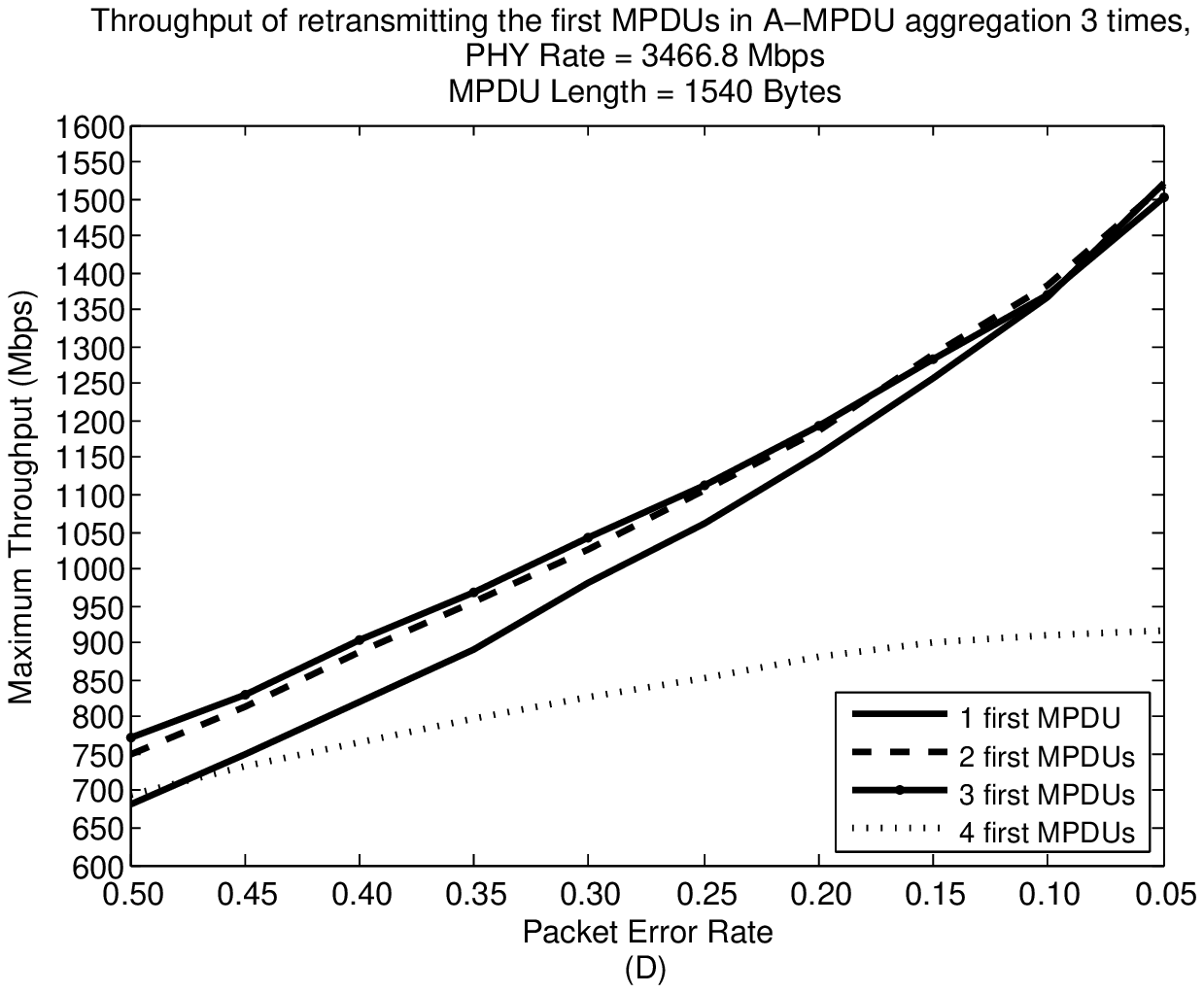}
\caption{Throughput comparison when retransmitting the first 1 to 5 MPDUs in
an A-MPDU 2 and 3 times, A-MPDU aggregation, MPDUs of 168 and 1540 bytes, PHY rate = 3466.8 Mbps }
\label{fig:fig02}
\end{figure}

In Figures~\ref{fig:fig01} and~\ref{fig:fig02}
we justify our decision to check the performance
of retransmitting only the first 1 to 4 MPDUs in A-MPDU frames.
The figures correspond to A-MPDU aggregation only,
but the same results also hold for the Two-Level aggregation.
In these figures we consider MPDUs of 168 and 1540 bytes.
In Figure~\ref{fig:fig01} we assume a PHY rate of 1299.9 Mbps
and in Figure~\ref{fig:fig02} we assume a PHY rate of 3466.8 Mbps.
We see that for MPDUs of 168 bytes, Figures~\ref{fig:fig01}(A), (C) it
is worthwhile to retransmit the first 4 MPDUs. Retransmitting the
5th achieves a smaller Throughput for all the PER values we consider.
On the other hand, for MPDUs of 1540 bytes, Figures~\ref{fig:fig01}(B),(D)
it is preferable to retransmit only the first 3 MPDUs.

These results can be explained as follows: The rationale
behind retransmitting the first MPDUs is to enable
the $TW$ to slide and thus contain new MPDUs. The probability
that the first 4 or 5 MPDUs will all corrupt is minimal;
on the other hand retransmitting MPDUs increases the
transmission time . For short MPDUs this increase is negligible. For
larger MPDUs, such as 1540 bytes, 
this increase is more significant and therefore,
retransmitting the 4th MPDU of this size is not efficient.

In Figure~\ref{fig:fig02} we show the same results for PHY rate
3466.8 Mbps. In this case the 'penalty' for retransmitting
MPDUs is lower, however the same relative results are still observed.

\subsection{Performance of {\it Set1-Set4}}

\begin{figure}
\vskip 16cm
\includegraphics{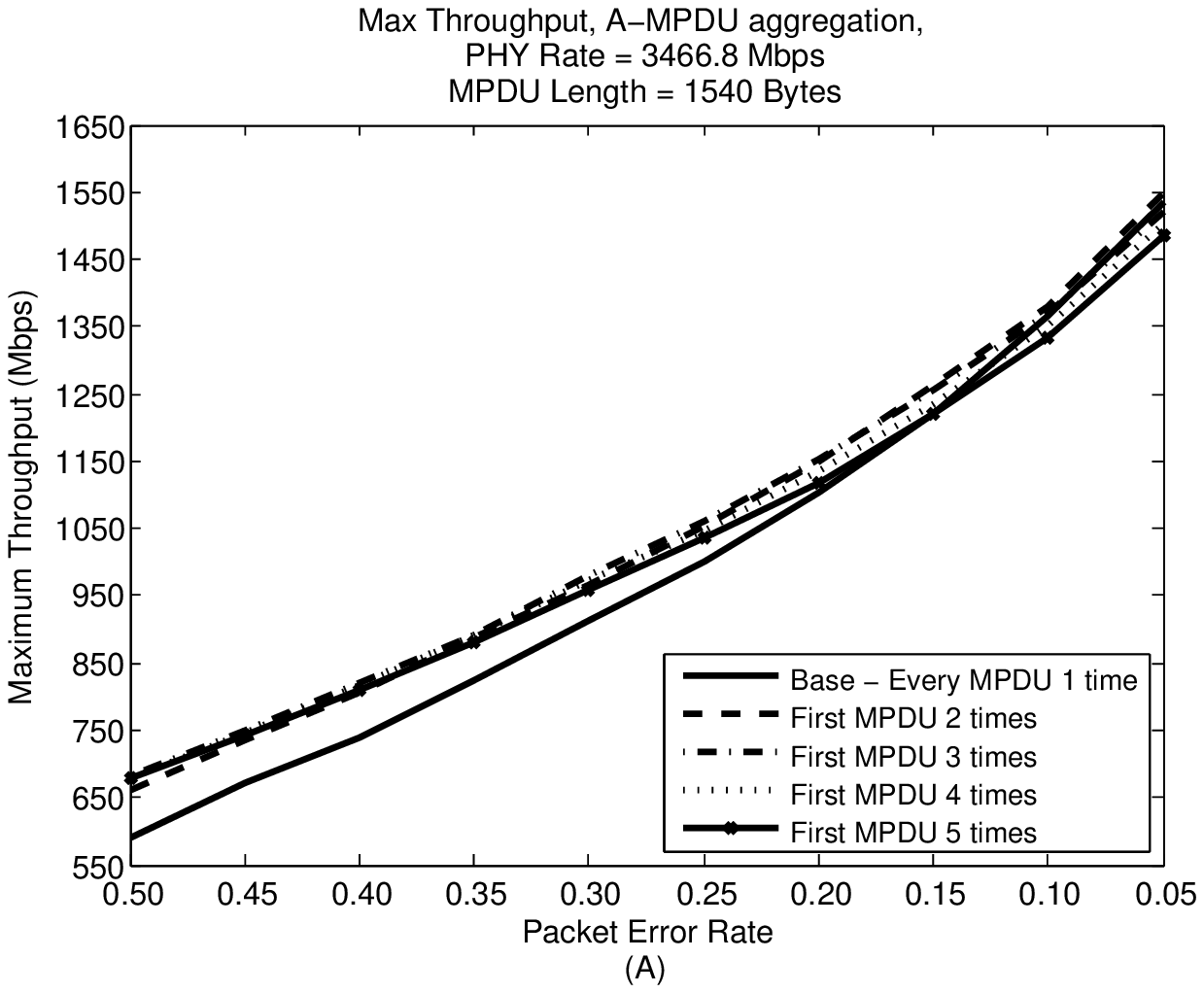}
\includegraphics{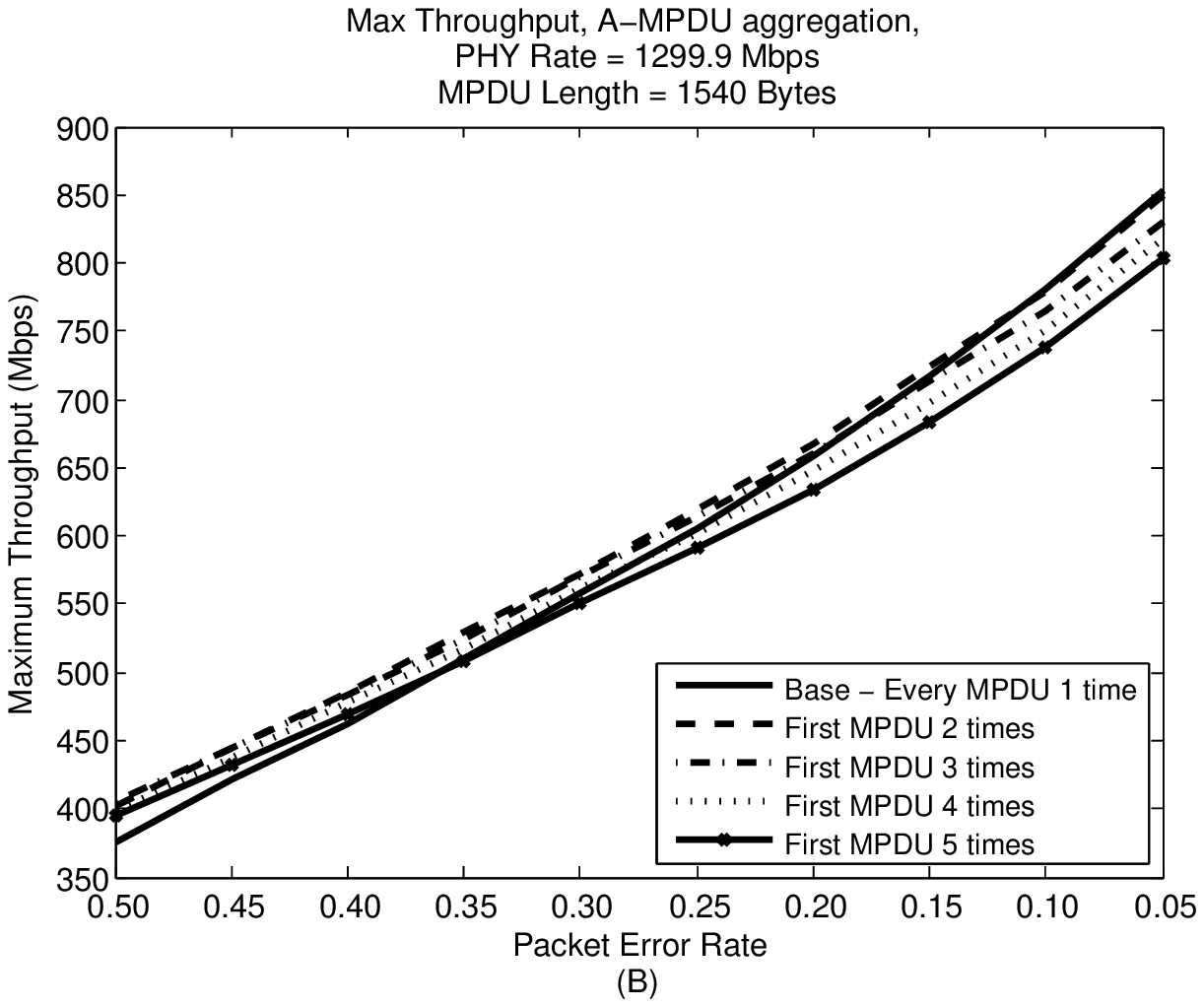}
\includegraphics{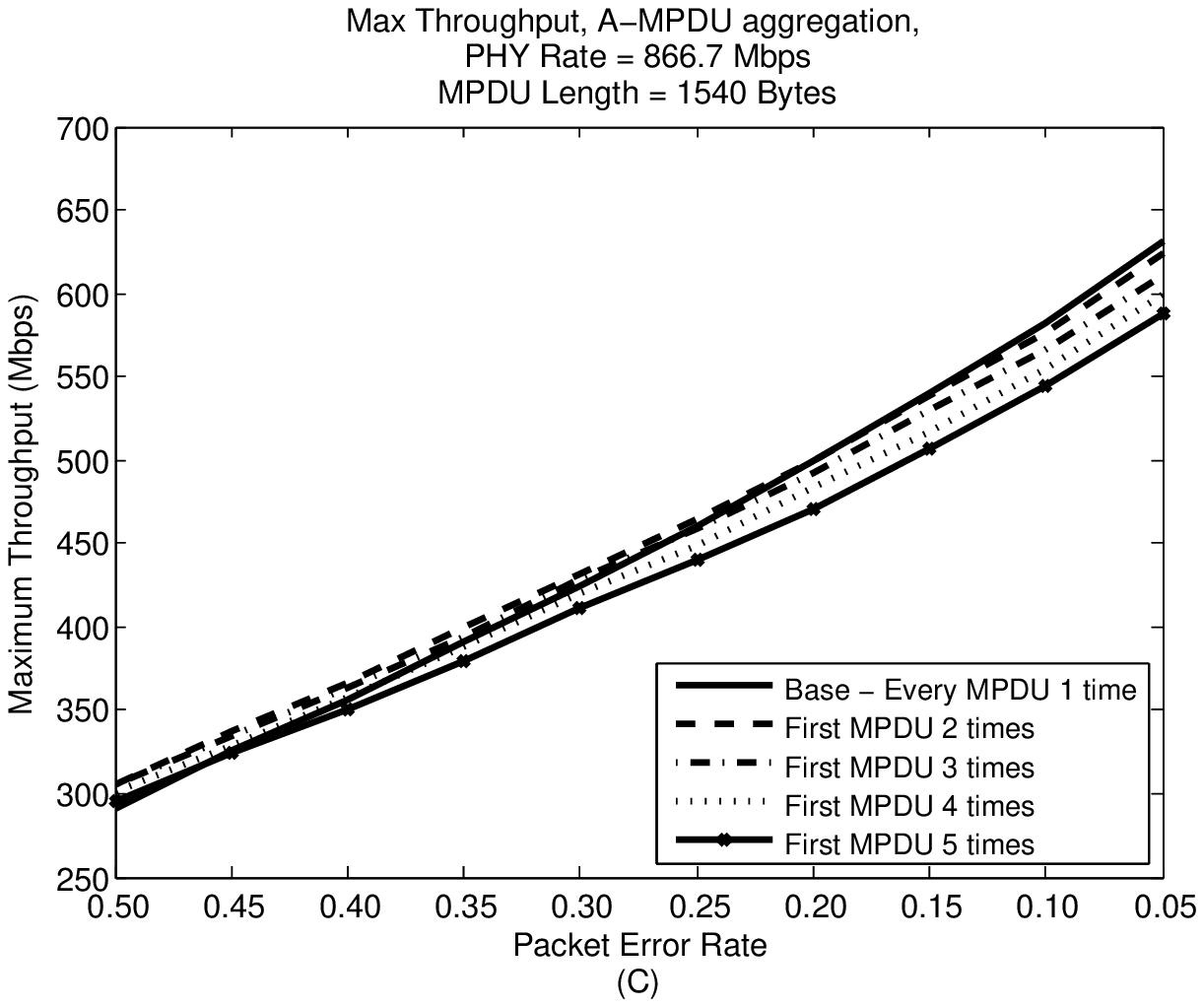}
\includegraphics{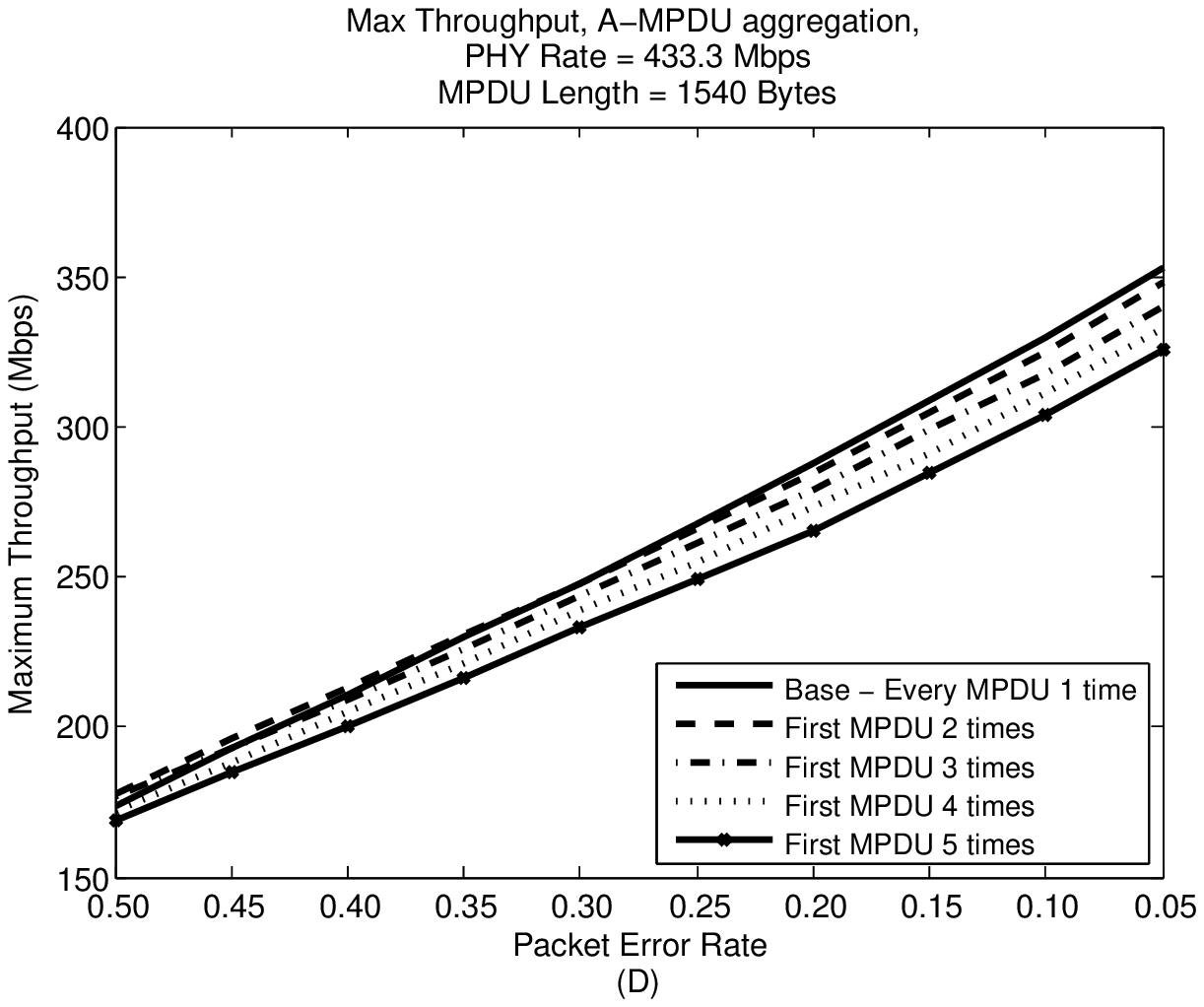}
\caption{Maximum Throughput when retransmitting the first MPDU in window,  A-MPDU aggregation, MPDU size 1540 bytes}
\label{fig:fig1}
\end{figure}

\begin{figure}
\vskip 16cm
\includegraphics{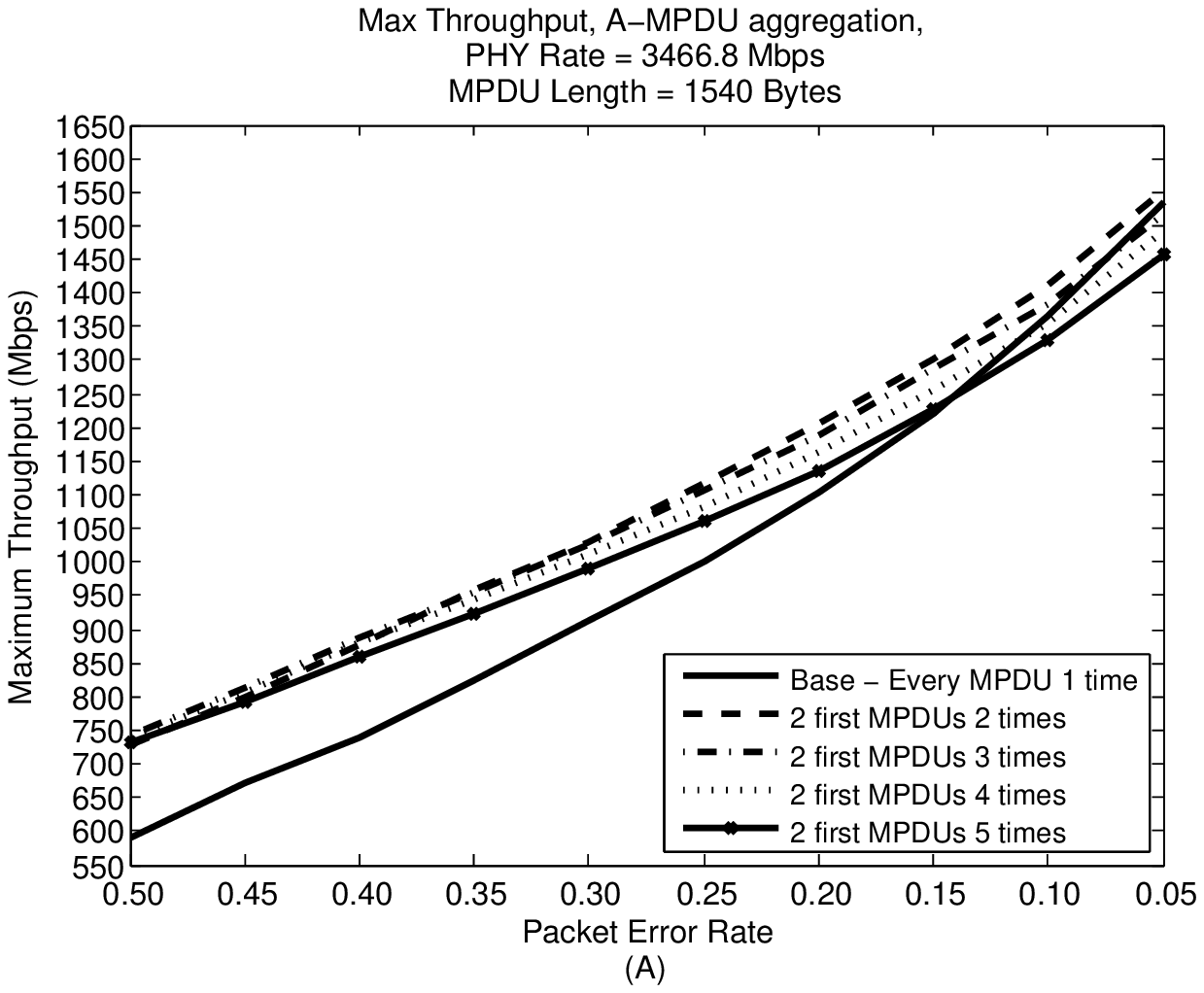}
\includegraphics{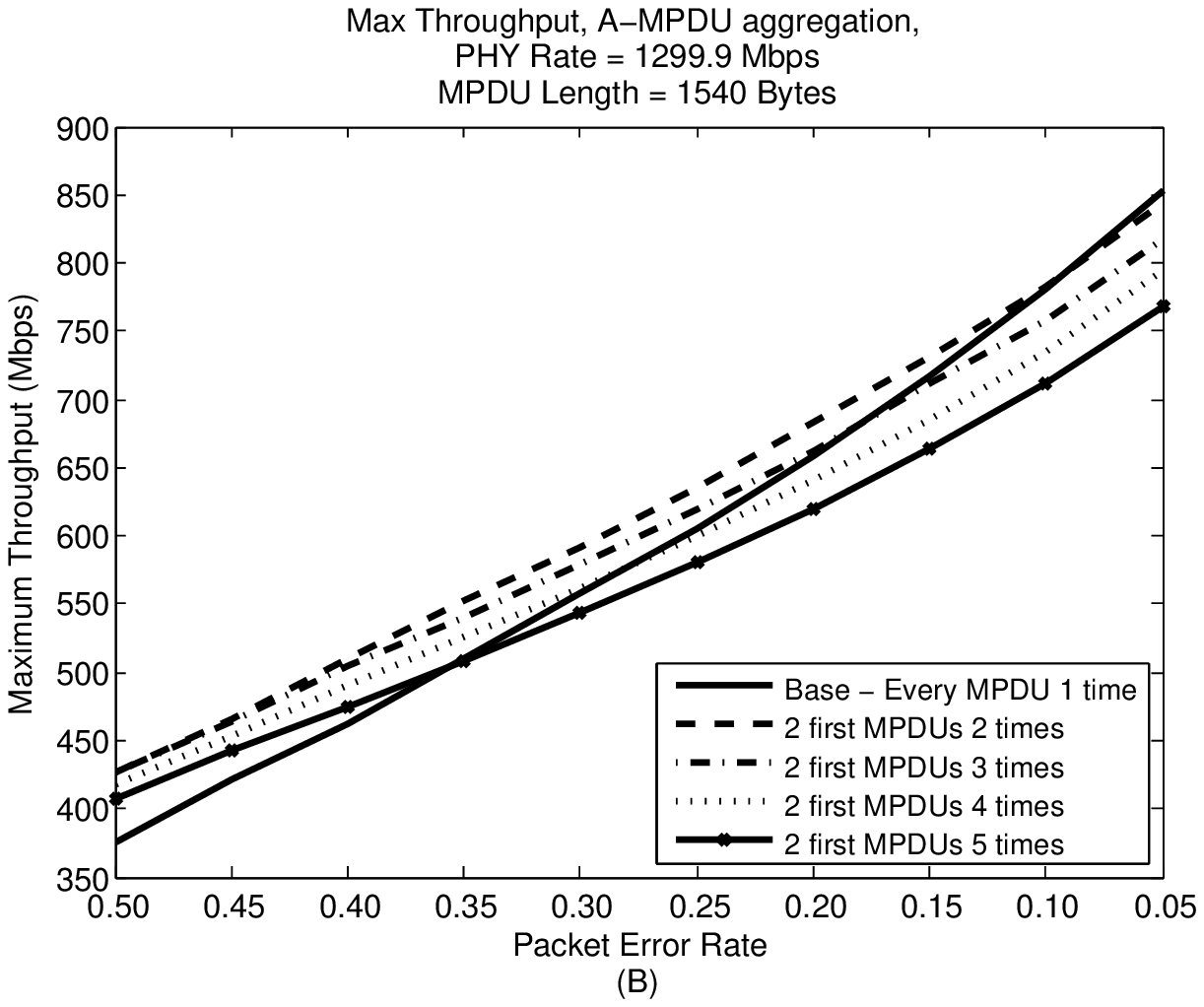}
\includegraphics{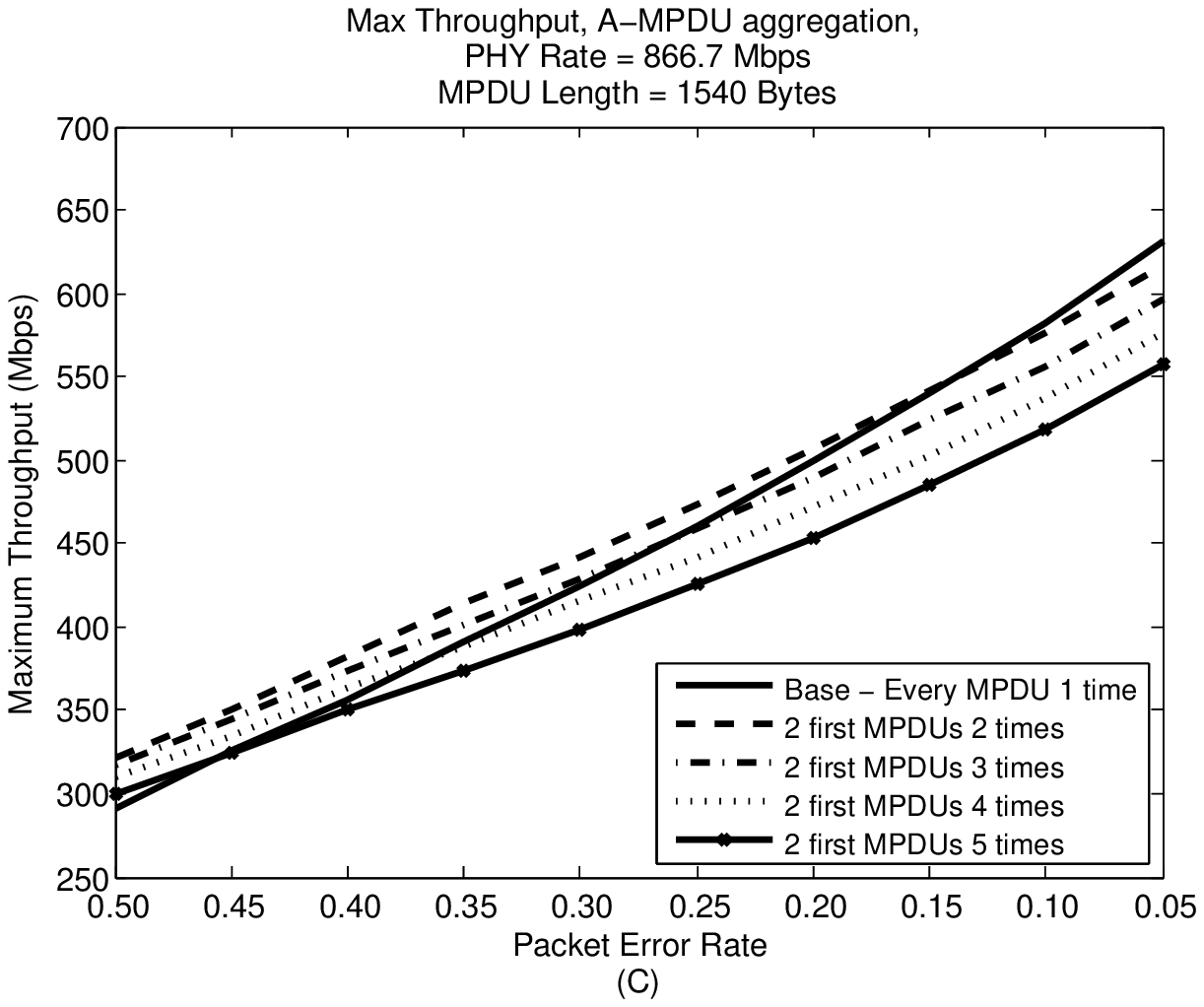}
\includegraphics{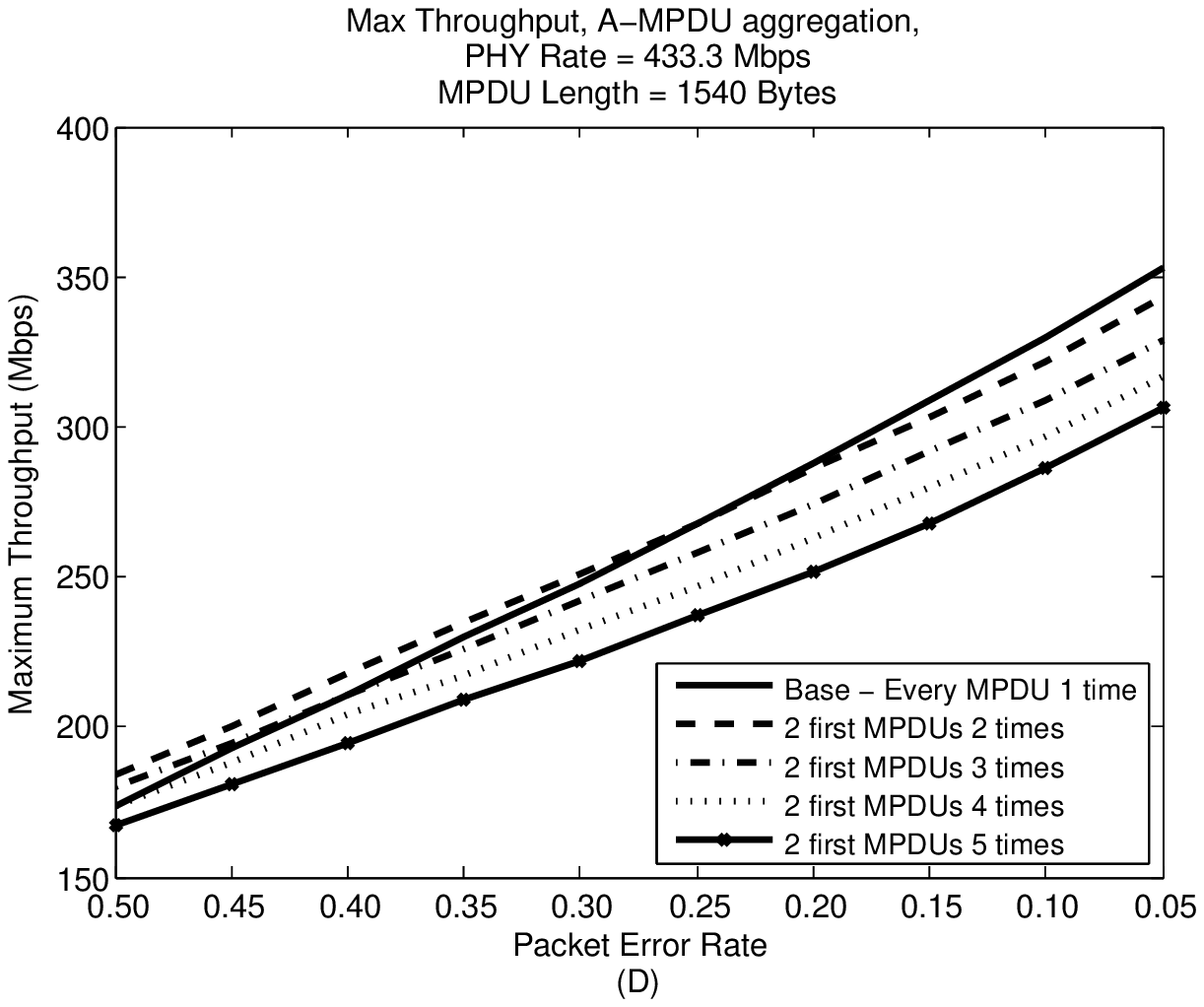}
\caption{Maximum Throughput when retransmitting the first 2 MPDUs in window,  A-MPDU aggregation, MPDU size 1540 bytes}
\label{fig:fig2}
\end{figure}

\begin{figure}
\vskip 16cm
\includegraphics{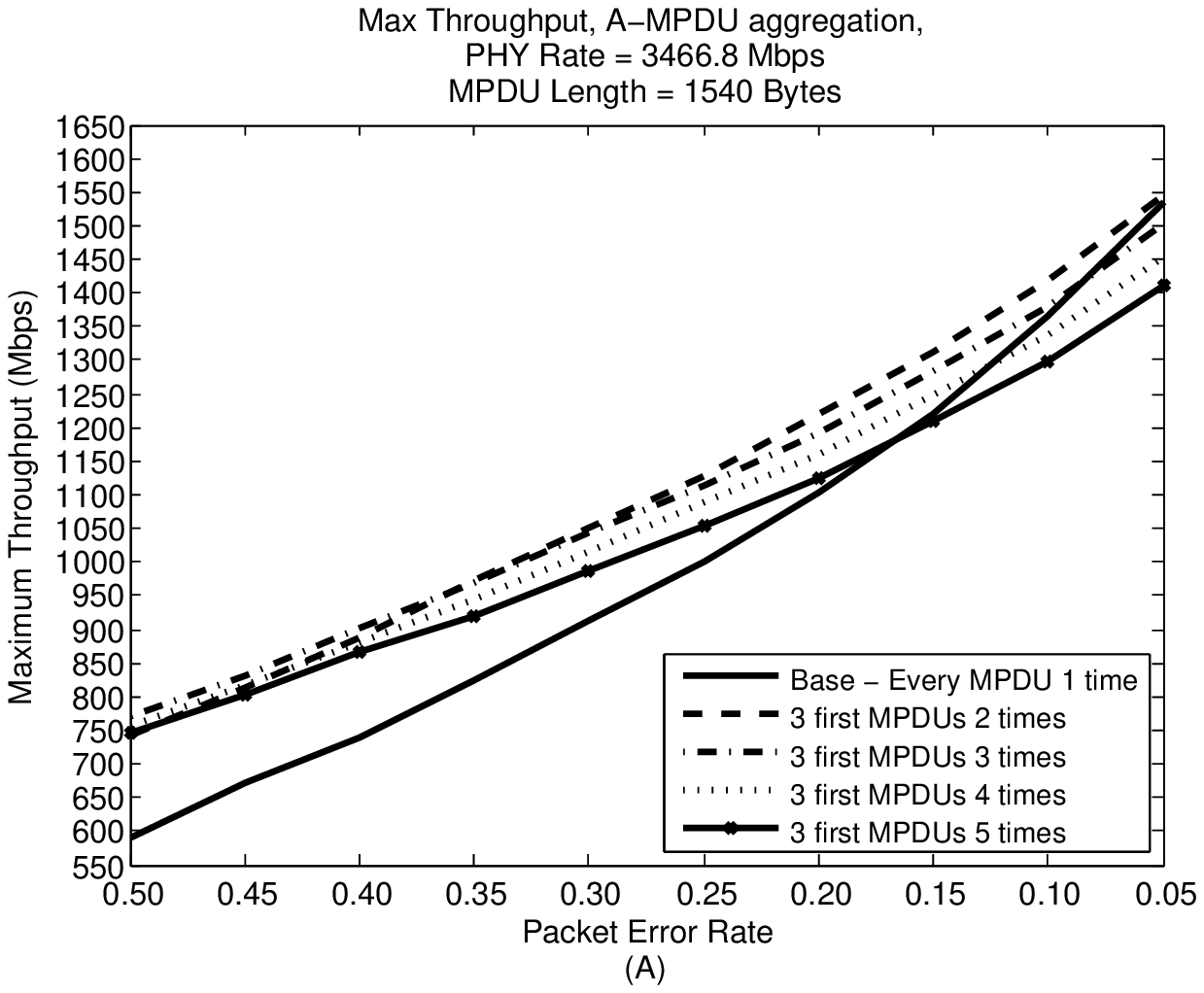}
\includegraphics{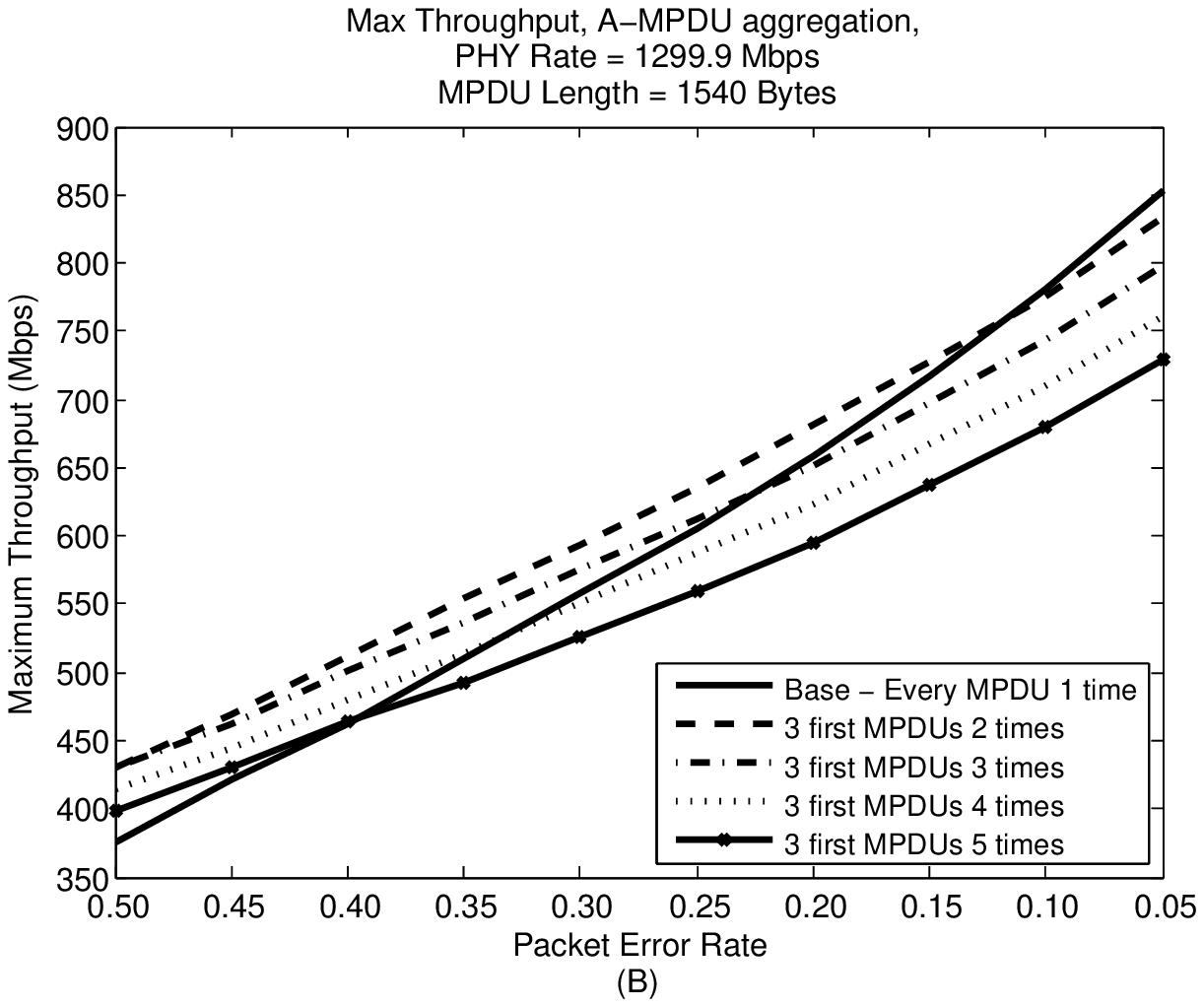}
\includegraphics{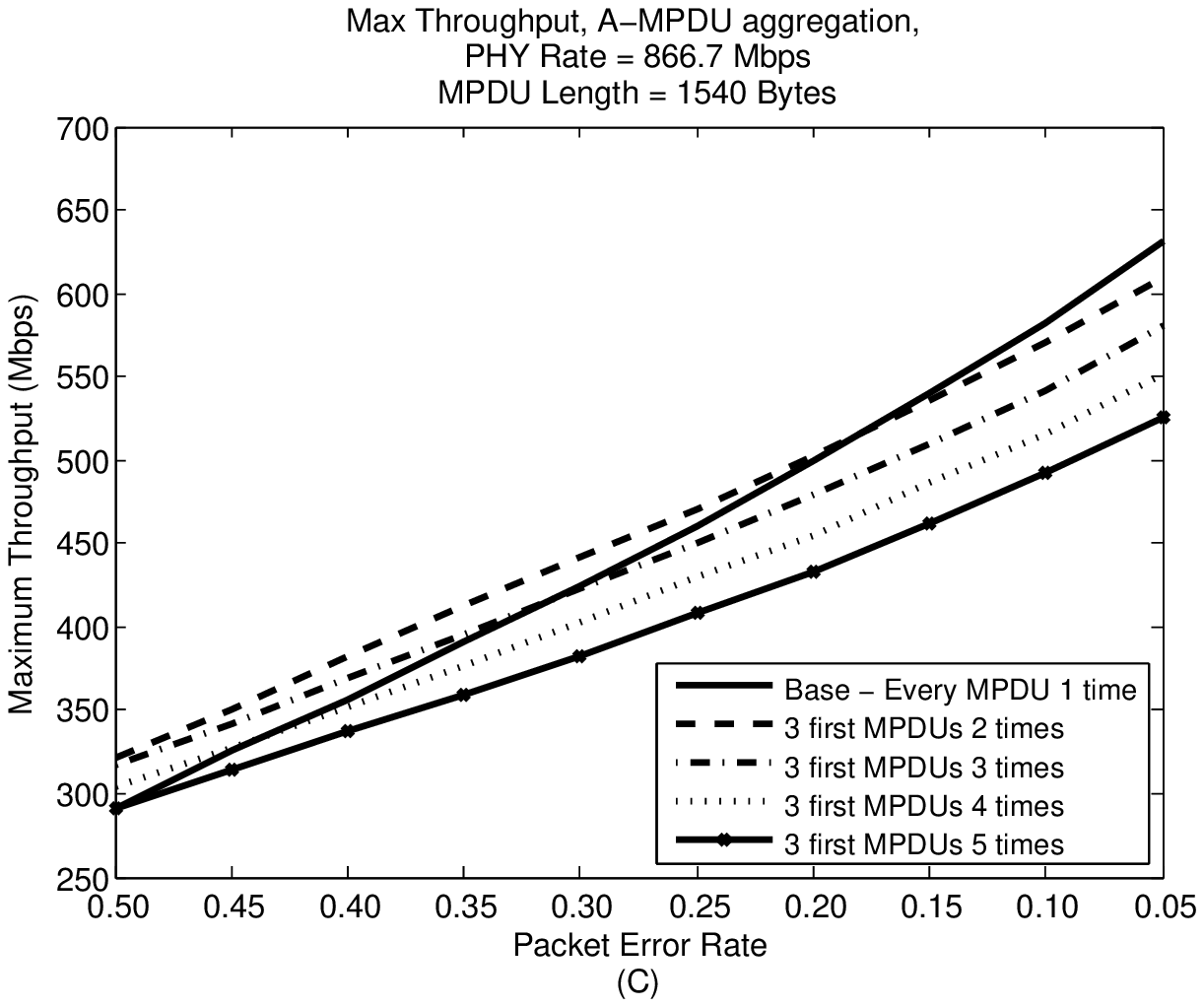}
\includegraphics{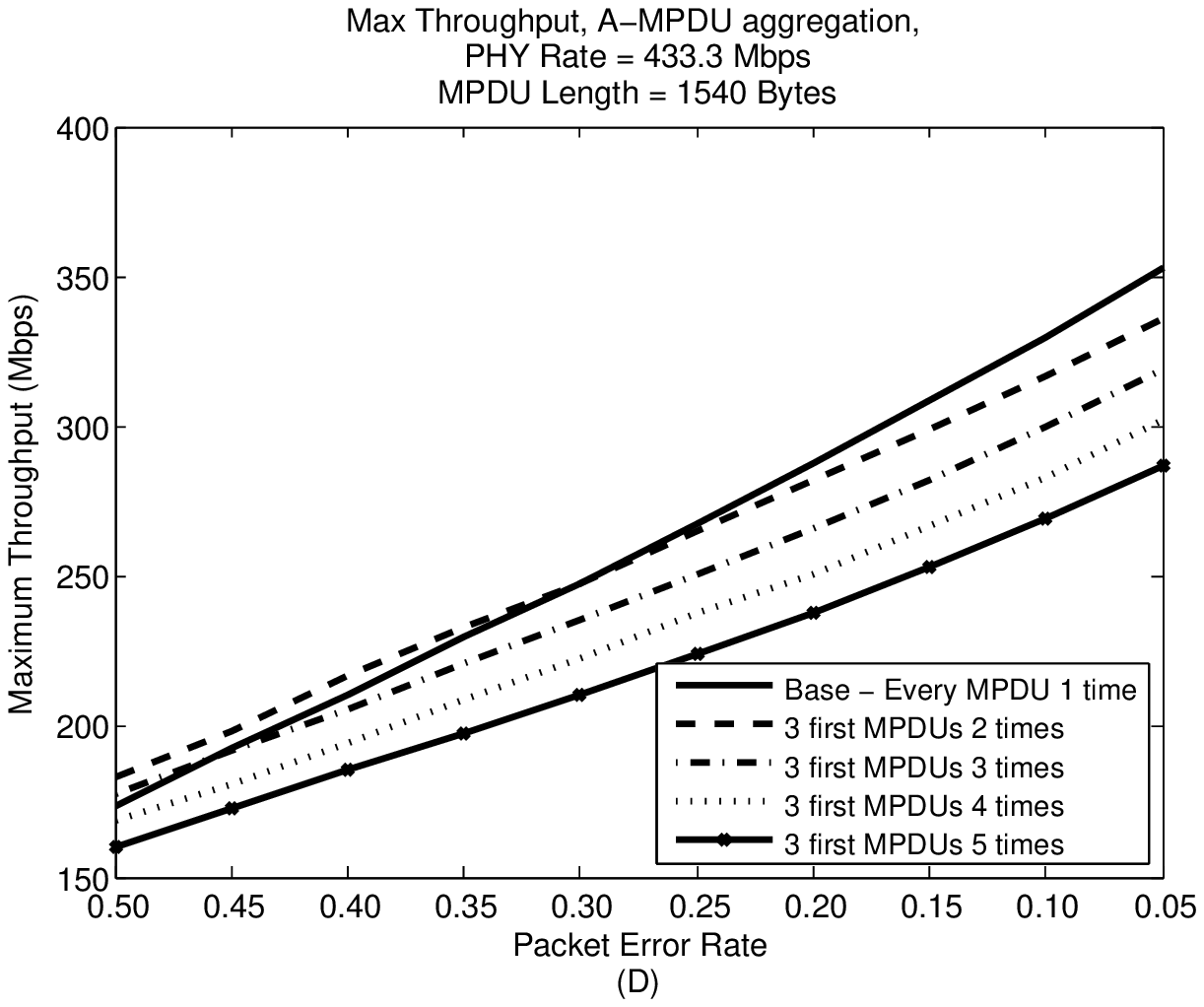}
\caption{Maximum Throughput when retransmitting the first 3 MPDUs in window, A-MPDU aggregation, MPDU size 1540 bytes}
\label{fig:fig3}
\end{figure}

The next set of results corresponds to the A-MPDU aggregation
scheme.
In Figure~\ref{fig:fig1} we consider the first set of
methods, {\it Set1}, in which only $MPDU_{min}$ is retransmitted. We assume
MPDUs of 1540 bytes.
We also consider 4 PHY rates: 3466.8, 1299.9, 866.7 and
433.3 Mbps, in Figures~\ref{fig:fig1}(A), (B), (C) and (D)
respectively.

The curve of the $Base$ method, in which every
MPDU is transmitted once, is the best
for PHY rates 866.7 and 433.3 Mbps except for
very large PERs.
Therefore, it can be concluded that
in these PHY rates, none of the methods
in {\it Set1} significantly improve 
the Throughput of the $Base$ method.
Notice that all the methods in {\it Set1} attempt to improve
the Throughput by transmitting $MPDU_{min}$ several times,
but this is achieved at the cost of larger PSDUs and
longer transmission times. In relatively small PHY rates
this cost is high, and is not worth the improvement
achieved by retransmitting $MPDU_{min}$.

For 3466.8 and 1299.9 Mbps {\it 1MPDU2} achieves the largest
Throughputs. In PER$=$0.5 the improvements over the $Base$
method are 12$\%$ and 5$\%$ respectively.

In Figure~\ref{fig:fig2} we show 
the same results as in Figure~\ref{fig:fig1},
but for {\it Set2} of methods. For PHY rates 866.7 and 433.3 Mbps
there is a more significant
improvement over the $Base$ method compared to {\it Set1}, but for
the larger rates of 3466.8 and 1299.9 Mbps the improvement
is much more significant: 25$\%$ and 15$\%$ for PER$=$0.5 respectively.
Again, the retransmission of size 2, i.e. {\it 2MPDU2} is the best method.

In Figure~\ref{fig:fig3} we show the results for {\it Set3} of methods.
The results are similar to those for {\it Set1} and {\it Set2}. Notice
that for PER$=$0.5 the improvements in Throughputs over the $Base$ method
are 30$\%$ and 17$\%$ in the PHY rates 3466.8 and 1299.9 Mbps
respectively.

\begin{figure}
\vskip 16cm
\includegraphics{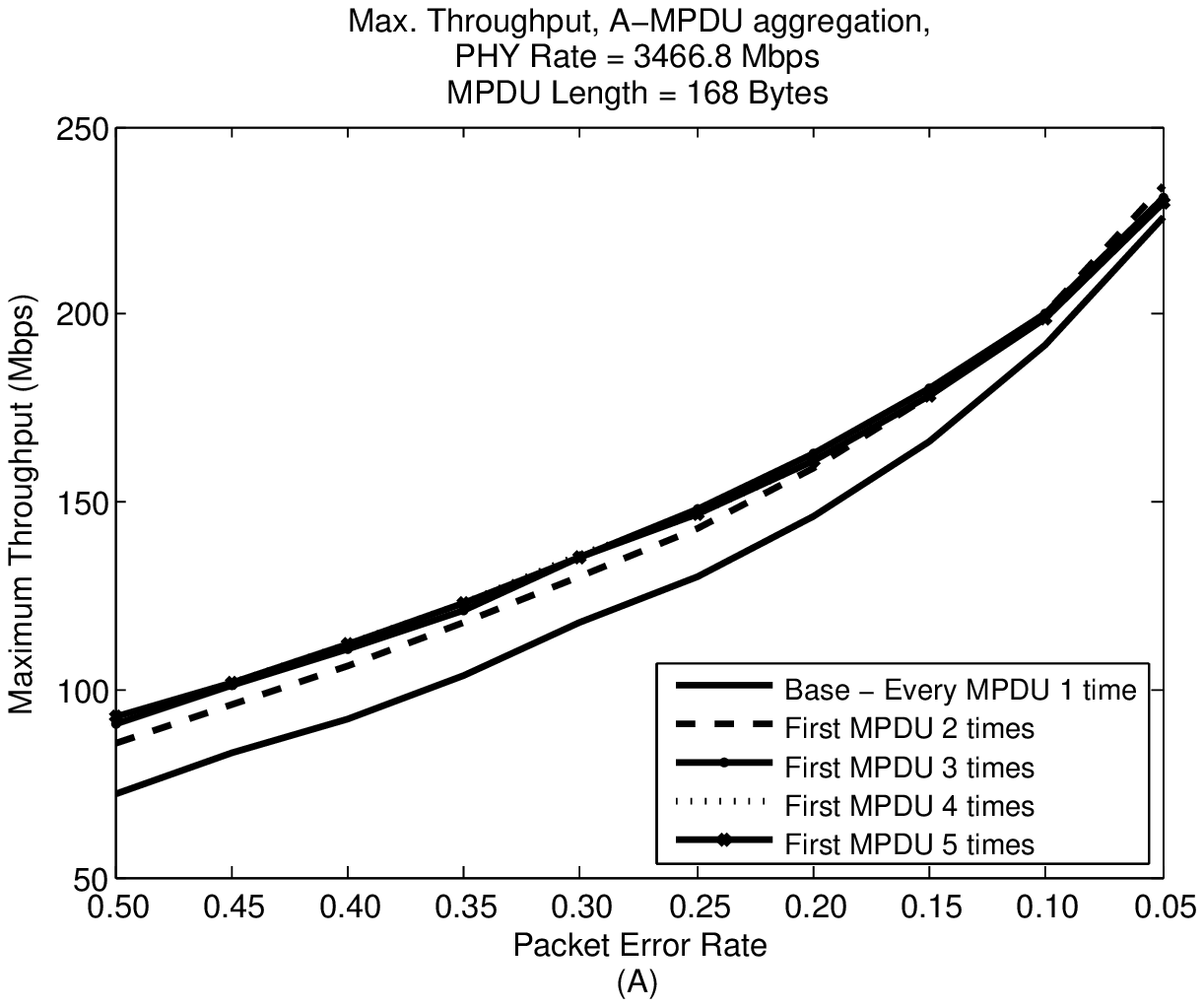}
\includegraphics{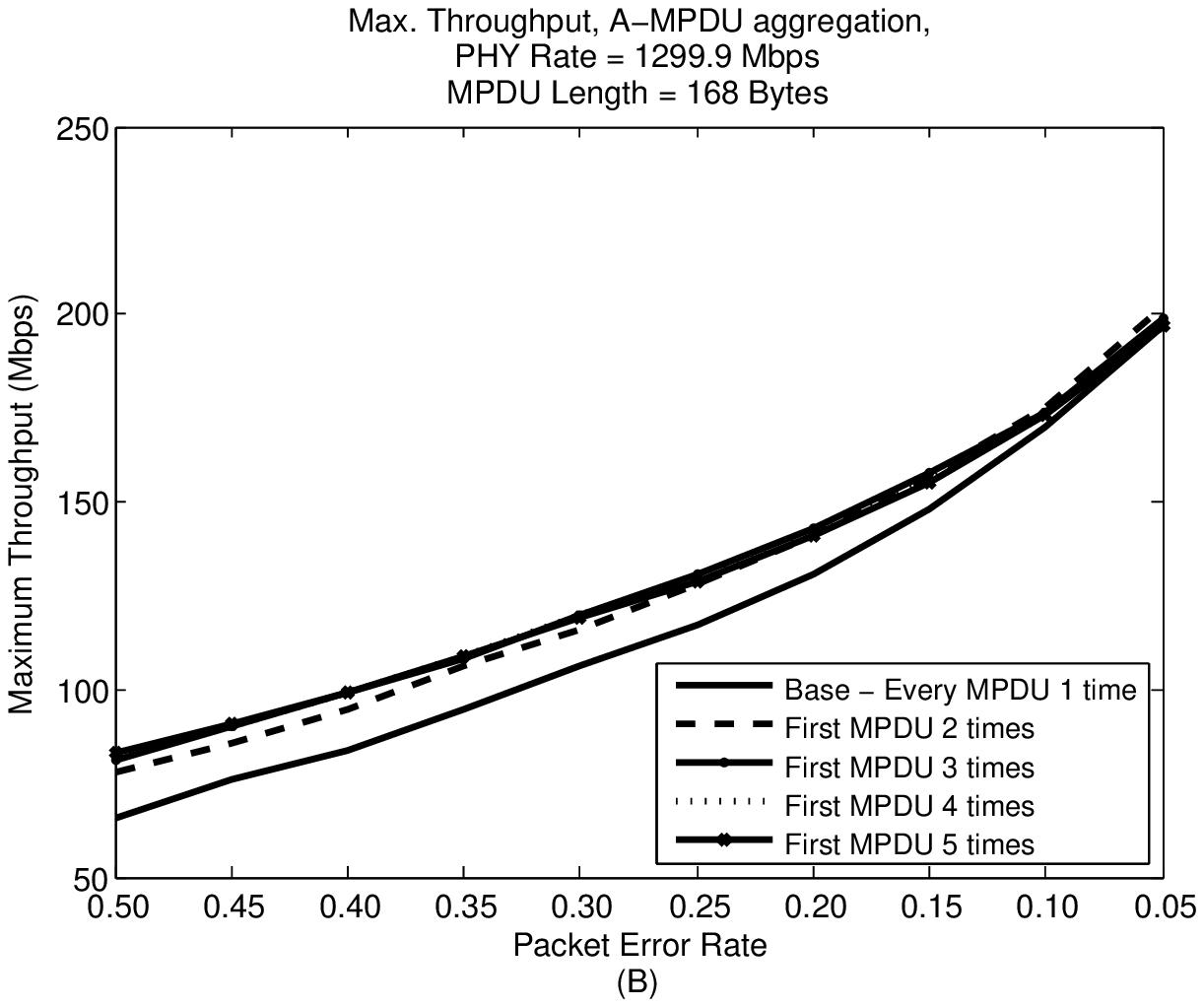}
\includegraphics{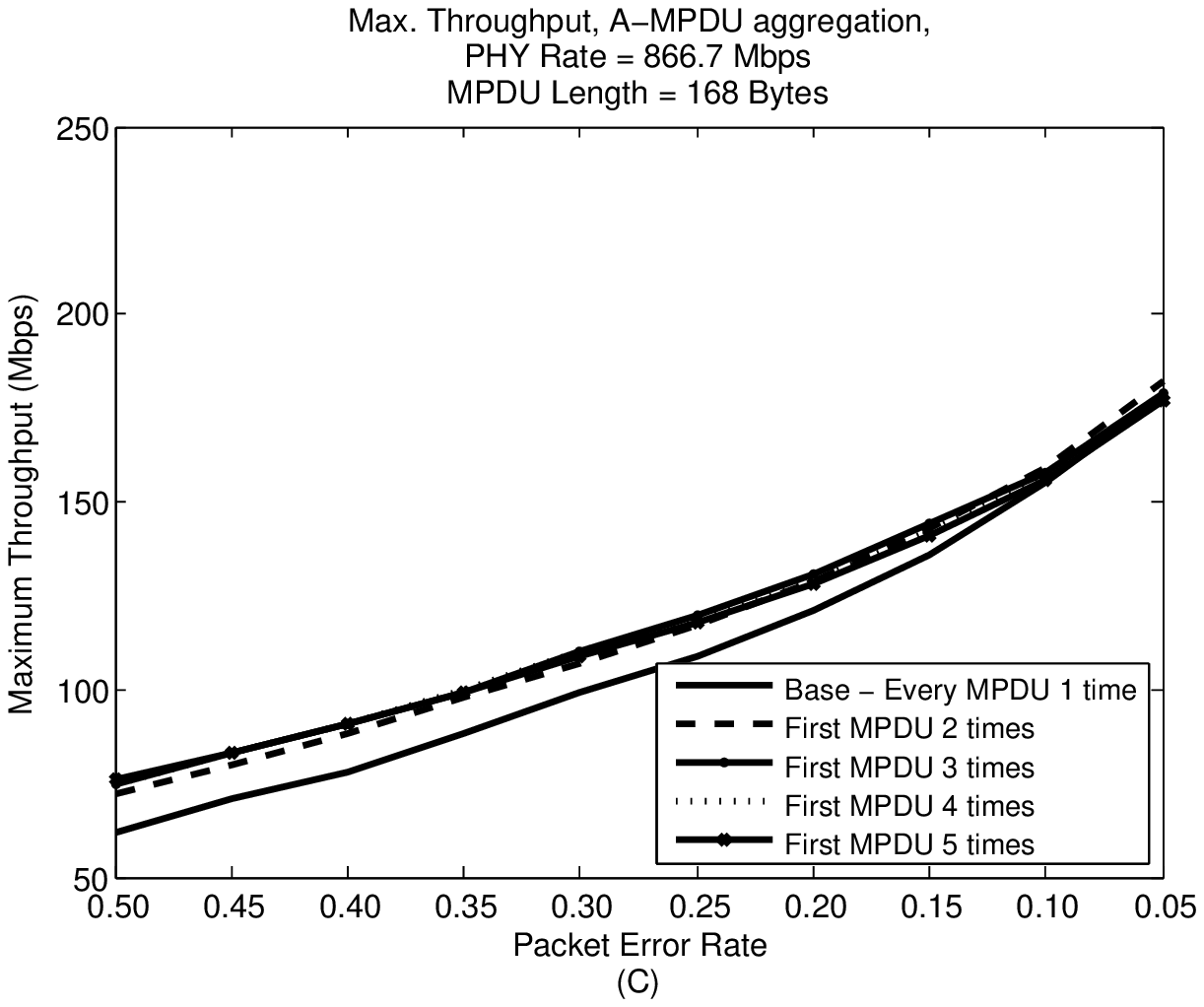}
\includegraphics{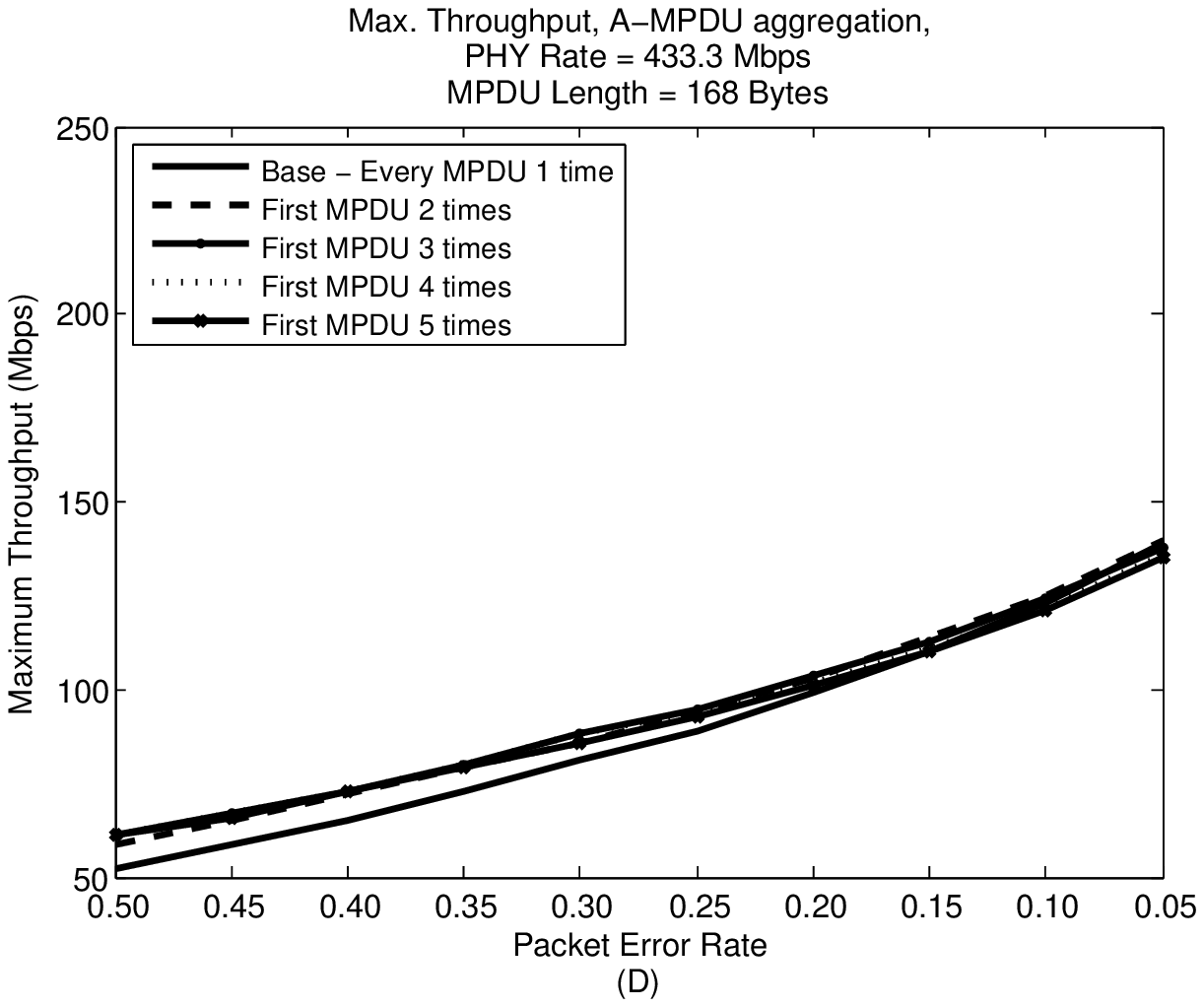}
\caption{Maximum Throughput when retransmitting the first MPDU in window, A-MPDU aggregation, MPDU size 168 bytes}
\label{fig:fig11}
\end{figure}

\begin{figure}
\vskip 16cm
\includegraphics{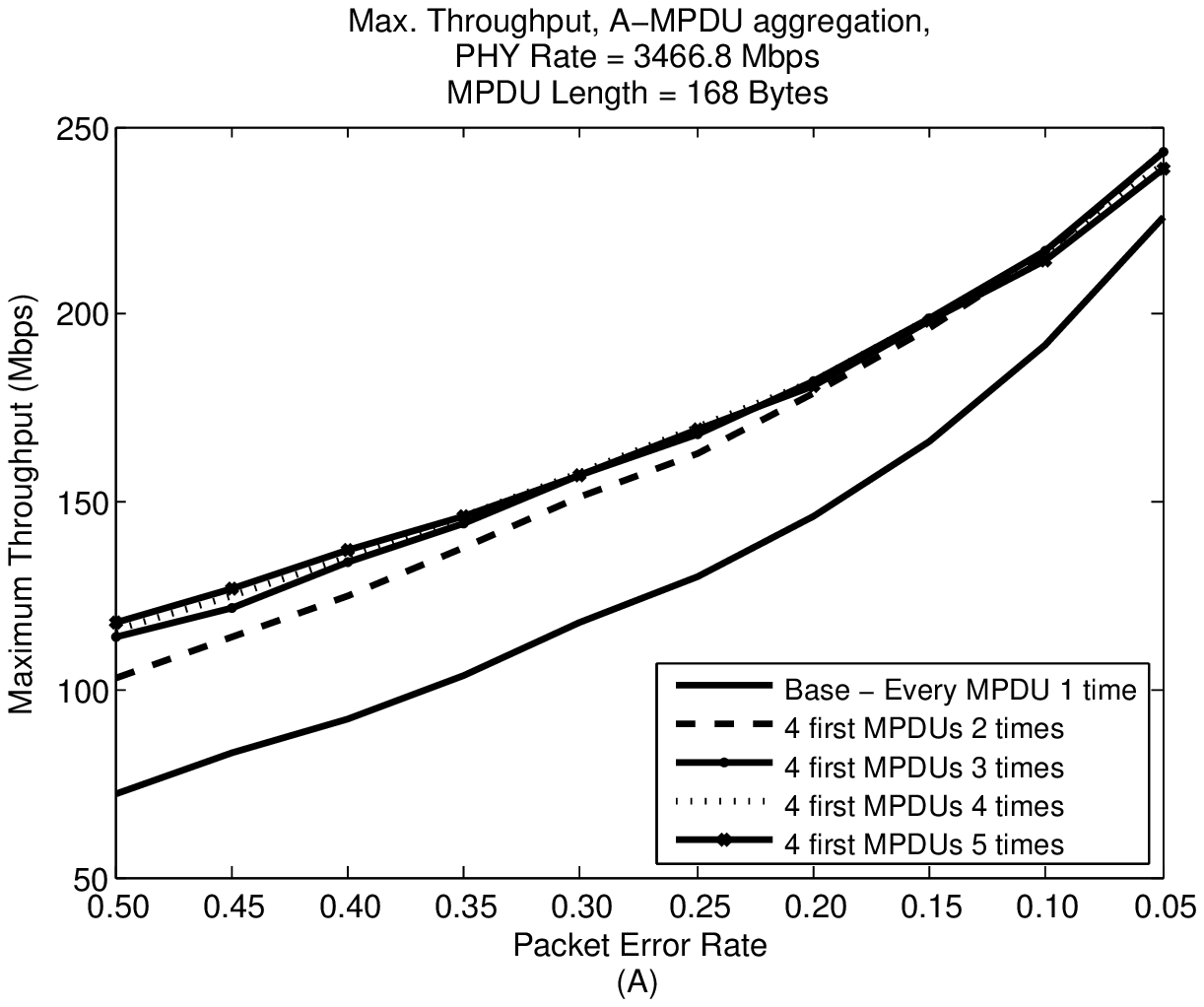}
\includegraphics{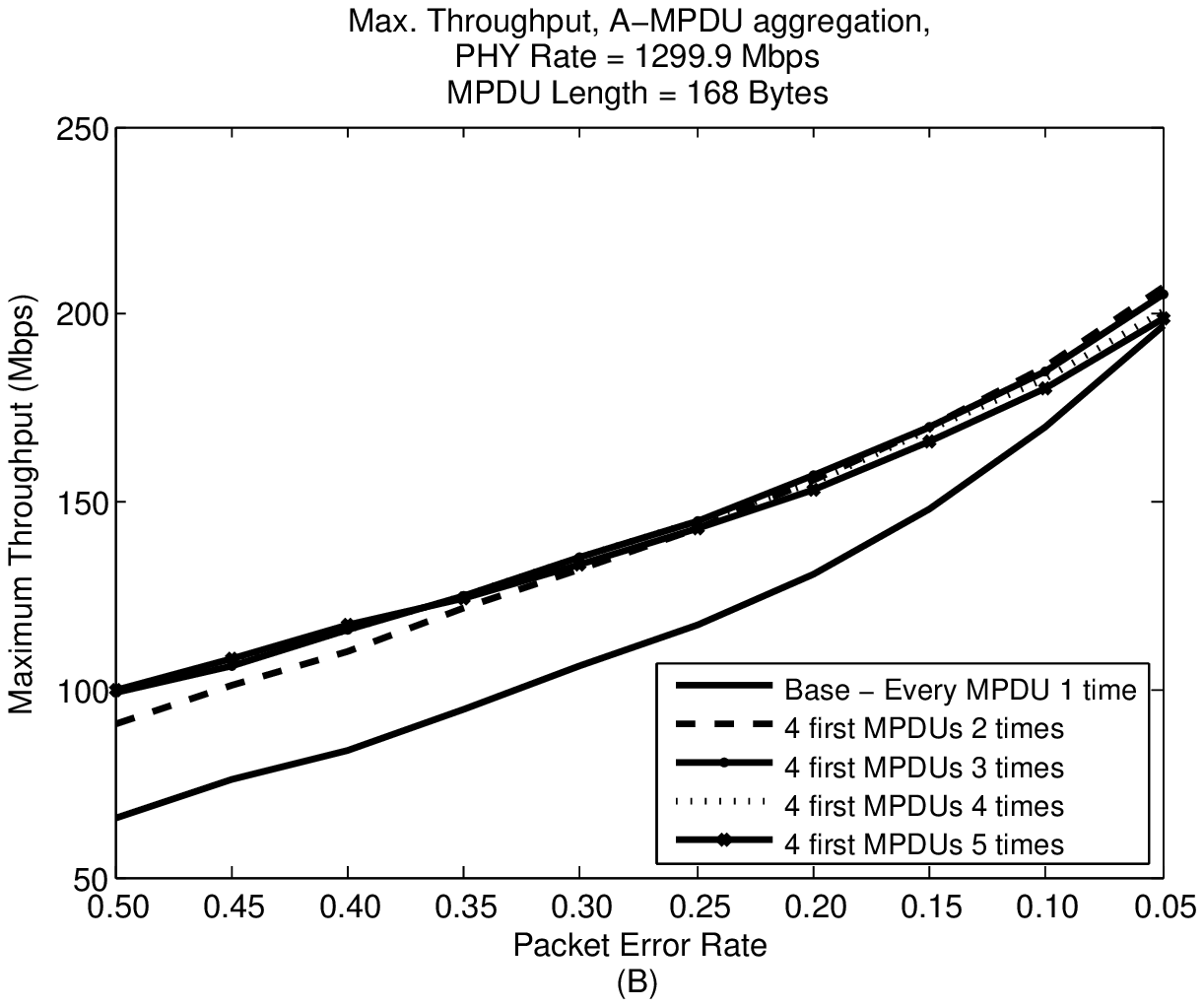}
\includegraphics{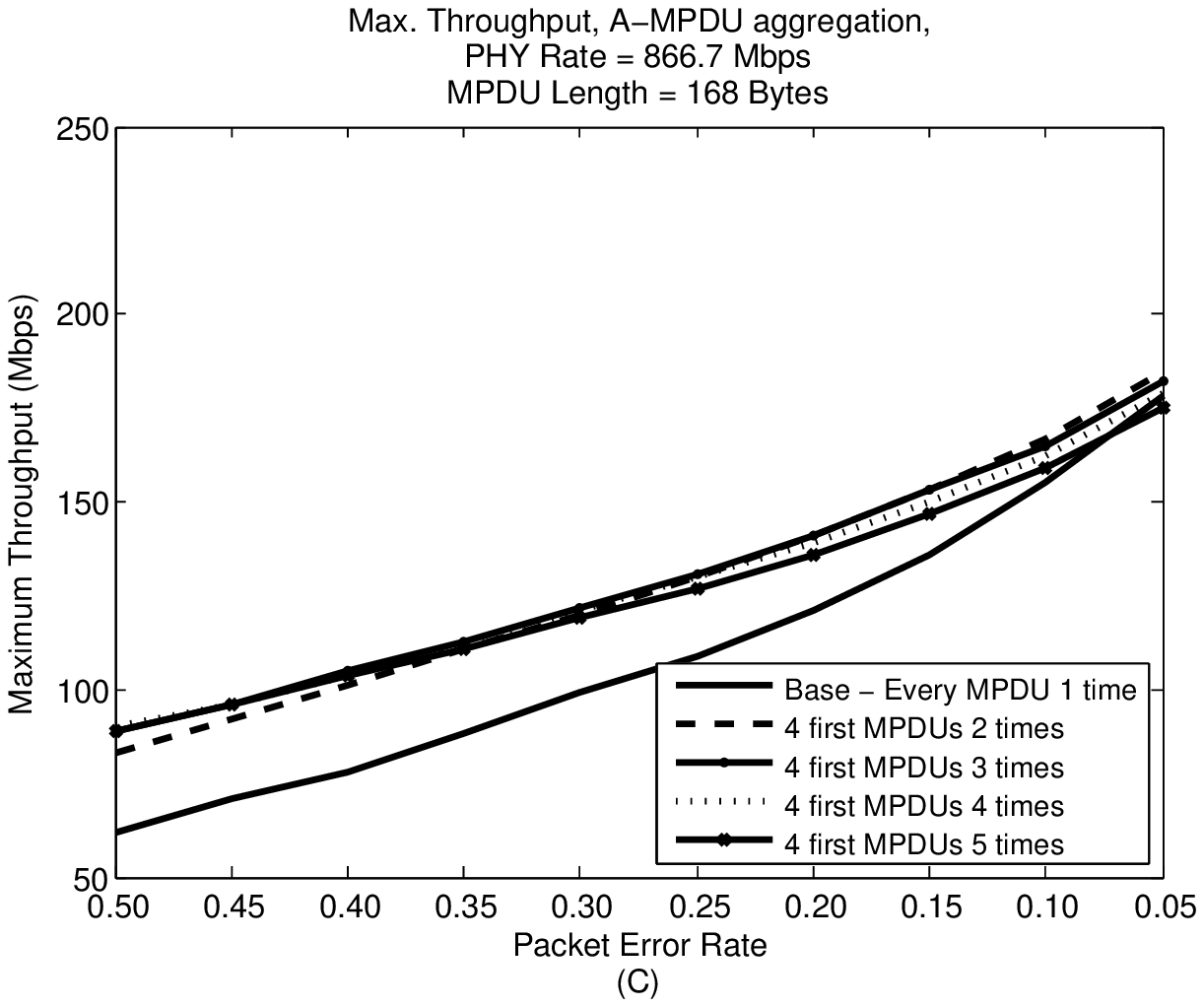}
\includegraphics{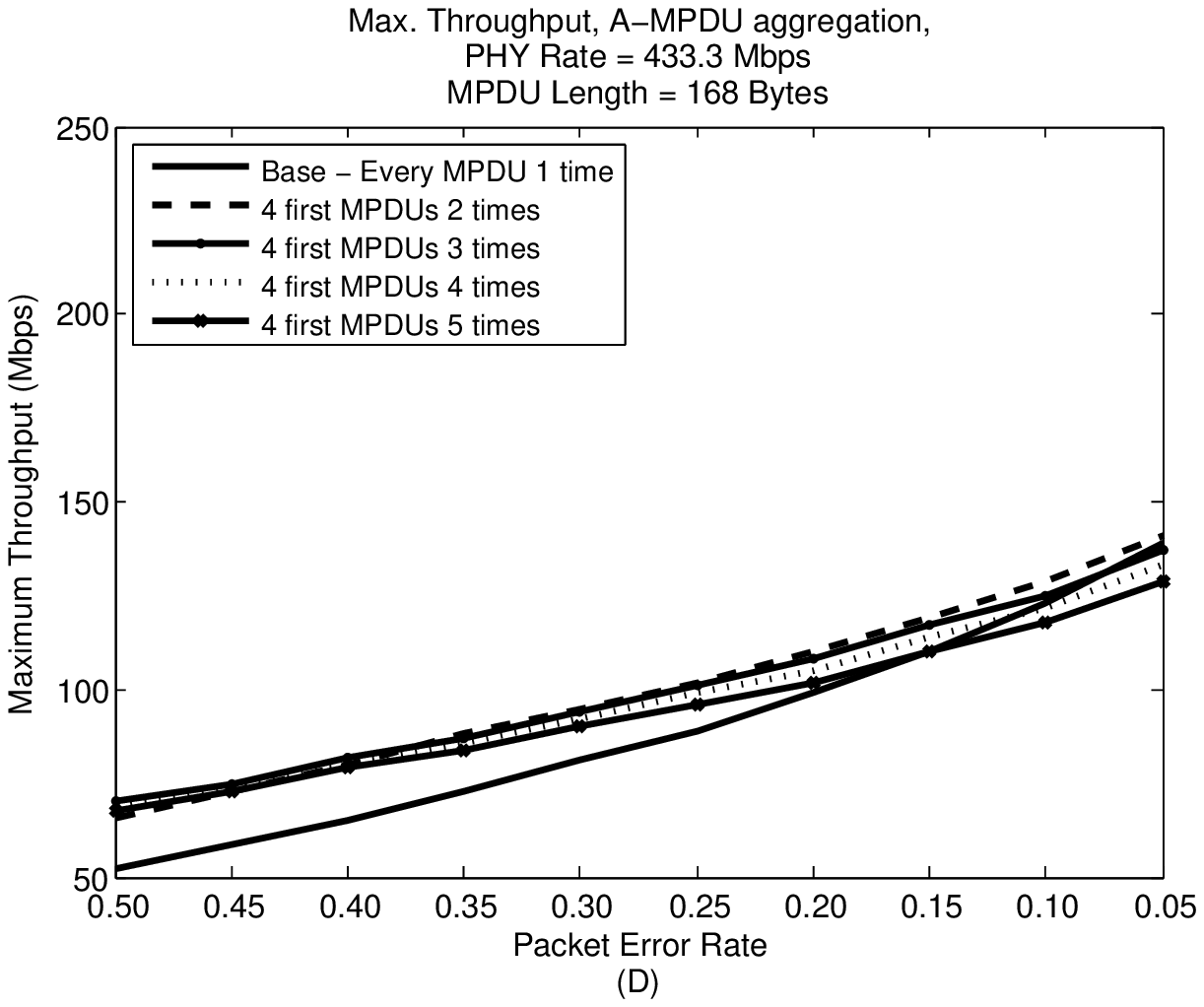}
\caption{Maximum Throughput when retransmitting the first 4 MPDUs in window, A-MPDU aggregation, MPDU size 168 bytes}
\label{fig:fig13}
\end{figure}

In Figures~\ref{fig:fig11} and~\ref{fig:fig13} we show the results
for {\it Set1} and {\it Set4} respectively, when the MPDU size is 168 bytes.
We omit the results for {\it Set3} and {\it Set4} as they fall between
{\it Set1} and {\it Set4}.

One can see that in all the PHY rates there is an improvement
in the Throughput over the $Base$ method because the MPDUs
are relatively short. Therefore, the penalty due to 
retransmissions is small compared to MPDUs of 1540 bytes.
In {\it Set1}, Figure~\ref{fig:fig11}, the maximum Throughput is always
received for {\it 1MPDU5}, i.e. it is worthwhile transmitting $MPDU_{min}$
several times. For PHY rates 3466.8 and 1299.9 Mbps, and
PER$=0.5$, the improvements in the Throughput over the $Base$
method are 29$\%$  and 25$\%$ respectively ! When considering {\it Set4}, 
Figure~\ref{fig:fig13}, it is not always worth transmitting the first
4 MPDUs 5 times when the PER is relatively low. The additional
transmission time in these cases does not warrant the improvement
in the Throughput. For low PERs it is worthwhile transmitting the first
4 MPDUs twice. For PER$=0.5$ and PHY rates 3466.8 and 1299.9 Mbps
the improvements in Throughput are 63$\%$ and 51$\%$  respectively !

The conclusion from this set of results are that small
size MPDUs benefit more from {\it Set1-Set4} than large size
MPDUs over a large scale of PHY rates, while large size
MPDUs are beneficial only in large PHY rates. As a rule,
transmitting 2 copies of MPDUs, i.e. {\it 1MPDU2, 2MPDU2, 3MPDU2} and
{\it 4MPDU2} is the most efficient method.

\subsection{Performance of {\it Set5}}

\begin{figure}
\vskip 16cm
\includegraphics{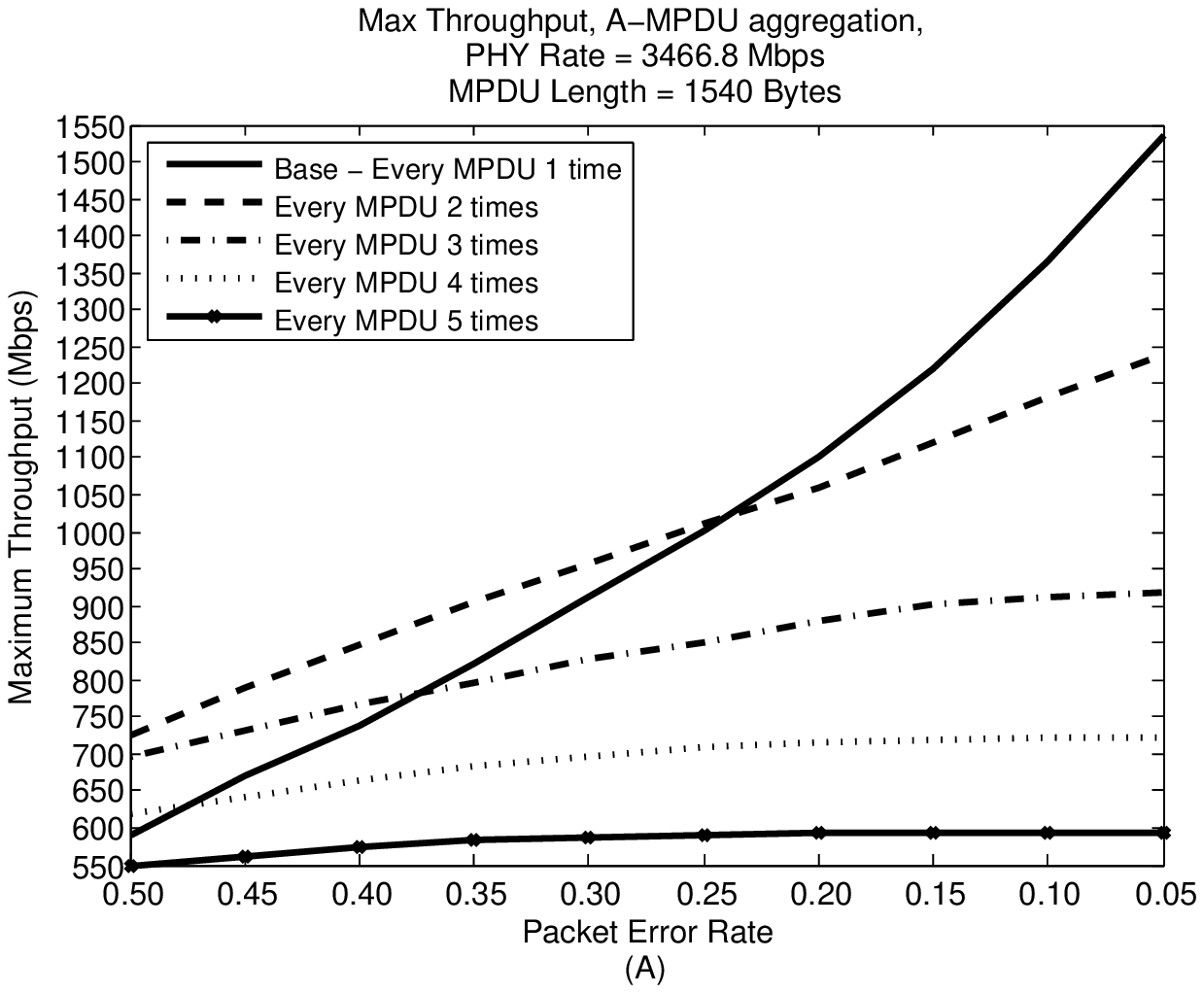}
\includegraphics{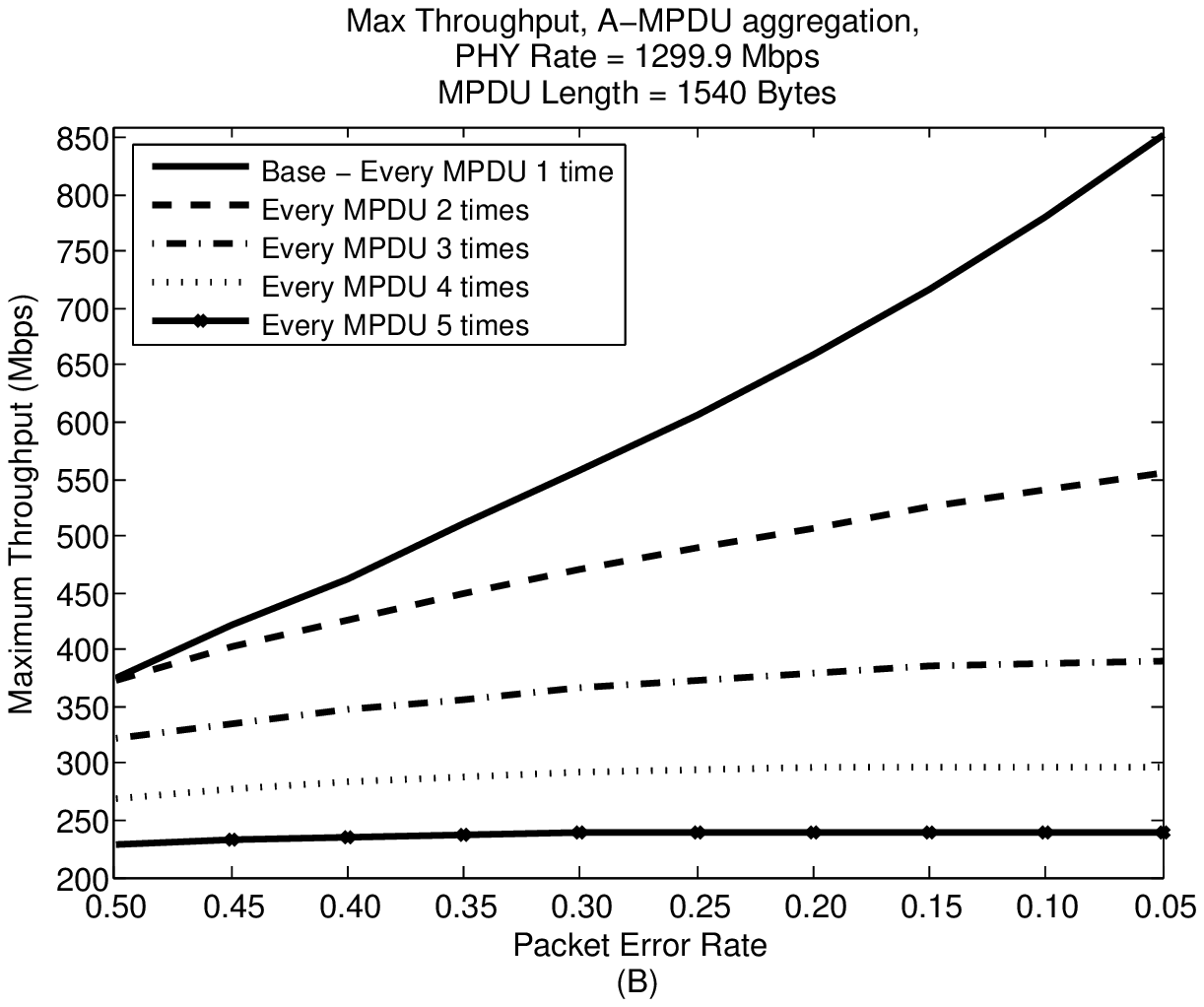}
\includegraphics{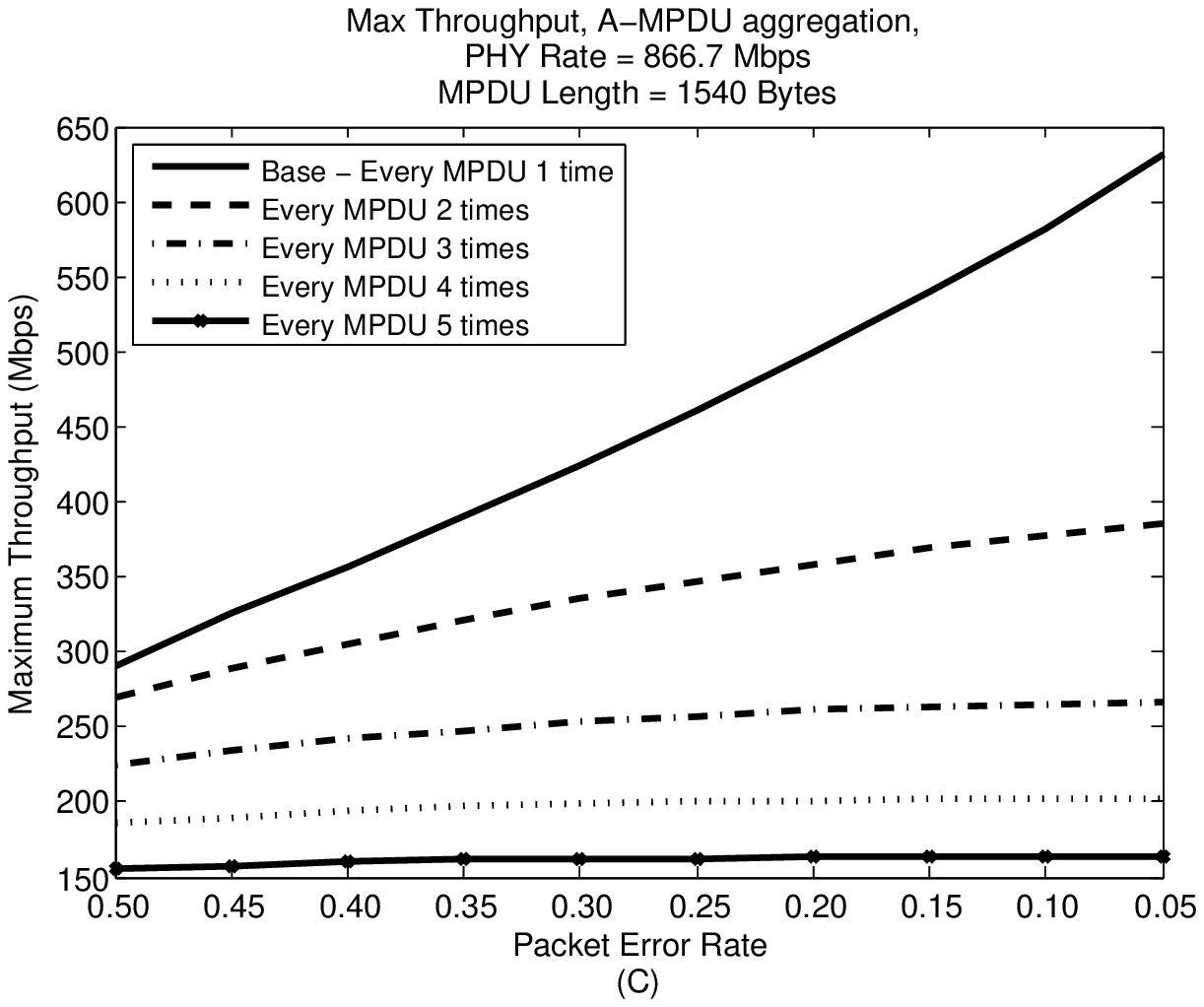}
\includegraphics{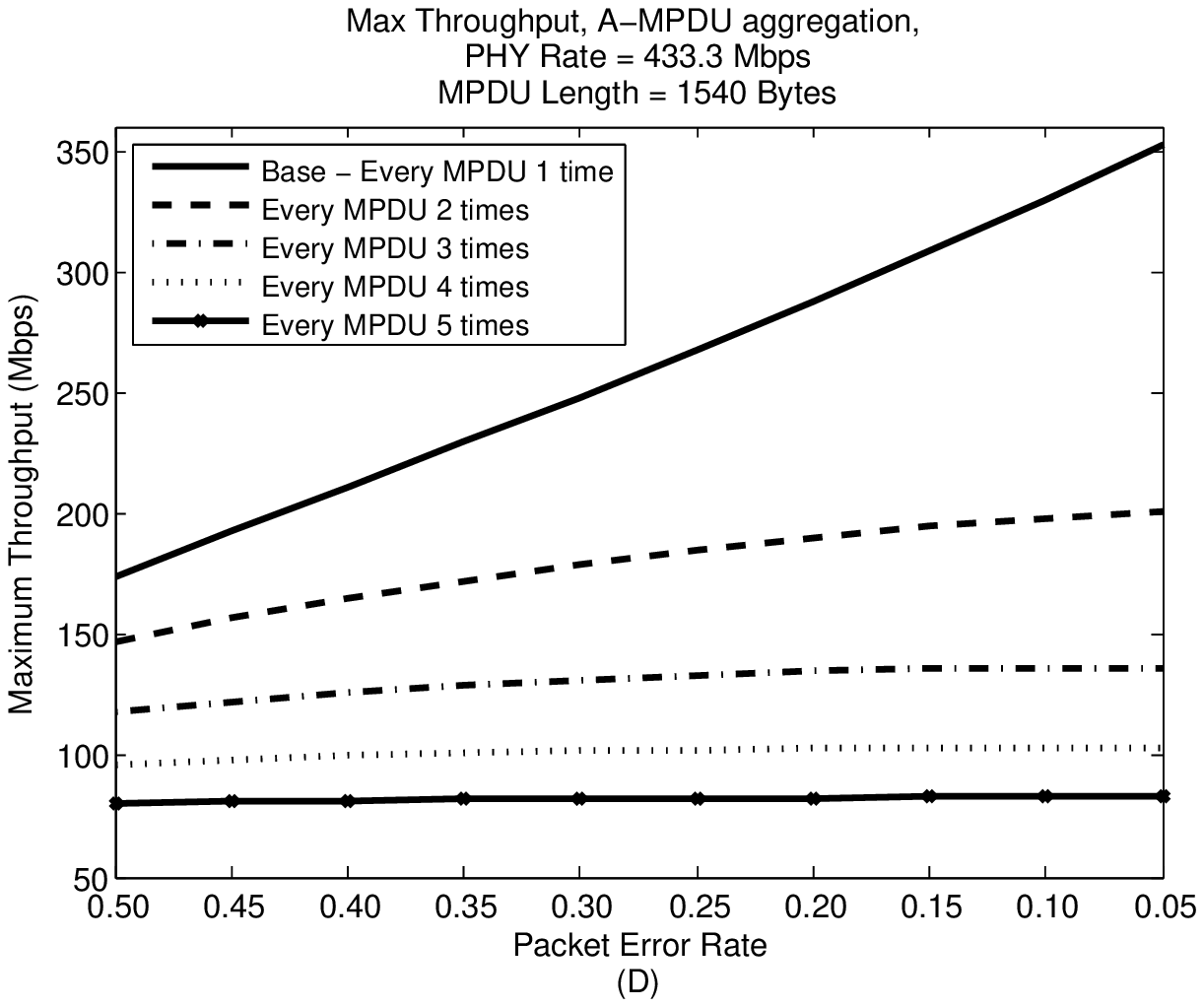}
\caption{Maximum Throughput when retransmitting every MPDU in window, A-MPDU aggregation, MPDU size 1540 bytes}
\label{fig:fig4}
\end{figure}

\begin{figure}
\vskip 16cm
\includegraphics{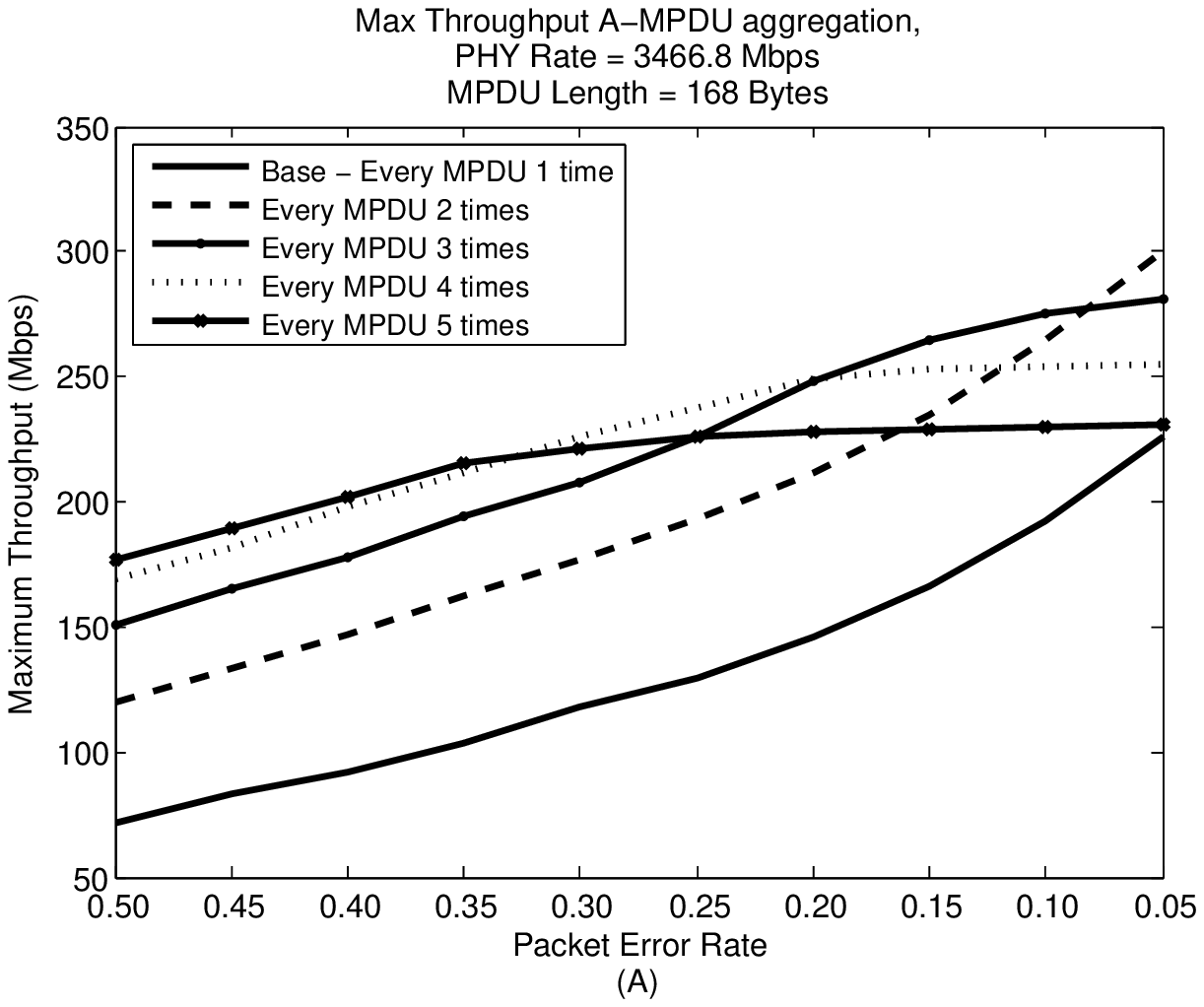}
\includegraphics{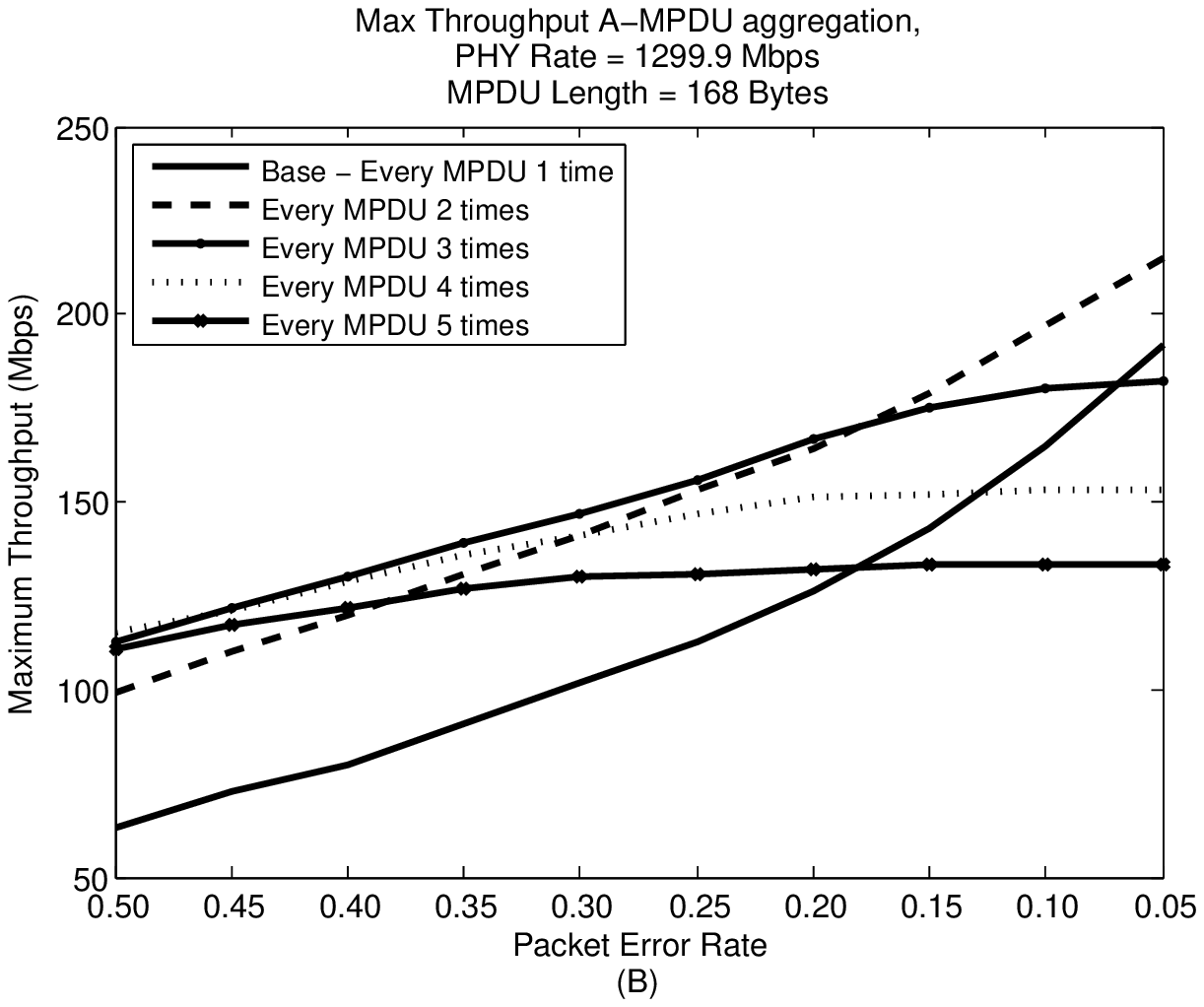}
\includegraphics{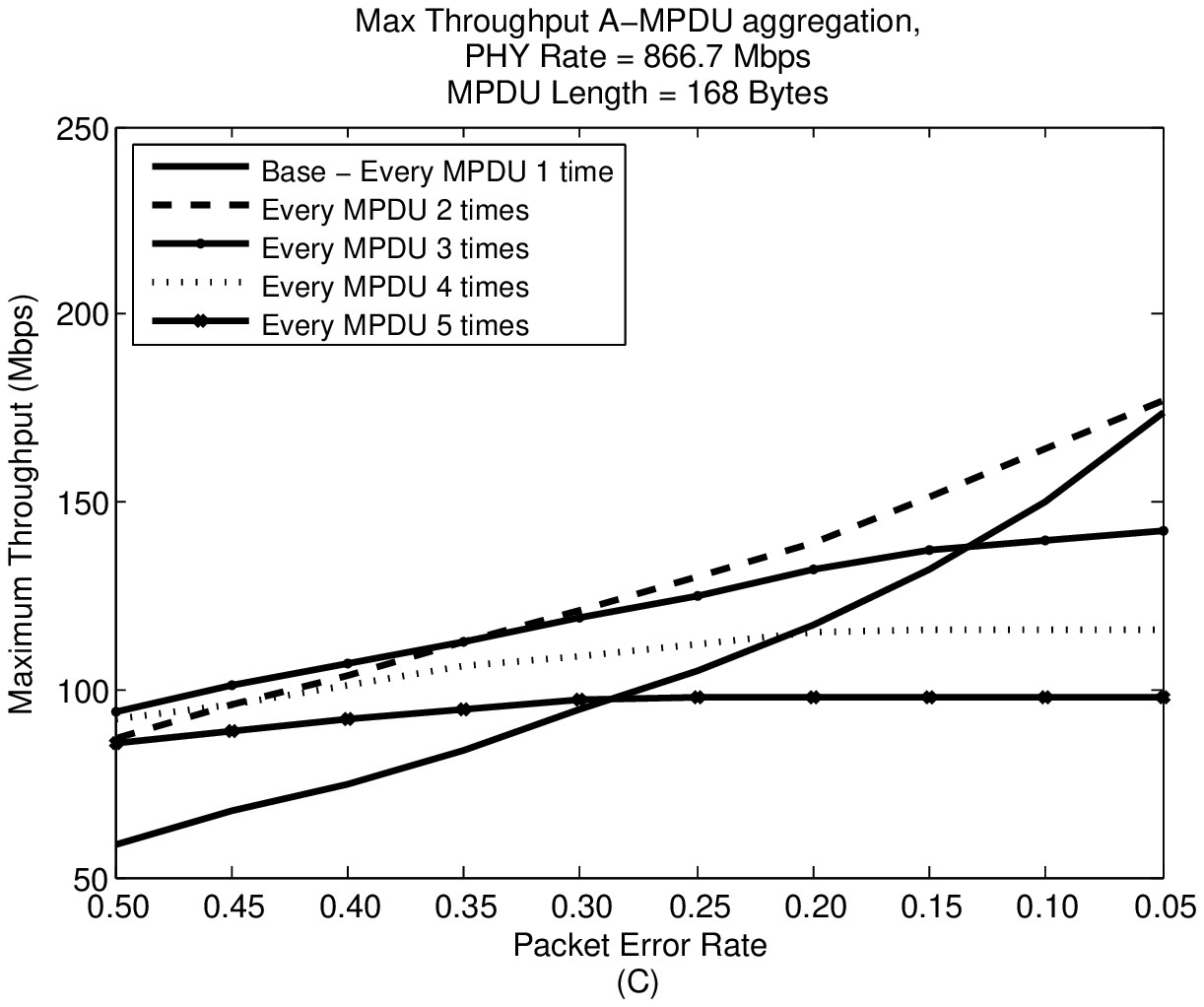}
\includegraphics{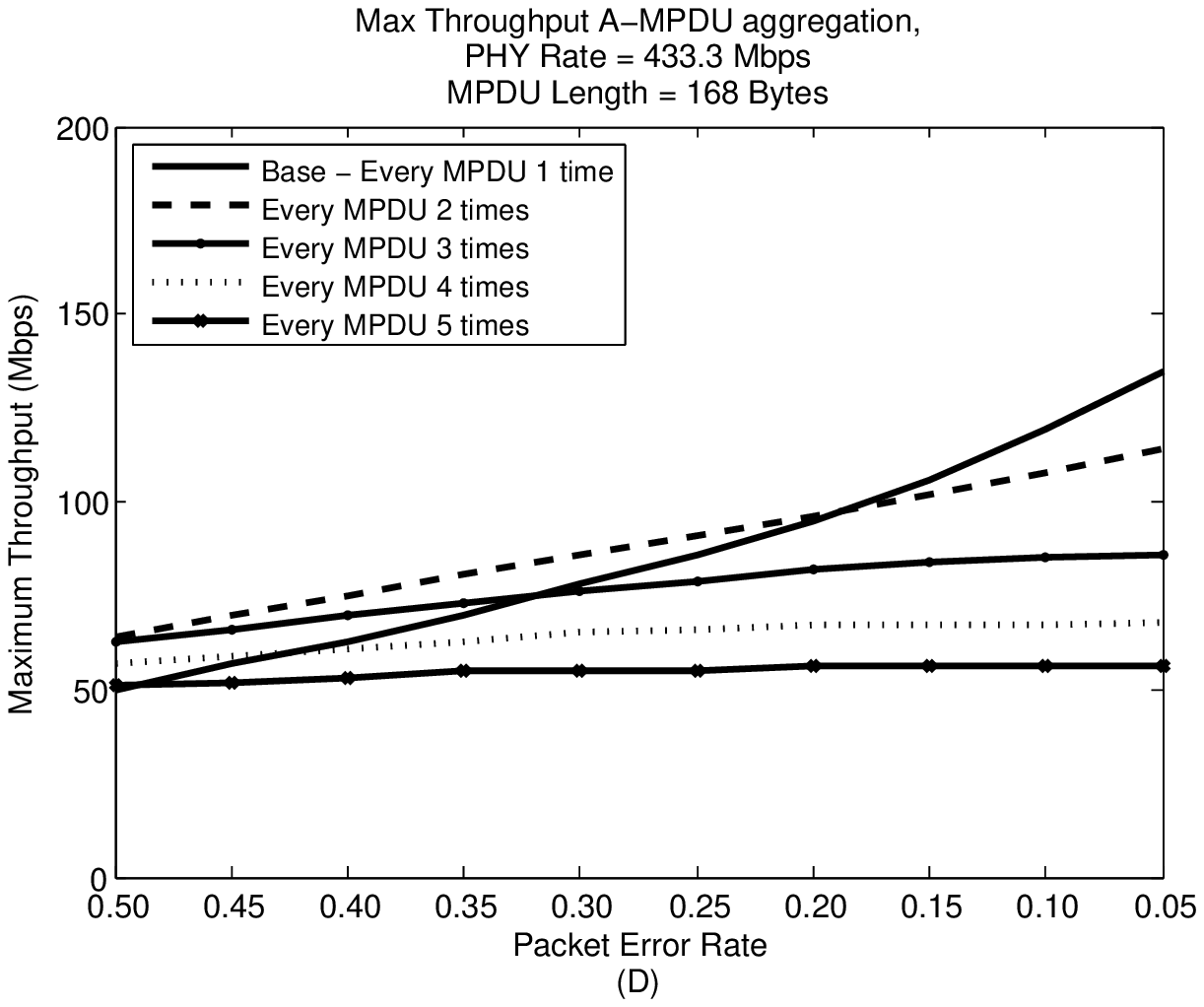}
\caption{Maximum Throughput when retransmitting every MPDU in window, A-MPDU aggregation, MPDU size 168 bytes}
\label{fig:fig41}
\end{figure}

In Figures~\ref{fig:fig4} and~\ref{fig:fig41}
we show the results for {\it Set5} 
for MPDUs of 1540 and 168 bytes
respectively, as well as for all the considered PHY rates. For MPDUs of
1540 bytes, Figure~\ref{fig:fig4}, a
significant improvement in the Throughput
over the $Base$ method appears only for a PHY rate
of 3466.8 Mbps, and  large PERs. For example,
an improvement of 24$\%$ is observed in the case of PER$=$0.5 .
In {\it Set5} all the MPDUs are retransmitted in the transmission;
thus the PSDU and its transmission time increase
significantly. This penalty is significant
in relatively small PHY rates, and it becomes evident that
only in the largest PHY rate the retransmission of all
the MPDUs becomes beneficial.

For MPDUs of 169 bytes, Figure~\ref{fig:fig41}, a significant
improvement over the $Base$ method is achieved for all the PHY rates. 
For small size
MPDUs the penalty in time, when retransmitting MPDUs, is minimal.
Therefore, the retransmission is beneficial.
Notice that for the small PHY rates, 433.3 and 866.7 Mbps, {\it ALL2}
is most efficient, while for the 1299.9 and 3466.8 Mbps PHY rates
{\it ALL2} and {\it ALL5} can also be efficient.
In these PHY rates the time penalty
of retransmissions is smaller. Thus, it is worth retransmitting
MPDUs an additional number of
times in order to increase their success probability.

\subsection{Overall Throughput improvement}

\begin{figure}
\vskip 4cm
\includegraphics{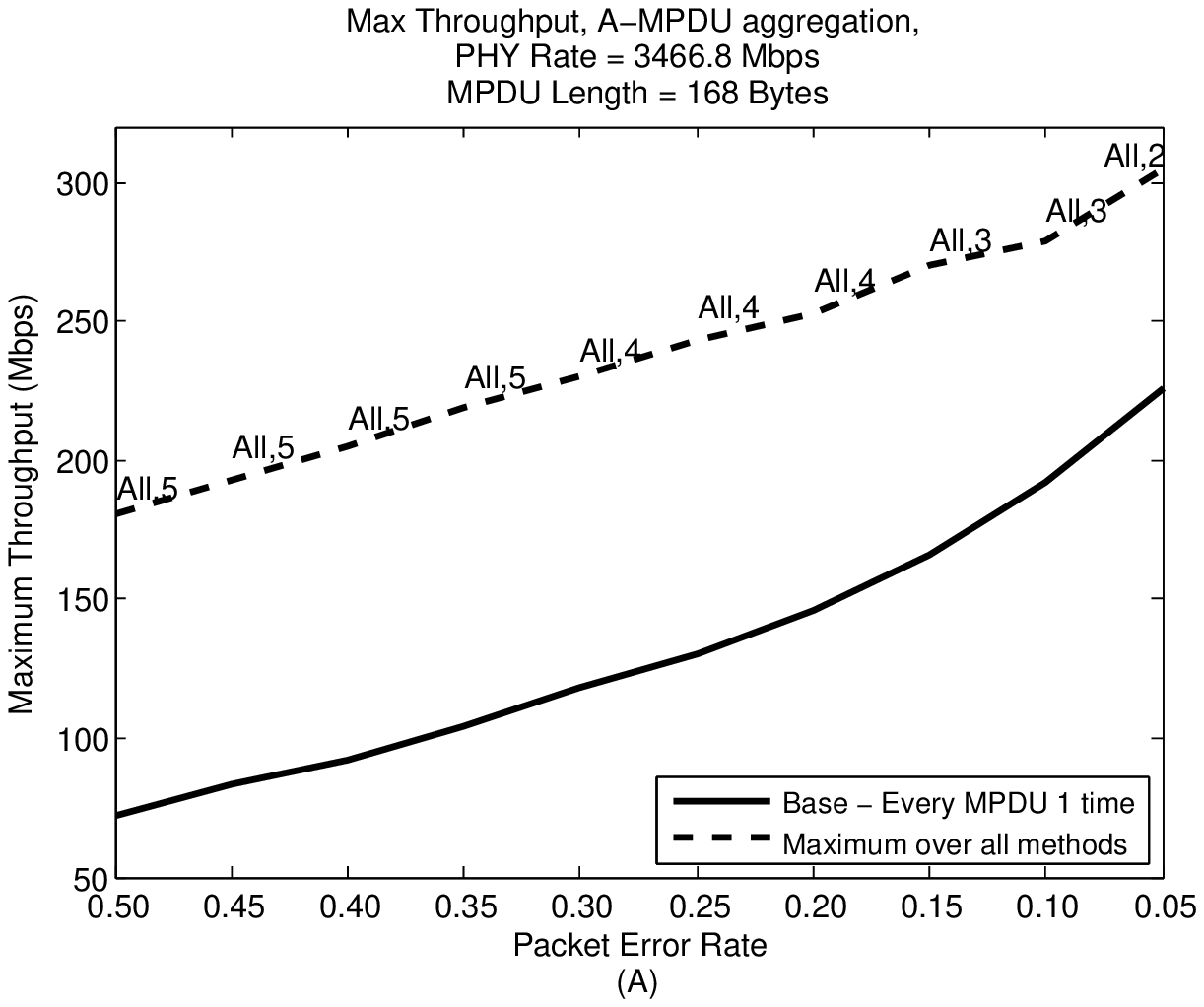}
\includegraphics{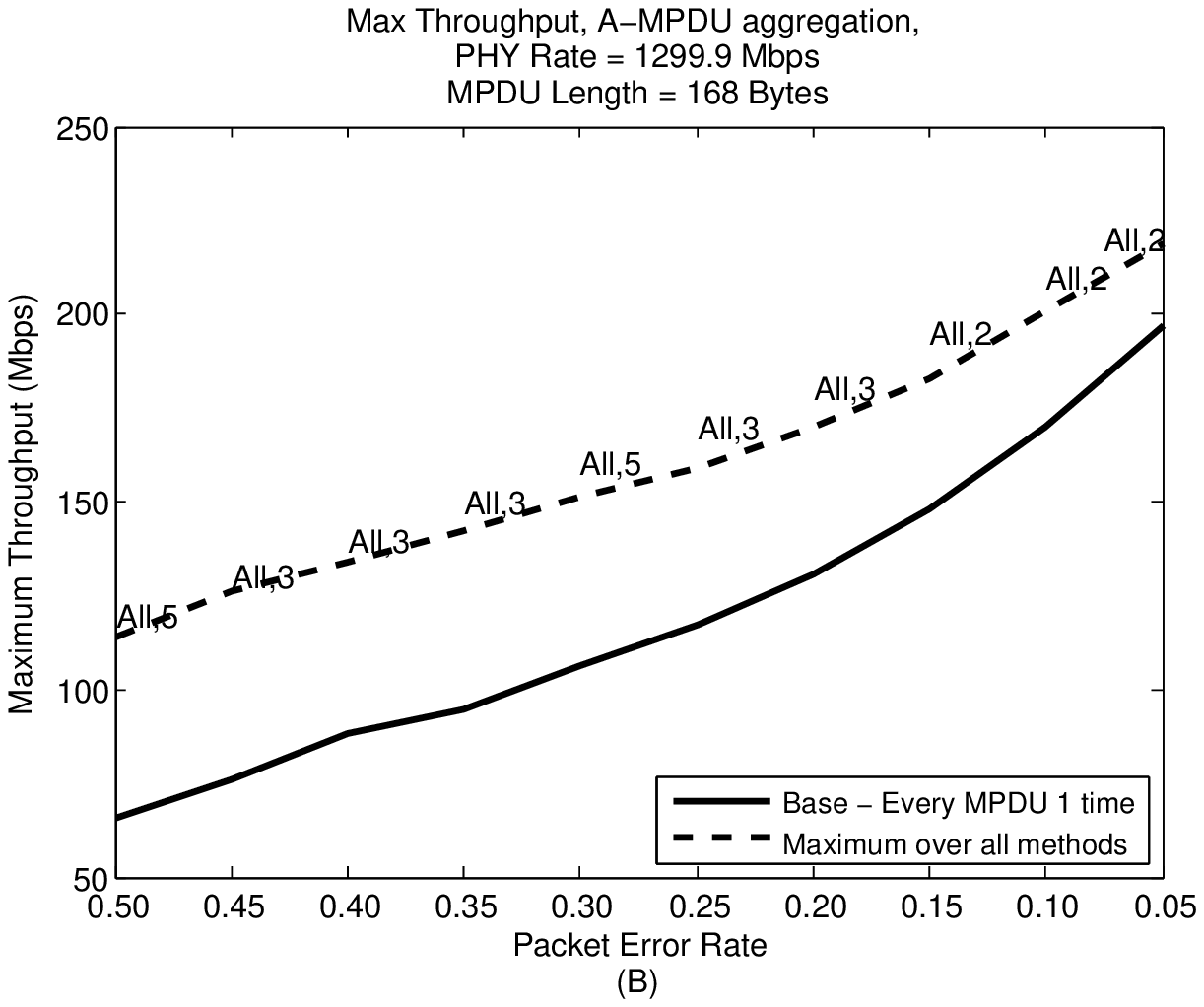}
\caption{Maximum Throughput over all methods vs. $Base$ method, A-MPDU aggregation, MPDU size 168 bytes}
\label{fig:fig5}
\end{figure}

\begin{figure}
\vskip 4cm
\includegraphics{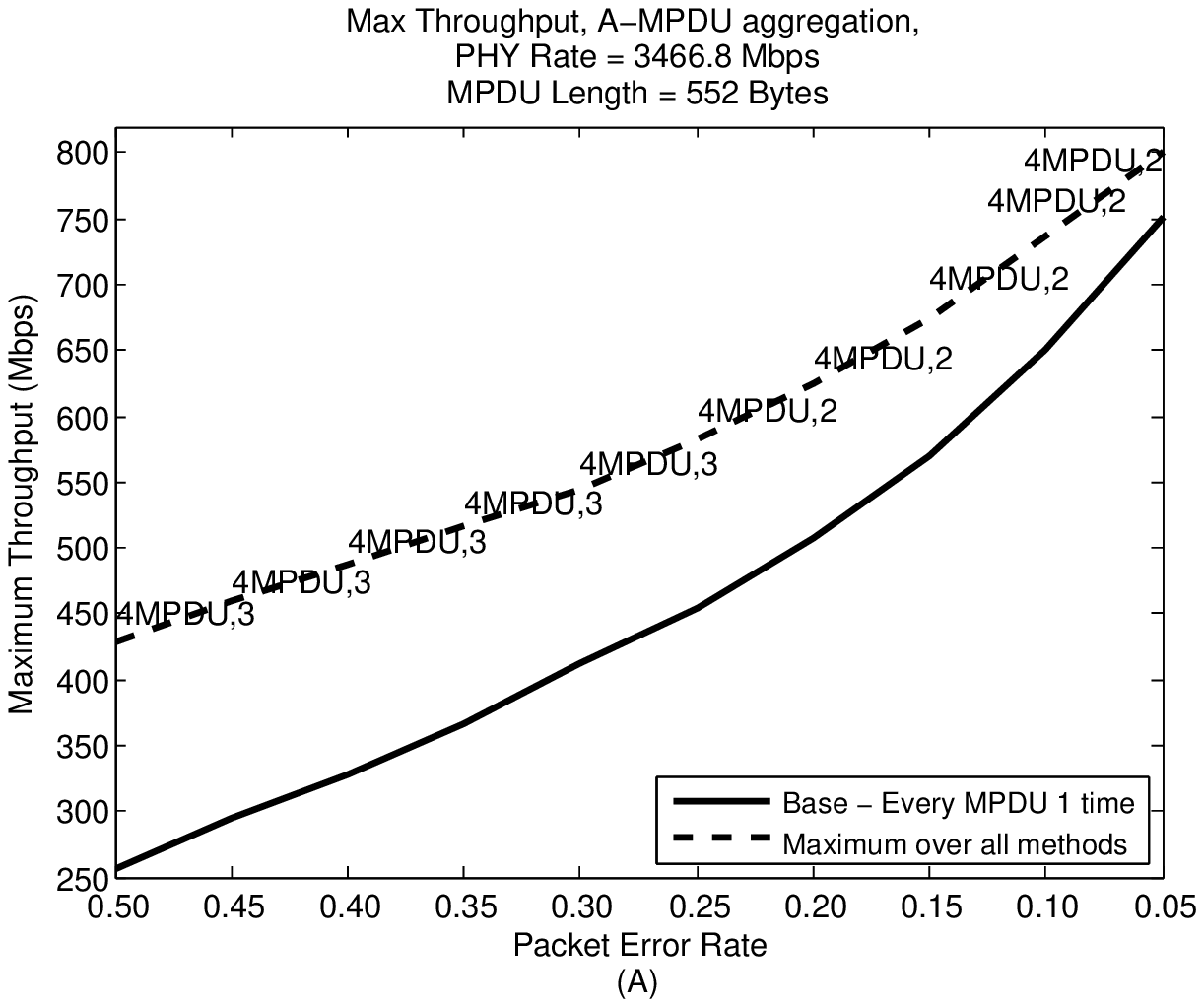}
\includegraphics{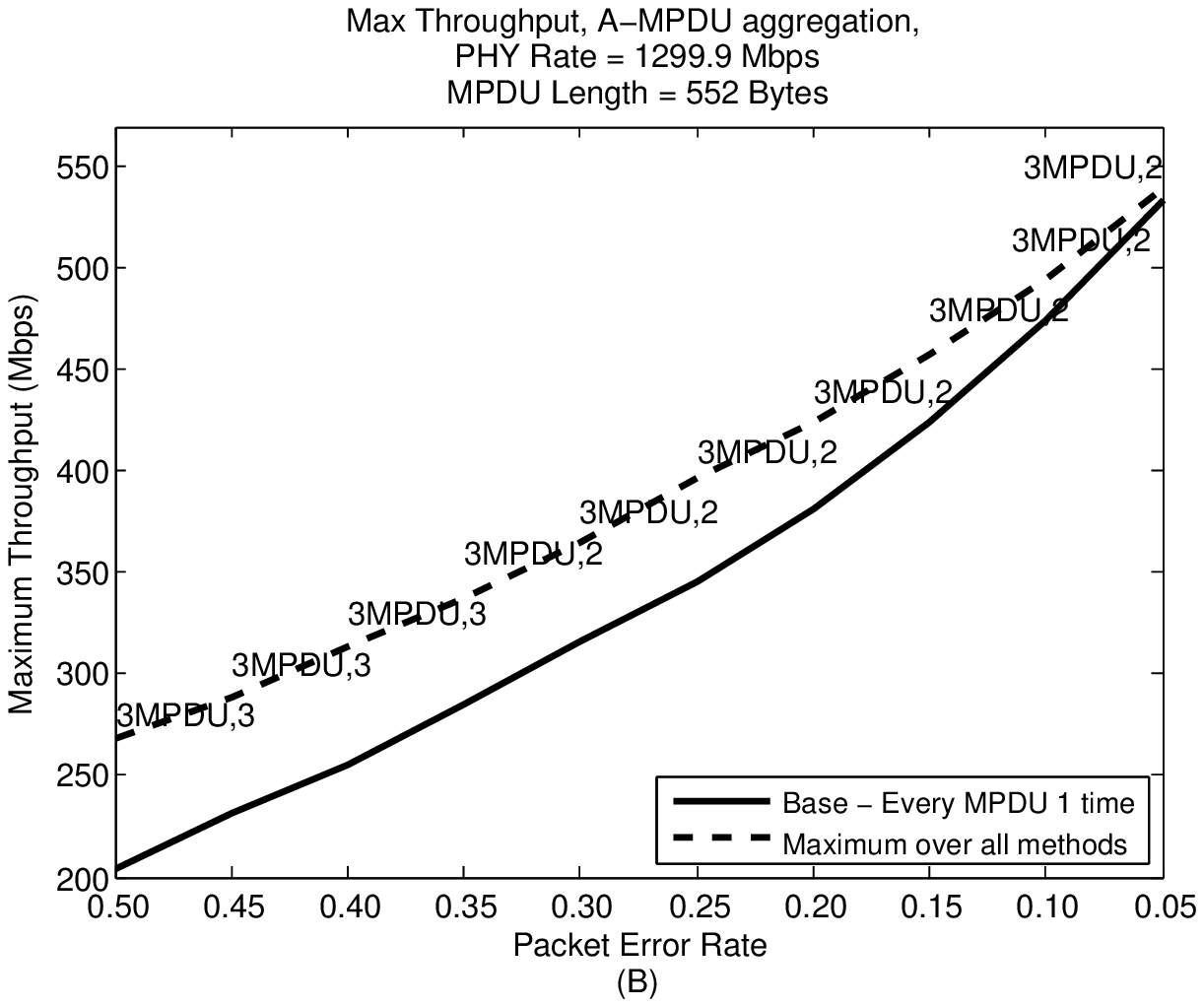}
\caption{Maximum Throughput over all methods vs. $Base$ method, A-MPDU aggregation, MPDU size 552 bytes}
\label{fig:fig6}
\end{figure}

\begin{figure}
\vskip 4cm
\includegraphics{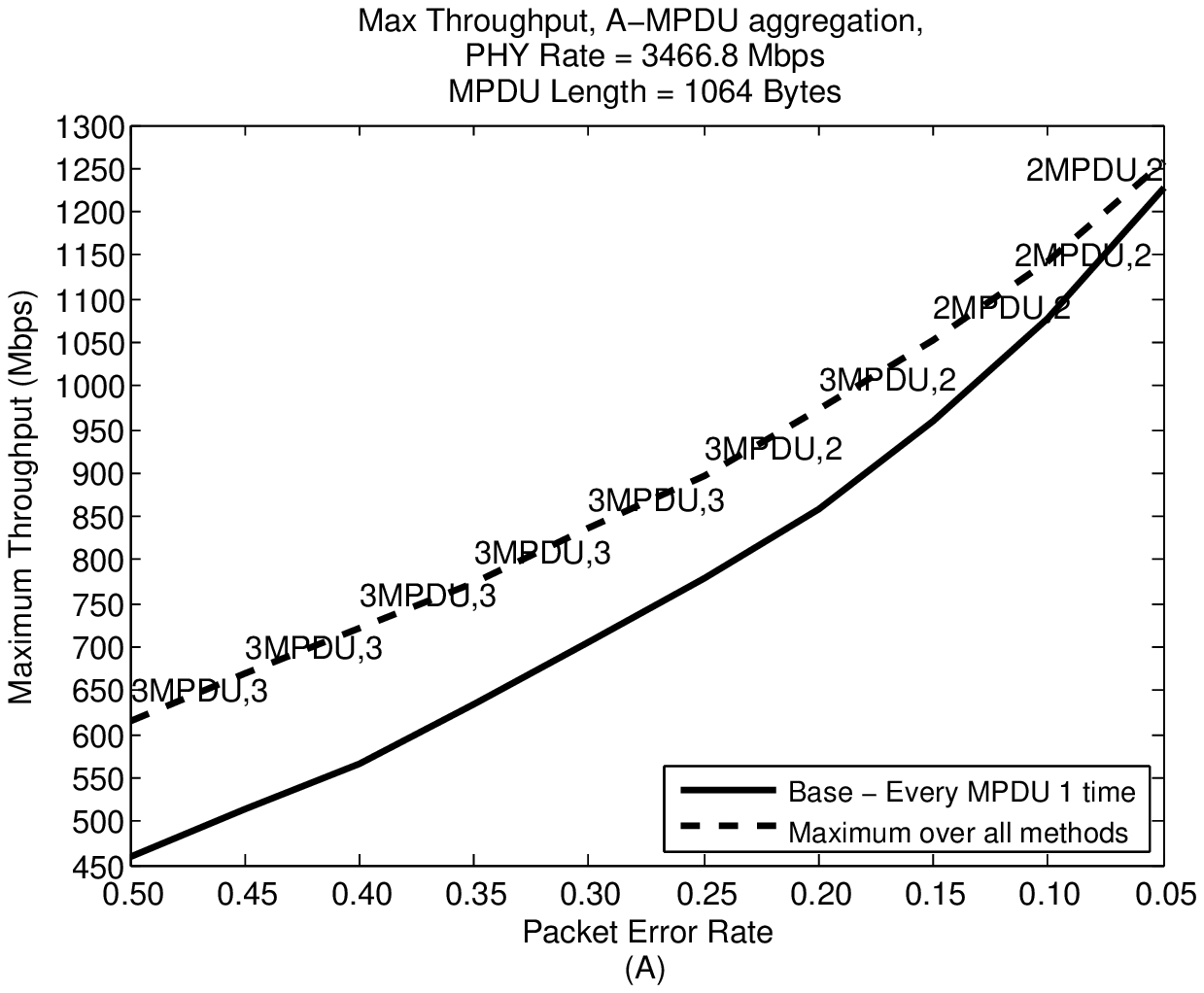}
\includegraphics{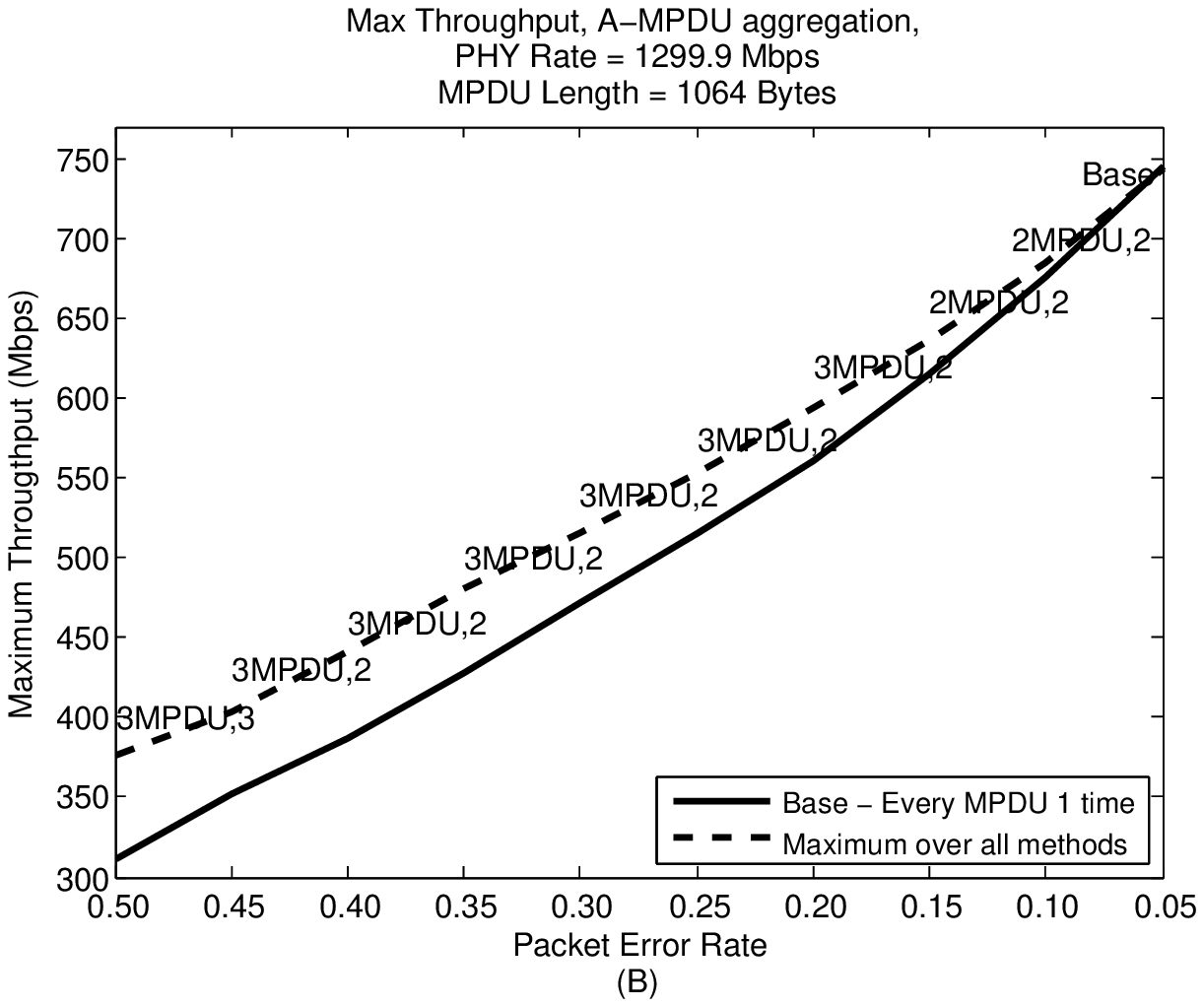}
\caption{Maximum Throughput over all methods vs. $Base$ method, A-MPDU aggregation, MPDU size 1064 bytes}
\label{fig:fig7}
\end{figure}

\begin{figure}
\vskip 4cm
\includegraphics{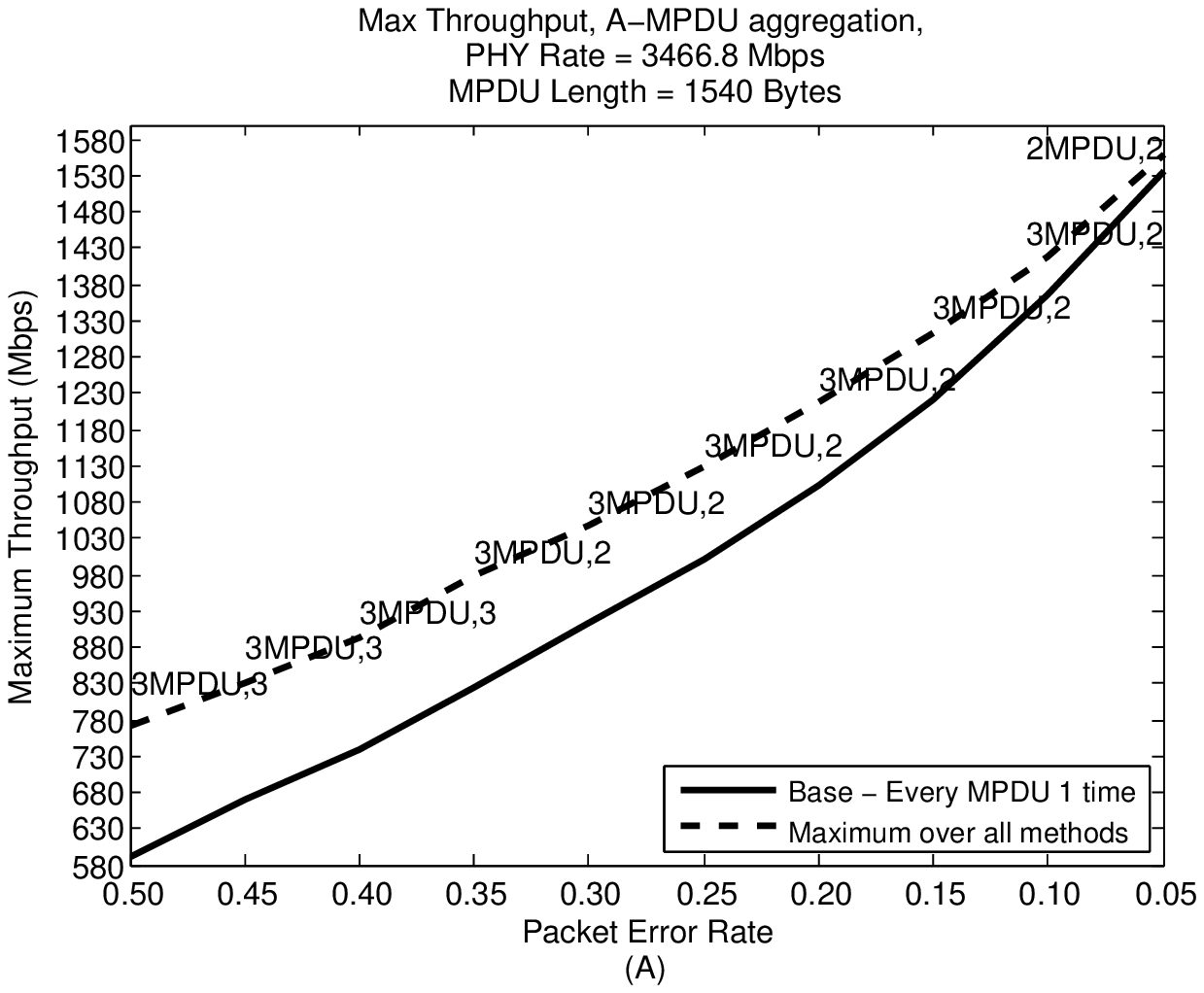}
\includegraphics{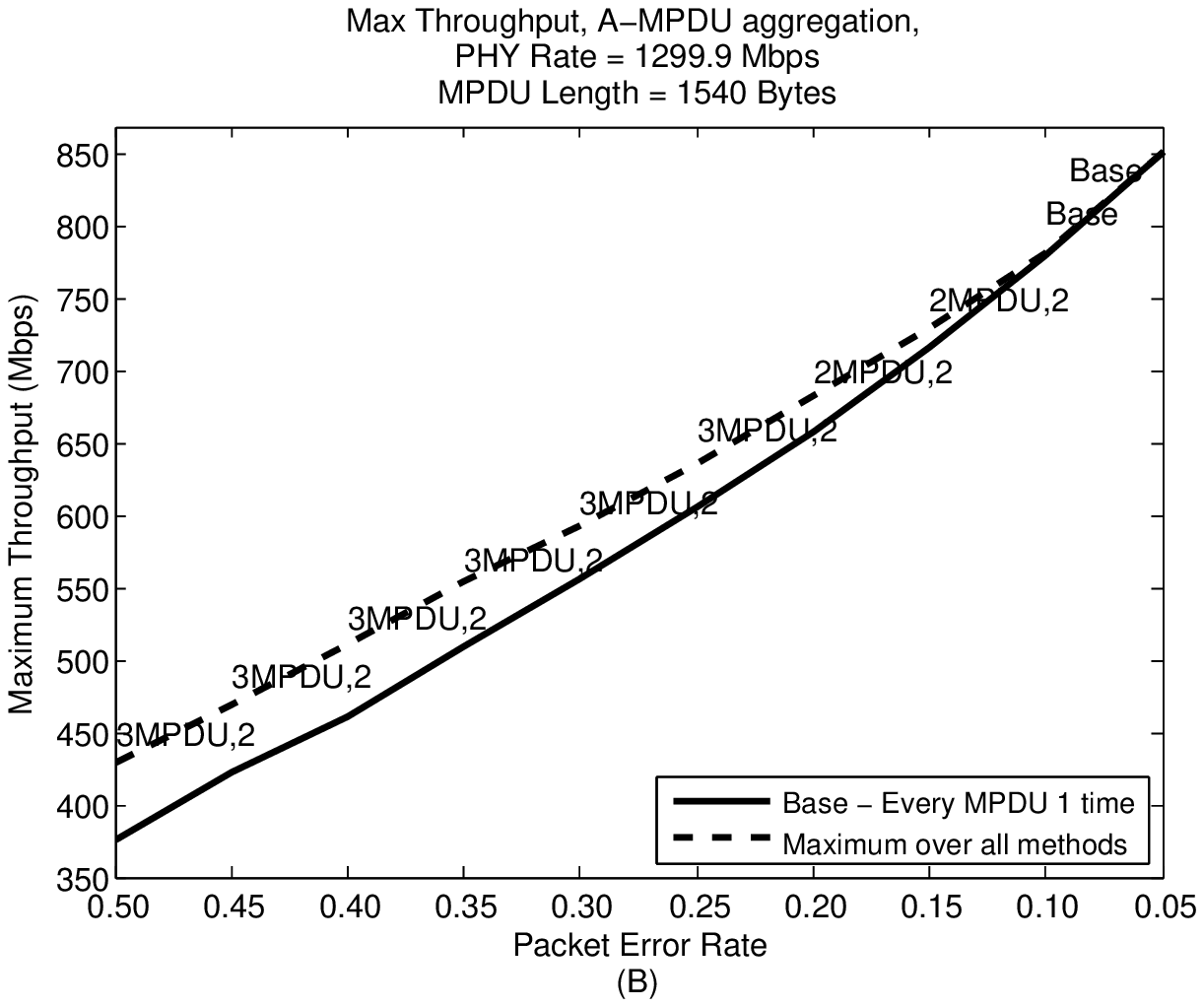}
\caption{Maximum Throughput over all methods vs. $Base$ method, A-MPDU aggregation, MPDU size 1540 bytes}
\label{fig:fig8}
\end{figure}

\begin{figure}
\vskip 5cm
\includegraphics{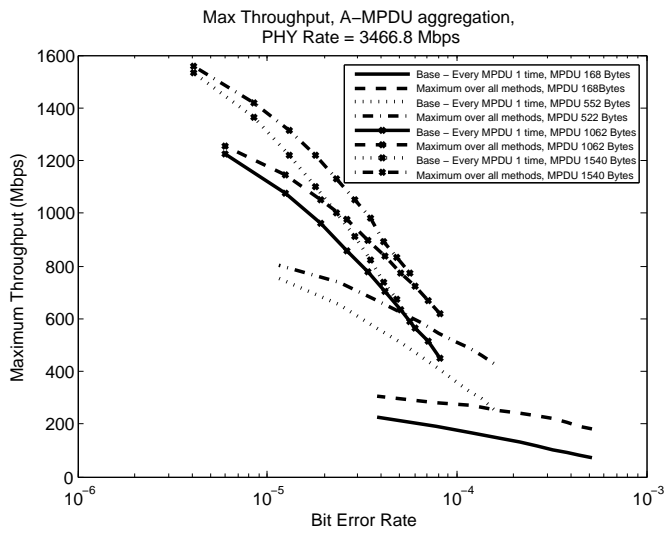}
\caption{Maximum Throughput over all methods and $Base$ method vs. BER, A-MPDU aggregation, MSDUs' size 128, 512, 1024 1500 bytes}
\label{fig:fig9}
\end{figure}

In Figures~\ref{fig:fig5}-\ref{fig:fig8} we consider 
MPDUs of length 168, 552, 1064 and
1540 bytes respectively. In every
figure we only consider PHY rates 3466.8 and 1299.9 Mbps.
We show 2 curves: one for the Throughput of the $Base$ method.
The second indicates, for every PER, the maximum achievable
Throughput considering all the new methods. We also
show the method that achieves the maximum Throughput. For
example, in Figure~\ref{fig:fig5}(A) and PER$=$0.5, it is 
most efficient to transmit each MPDU 5 times.
At the same point in Figure~\ref{fig:fig6} it is best
to use {\it 4MPDU3}, i.e. to transmit the first 4 MPDUs 3 times.

From Figures~\ref{fig:fig5}-~\ref{fig:fig8} one can see that
for MPDUs
of 168 bytes, the percentage of the Throughput
improvement is the largest. For example, for PHY rate
3466.8 Mbps and PER$=$0.5, it is 257$\%$ ! For PER$=$0.05 it is 33$\%$.

Clearly, as the PER decreases, the improvement
in the Throughput decreases as well. The channel
becomes reliable, the need for retransmissions is smaller
and the penalty of longer PSDUs' transmission times
is therefore more significant.
The reason that smaller MSDUs show greater improvement
(in percentage ) in the Throughput is due
to the overhead associated with every
transmission. As the MSDU's length is smaller, the 
transmission time of an MSDU is smaller, and the
size of the overhead ($C_1$ in Eq.~\ref{equ:equ1})
is more significant. When the size
of the overhead is more significant, the penalty of
transmitting the same MSDU several times is relatively
lower and therefore, the improvement in percentage is
more significant.
When later discussing the Two-Level aggregation method we provide an
analytical explanation to the argument above.

Overall, from Figures~\ref{fig:fig5}-\ref{fig:fig8} one can see
that the new methods improve the Throughput considerably, mainly
in PERs in the range 0.10-0.50  .

In Figure~\ref{fig:fig9} we show part (A) of
Figures~\ref{fig:fig5}-\ref{fig:fig8} in one Figure,
but this time the x-axis is the BER. We normalized
the PER in Figures~\ref{fig:fig5}-\ref{fig:fig8} into
BER according to Eq.~\ref{equ:failureprobbasic} in order
to show that the improvement in the Throughput
is achieved over a wide range of BER values.

\subsection{Throughput improvement in Two-Level aggregation}

\begin{figure}
\vskip 11cm
\includegraphics{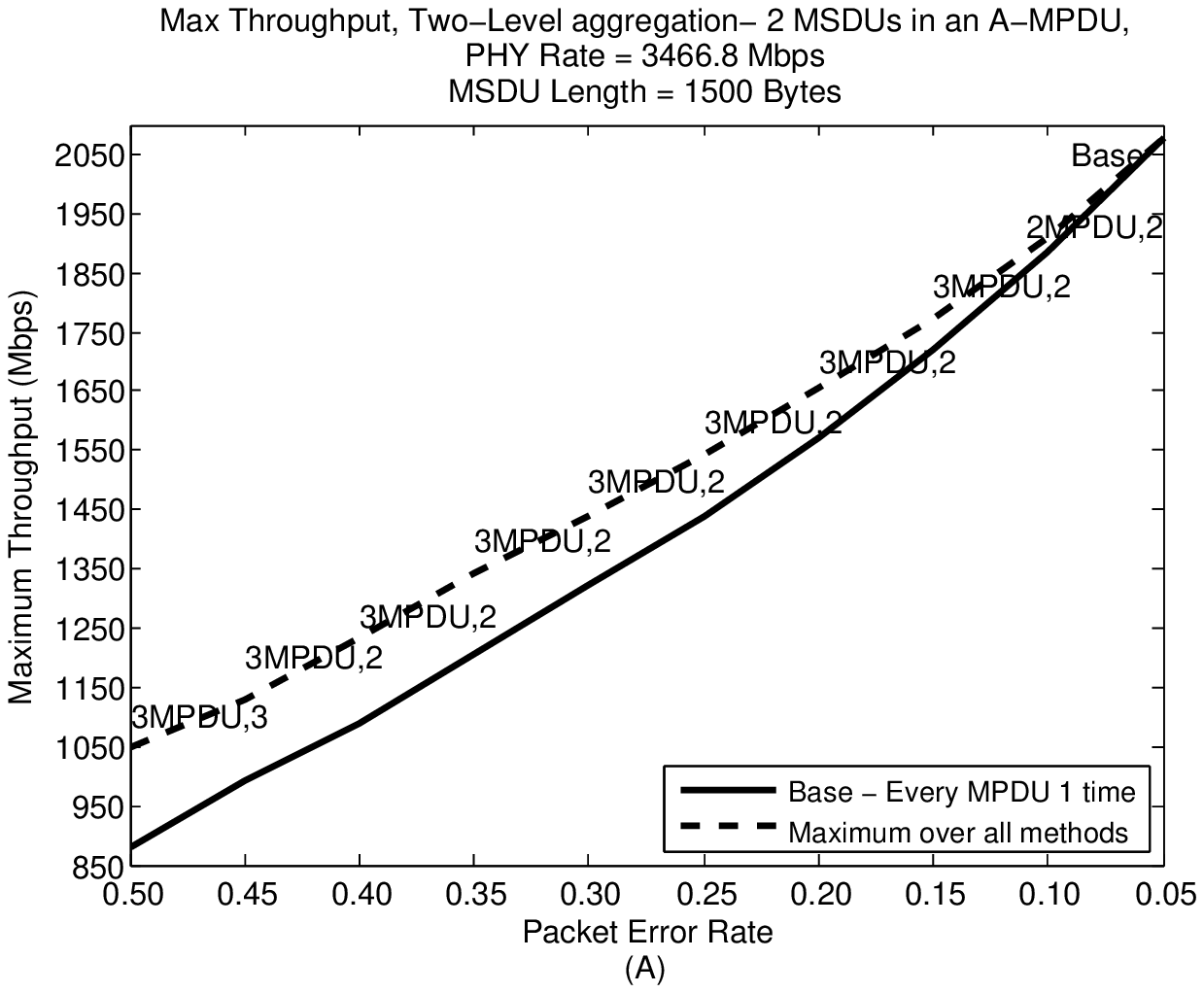}
\includegraphics{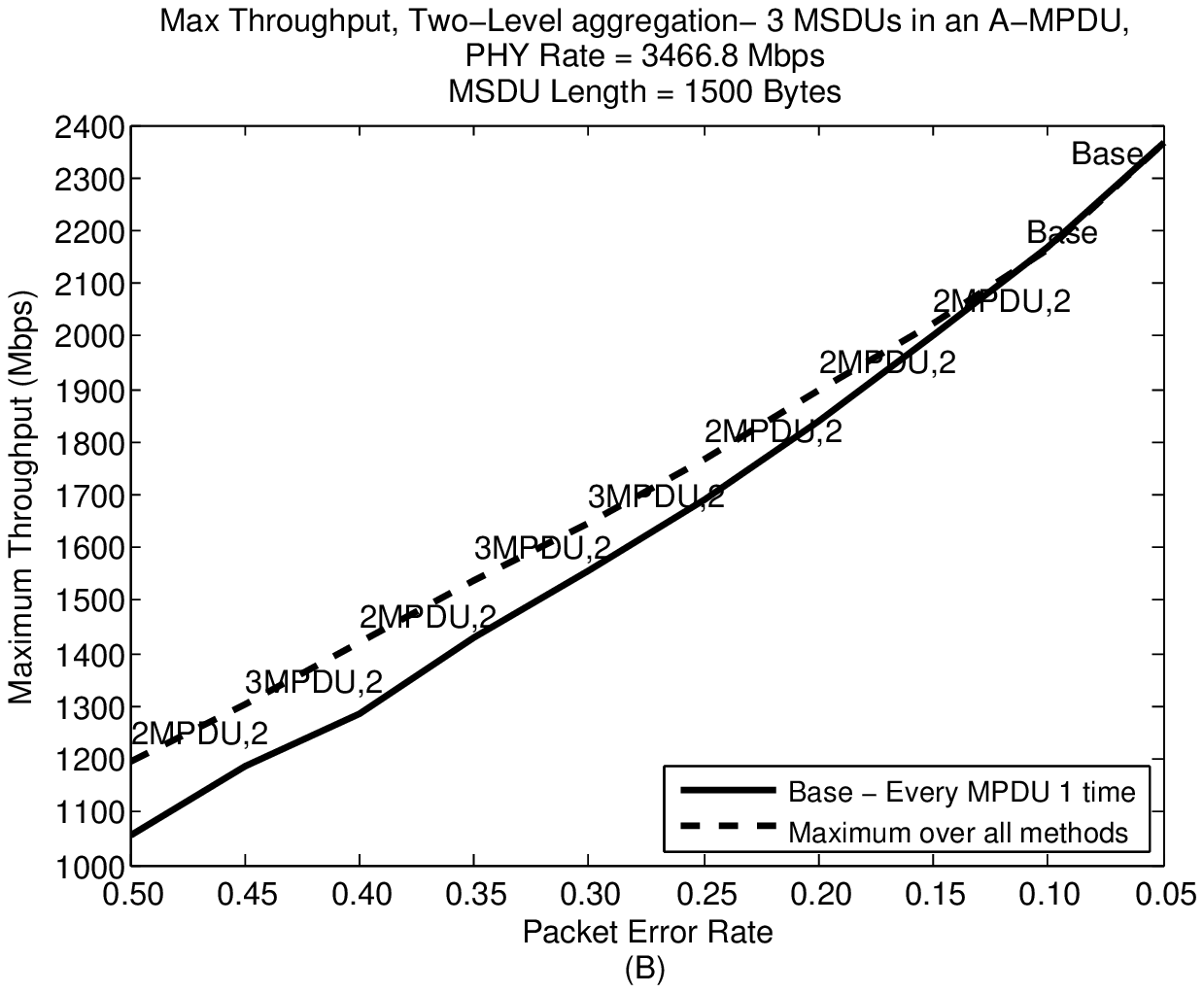}
\includegraphics{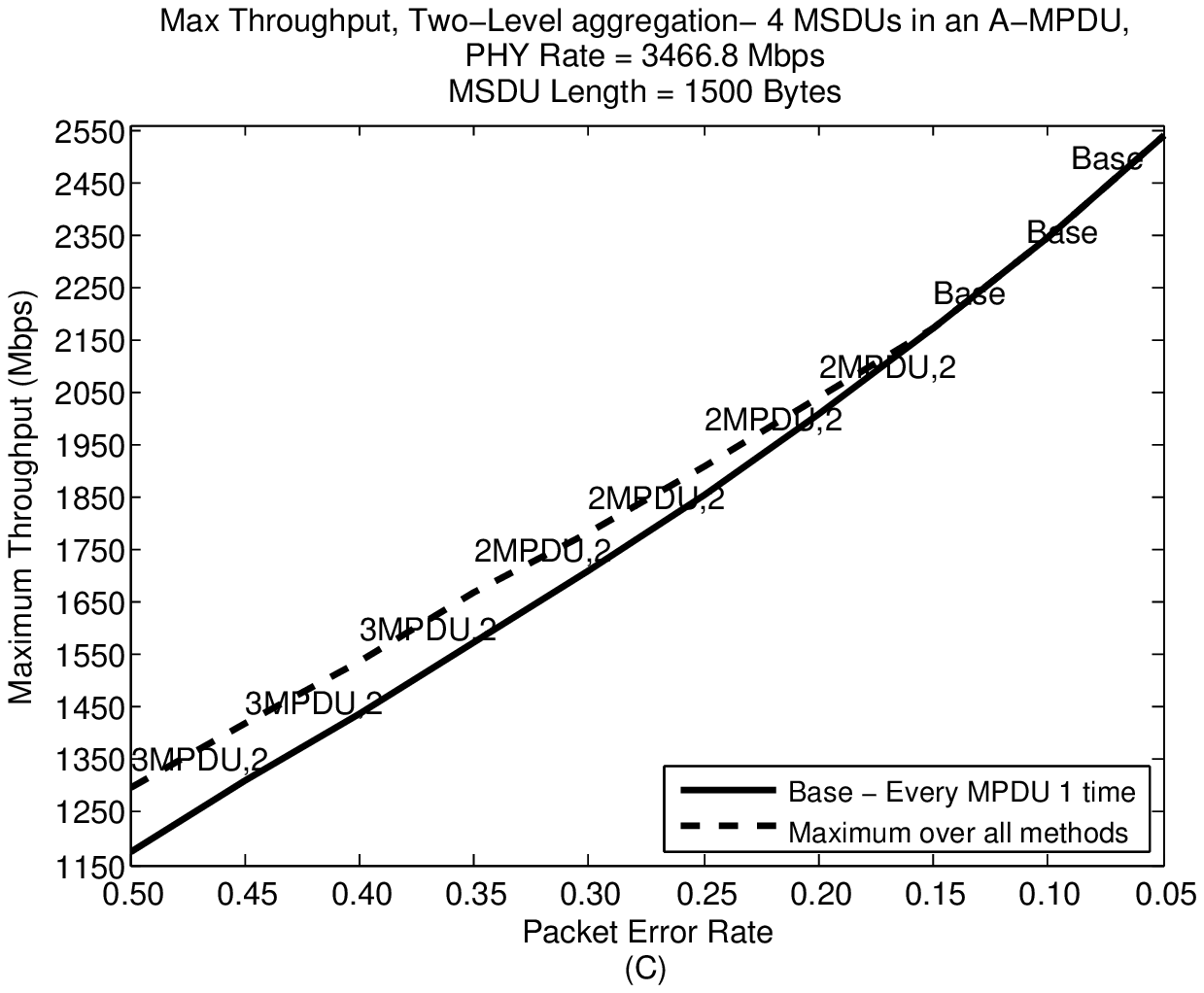}
\includegraphics{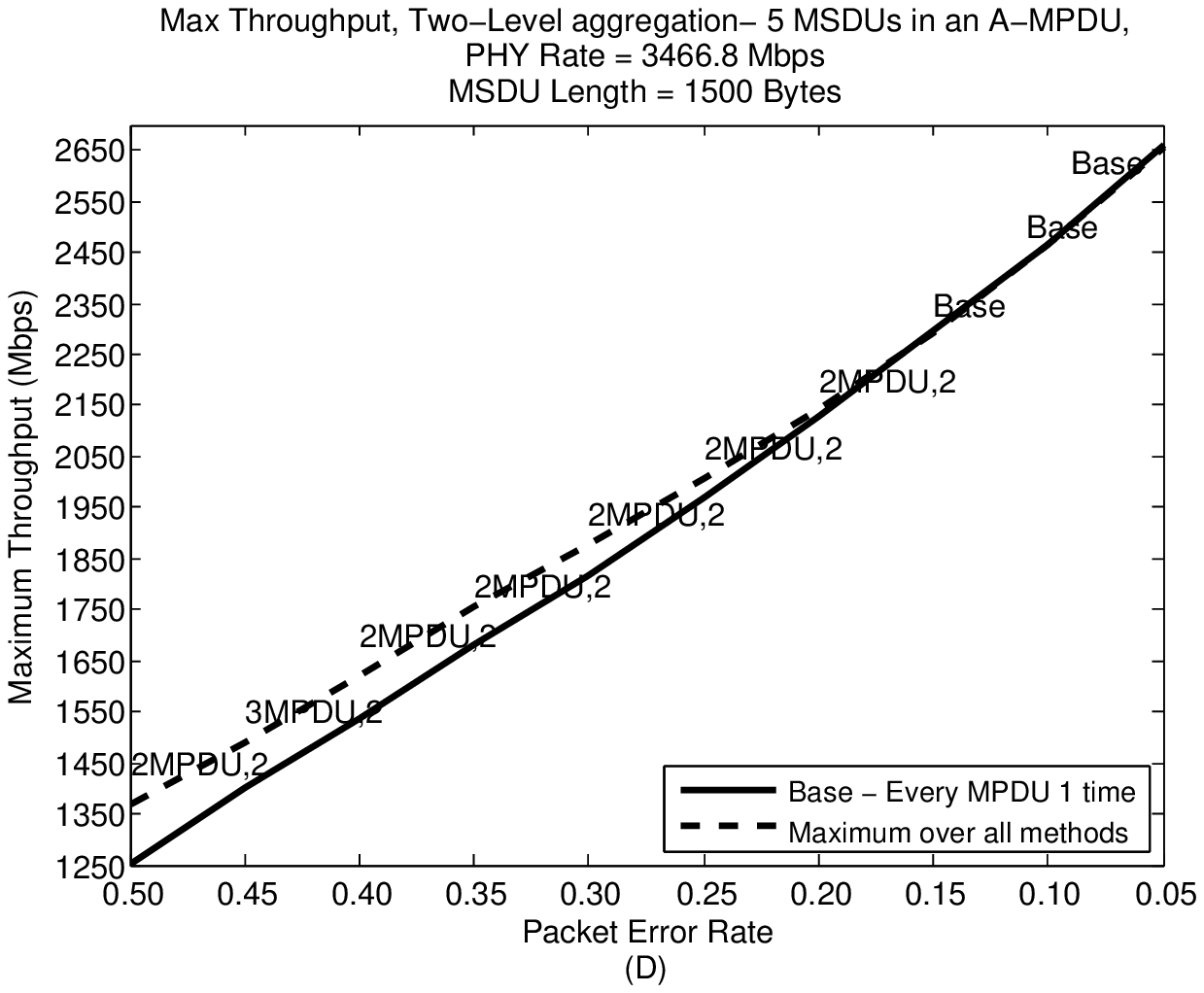}
\includegraphics{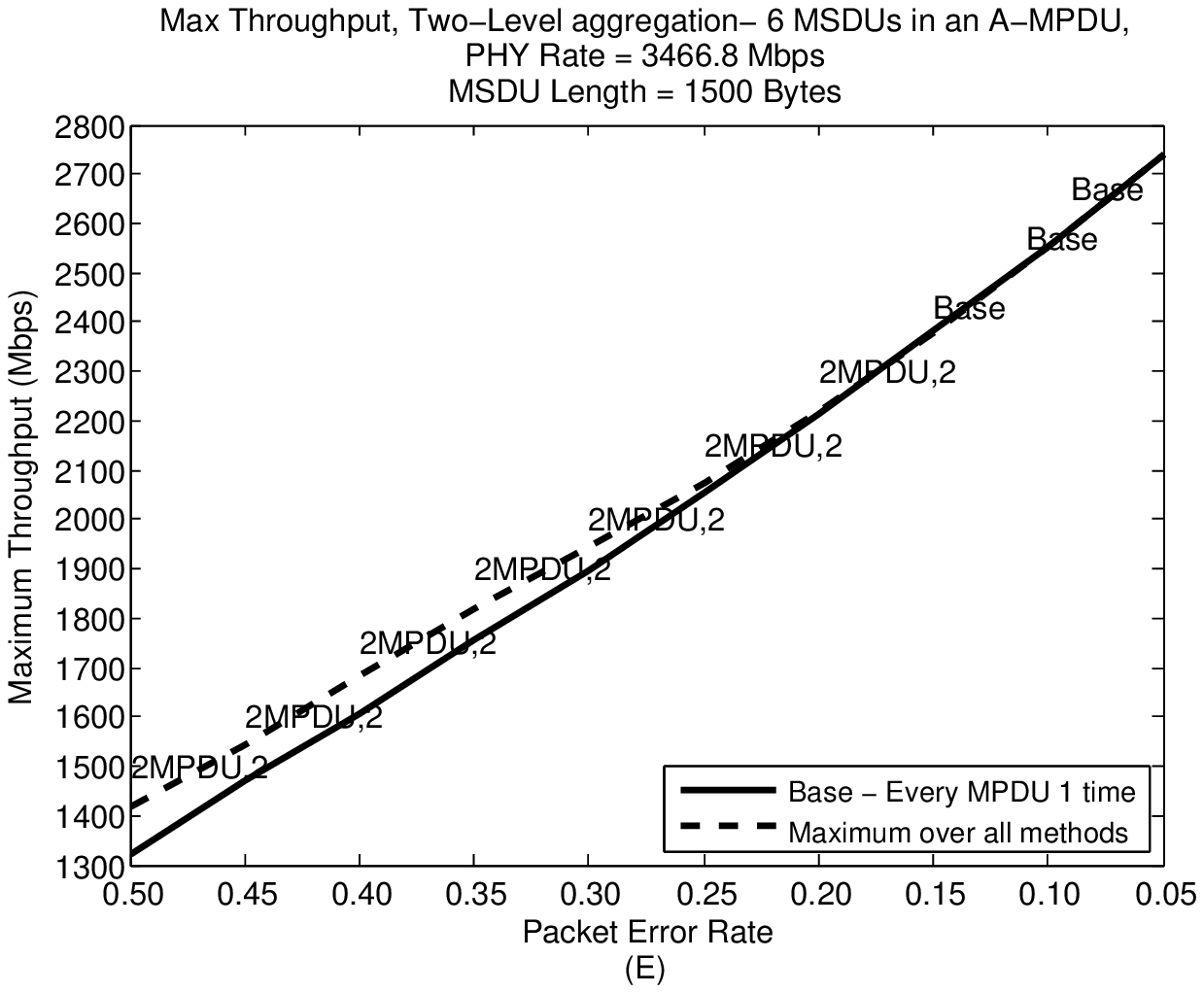}
\includegraphics{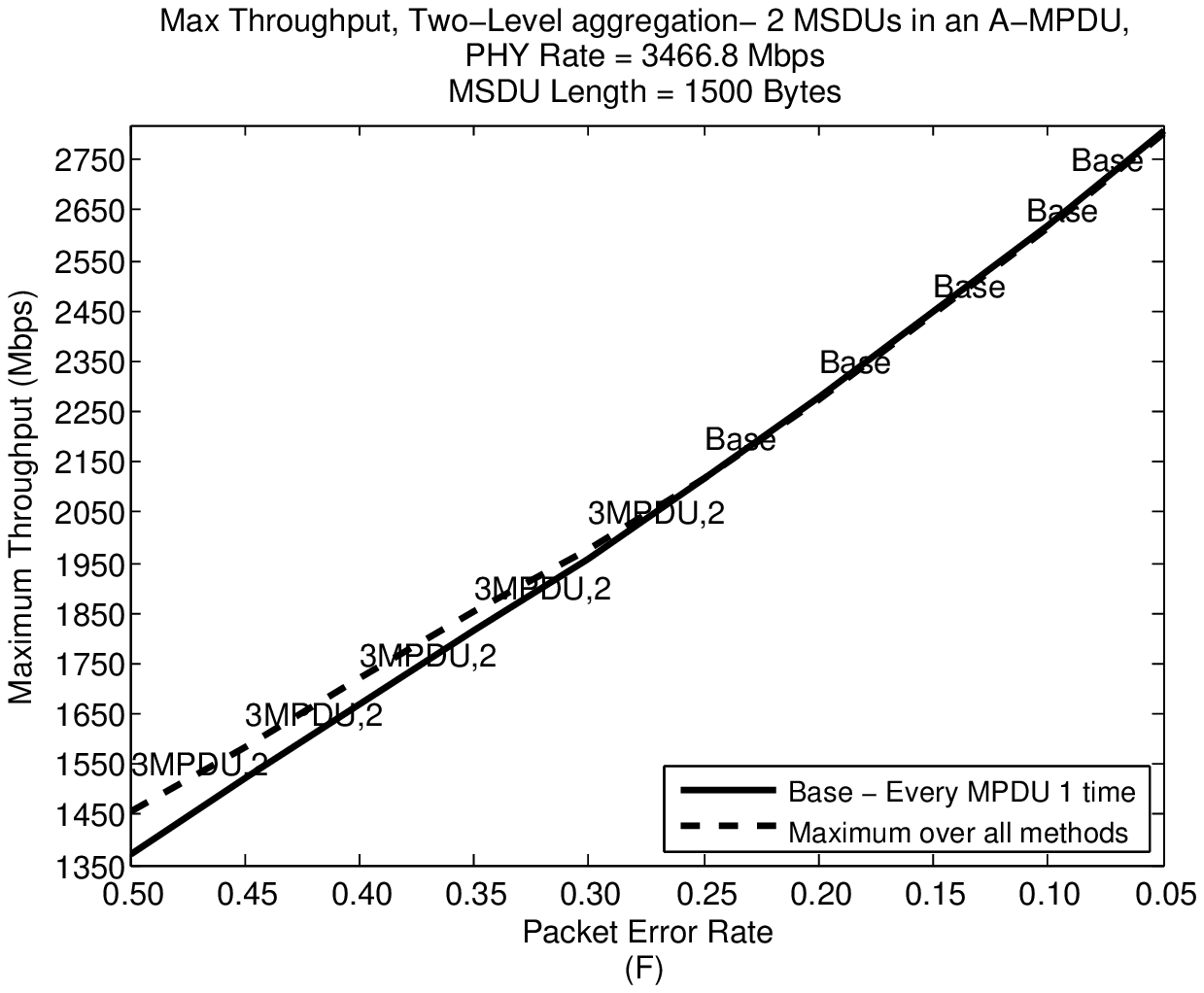}
\caption{Maximum Throughput over all methods vs. $Base$ method , Two-Level aggregation, 2-7 MSDUs per A-MPDU frame, MSDU size 1500 bytes}
\label{fig:fig10}
\end{figure}

In Figure~\ref{fig:fig10} we consider the Two-Level aggregation,
an MSDU's size of 1500 bytes and 2-7 MSDUs per A-MPDU frame
in Figure~\ref{fig:fig10}(A)-(F) respectively. We again
compare between the Throughput of the $Base$ method to the maximum
achieved by all the new methods.

One can see in Figure~\ref{fig:fig10} that as the number
of MSDUs in an A-MPDU frame increases, the percentage of improvement
by the new methods decreases. For PER$=$0.5 it is 20$\%$, 14$\%$,
10$\%$, $9\%$, $7.5\%$ and $5.4\%$ for 2 to 7 MSDUs per A-MPDU
respectively.
This phenomena can be explained by some approximation
as follows: Let's assume
the case of 2 MSDUs per A-MPDU and let $C_1$ be the overhead
as defined in Eq.~\ref{equ:equ1}. In the $Base$ method let
$T$ be the transmission time of an average
length PSDU and $B$ the average number of bits that arrive successfully
at the receiver.

Let's assume that the same new method $S$ is best
for all MSDUs in an A-MPDU. Notice that
in $S$, and again assuming 2 MPDUs per
A-MPDU frame, the transmission time of an average PSDU is now
$T+T_s$ because in $S$ several MPDUs are transmitted several
times. Let $B_s$ be the average number
of bits that arrive successfully at the receiver.
Then, the percentage of the improvement of $S$ over
the $Base$ method, for 2 MSDUs per A-MPDU frame, appears in Eq.~\ref{equ:thr1}:

\begin{equation}
\label{equ:thr1}
\frac{\frac{B_s}{C_1+T+T_s}}{\frac{B}{C_1+T}}
\end{equation}

Lets assume now $X$ MSDUs per A-MPDU frame, where $3 \le X \le 7$
and $X = 2 \cdot \alpha$, $\alpha > 1$. In this case, for
the $Base$ method $B$ becomes $\alpha B$ and $T$ becomes $\alpha T$.
Recall that we consider the same PER !
For $S$ it turns out that $B_s$ becomes $\alpha B_s$, and $T+T_s$ becomes
$\alpha (T+T_s)$. Now the percentage of the improvement is
as shown in Eq.~\ref{equ:thr2}:

\begin{equation}
\label{equ:thr2}
\frac{\frac{\alpha B_s}{C_1+\alpha (T+T_s)}}{\frac{\alpha B}{C_1+\alpha T}}
\end{equation}

One can easily verify that for $\alpha >1$ the percentage
of improvement is lower, and it decreases as $\alpha$ increases.
Intuitively, given the same overhead $C_1$, multiplying the
PSDU's transmission time at $S$ by $\alpha$ is more
significant than at the $Base$ method. This is
because in $S$ the PSDU's transmission time is larger.
This reduces the attractiveness of $S$ as $\alpha$ increases.

\clearpage

\clearpage

\clearpage

%% file: summary.tex
\clearpage

\section{Summary and further research}

We propose several methods to improve the Throughput
of a WiFi channel based on aggregation and blindly retransmitting
several copies of the same MPDU(s) within a single
transmission attempt. We examine the performance of retransmitting
the first 1, 2, 3 or 4 MPDUs in the transmission attempt, and the case
in which all the MPDUs in the transmission
attempt are retransmitted the same number of copies.
We show that there are cases, especially
in large PERs and small size MPDUs,
where a significant improvement in the Throughput
is achieved.

This research is only a first step in 
exploring the retransmission of MPDUs within aggregation.
The aim of this paper is mainly to introduce the idea and
show its feasibility.
It is possible now to implement the idea and evaluate its performance
in any WiFi transmission scenario
based on aggregation, e.g. in a WiFi cell where collisions
are possible, and/or in TCP traffic where a single
pair of transmitter/receiver can collide due to the transmission
of above$-$Layer 2 Acks. Another research direction is
to integrate this idea with the
use of Link Adaptation.

%% file: main.bbl
\begin{thebibliography}{10}

\bibitem{IEEEBase1}
\newblock{IEEE P802.11-REVmc$^{TM}$/D5.2 },
\newblock{Draft Standard for Information Technology - 
Telecommunications and Information Exchange between Systems - Local
and Metropolitan Area Networks - Specific Requirements. Part 11:
Wireless LAN Medium Access Control (MAC) and Physical Layer (PHY)
Specifications},
\newblock{IEEE, NewYork, March 2016}

\bibitem{IEEEac}
\newblock{IEEE Std. 802.11ac$^{TM}$-2013},
\newblock{IEEE Standard for Information Technology - 
Telecommunications and Information Exchange between Systems - Local
and Metropolitan Area Networks - Specific Requirements. Part 11:
Wireless LAN Medium Access Control (MAC) and Physical Layer (PHY)
Specific requirements. Amendment 4: Enhancements for Very
High Throughput for Operation in Bands below 6 GHz},
\newblock{IEEE, NewYork, 2013}

\bibitem{IEEEBase}
\newblock{IEEE Std. 802.11$^{TM}$ - 2012 },
\newblock{ Standard for Information Technology - 
Telecommunications and Information Exchange between Systems - Local
and Metropolitan Area Networks - Specific Requirements. Part 11:
Wireless LAN Medium Access Control (MAC) and Physical Layer (PHY)
Specifications},
\newblock{IEEE, NewYork, 2012}

\bibitem{UM}
R. de Vegt,
\newblock{802.11ac Usage Models Document},
\newblock{Doc.:IEEE 802.11-09/0161r2, 2009}

\bibitem{XR}
Y. Xiau, J. Rosdahl,
\newblock{Throughput and Delay Limits of IEEE 802.11}, 
\newblock{IEEE Communication letters, 6(8), 
(2002), 355-357}

\bibitem{L}
H.C. Lee,
\newblock{A MAC Throughput in the Wireless LAN},
in Book 
\newblock{Advanced Wireless LAN, Editor Dr. Song Guo,
Intech, 2012, pp. 23-69}

\bibitem{B}
G. Bianchi,
\newblock{Performance analysis of the IEEE 802.11 distributed coordination
function}, 
\newblock{IEEE Journal on Selected Areas in Communication, 18(3), (2000) pp. 
535-547}


\bibitem{CBV}
P. Chatzimisios, A. C. Boucouvalas, V. Vitsas,
\newblock{IEEE 802.11 Wireless LANs: Performance Analysis and Protocol
Refinement}, 
\newblock{EURASIP Journal on Applied Signal Processing (2005), pp. 67-78}

\bibitem{CA}
M. M. Carvalho, J. J. Garcia-Luna-Aceves,
\newblock{Delay Analysis of IEEE 802.11 in single-hop networks}, 
\newblock{Proc. 11th IEEE Int. Conf. on Network Protocols (ICNNP), (2003), 
pp. 146-155}



\bibitem{WLI}
Y.F. Wen, F.Y.S. Lin, K.W. Lai,
\newblock{Access Delay and Throughput Evaluation of Block Ack
under 802.11 WLAN}, 
\newblock{Proc. IASTED Communication and Computer networks conf (2004)}

\bibitem{TC}
I. Tinnirello, S. Choi,
\newblock{Efficiency Analysis of Burst Transmissions with Block Ack
in Contention-Based 802.11e WLANs},
\newblock{Proc.IEEE International Conference on 
Communication (ICC), (2005), 3455-3460} 

\bibitem{CSV}
O. Cabral, A. Segarra, F.J. Velez,
\newblock{Implemetation of IEEE 802.11e Block Acknowledgement
Policies}, 
\newblock{Proc. of the World Congress 
on Engineering (WCE), (2008), 741-746}


\bibitem{GPE}
J. Gross, O. Punal, M. Emmelmann,
\newblock{Multi-User OFDMA Frame Aggregation for Future Wireless
Local Area Networking}, 
\newblock{Proc. of IFIP Networking conf., (2009), 220-233}

\bibitem{C2}
H. Chen,
\newblock{Throughput Analysis of Block-Ack in IEEE 802.11n}, 
\newblock{The 2nd International Conference on Computer Application
and System Modeling (ICCASM), (2012), 956-959}


\bibitem{SC}
A. Sidelnikov, J., S. Choi,
\newblock{Fragmentation/Aggregation Scheme for Throughput
Enhancement of IEEE 802.11n WLAN}, 
\newblock{Proc. IEEE Asia Pacific Wireless Communications
Symposium (APWCS), (2006), 24-25}

\bibitem{LW}
Y. Lin, V.W.S. Wong,
\newblock{Frame Aggregation and Optimal Frame Size Adaptation for
IEEE 802.11n WLANs},
\newblock{IEEE Global Telecommunications Conferemce
(GLOBECOM), (2006), 1-6}

\bibitem{GK}
B. Ginzbyrg, A. Kesselman,
\newblock{Performance Analysis of A-MPDU and A-MSDU
Aggregation in IEEE 802.11n},
\newblock{IEEE Sarnoff Symposium, (2007), 1-5}

\bibitem{SNCSKJ}
D. Skotdoulis, Q. Ni, H.H, Chen, A.P. Stephens, C. Kiu, A. Jamalipour,
\newblock{IEEE 802.11n MAC Frame Aggregation Mechanisms For
next-Generation High-Throughput WLANS}, 
\newblock{IEEE Wireless Communications, 15(1), (2008), 40-47}

\bibitem{KHS}
B.S. Kim, H.Y. Huong, D.K. Sung,
\newblock{Effect of Frame Aggregation on the Throughput Performance
of IEEE 802.11n}, 
\newblock{Wireless Communications and Networking
Conf. (WCNC), (2008), 1740-1744}

\bibitem{C1}
K. Chan,
\newblock{Evaluation and Enhancements in 802.11n WLANs - Error-Sensitive
Adaptive Frame Aggregation}, 
\newblock{Master Thesis, San Jose State University, (2009), 
http://schlarworks,sjsu.edu/edt-projects/79/ (Accessed, March 2013)}

\bibitem{WW}
C.Y. Wang, H.Y. Wei,
\newblock{IEEE 802.11n MAC Enhancement and Performance Evaluation}, 
\newblock{Mobile Networks and Applications, 14(6), (2009),
760-771}

\bibitem{SS}
T. Selvan, S. Srikanth,
\newblock{A Frame Aggregation Scheduler for IEEE 802.11n}, 
\newblock{National Conference on Communication (NCC), (2010), 1-5}

\bibitem{Z}
B. Zielinski,
\newblock{Efficiency analysis of IEEE 802.11 protocol with block acknowledge
and frame aggregation}, 
\newblock{Bulletin of the Polish Academy of Sciences Technical Sciences,
59(2), (2011), 235-243}

\bibitem{DAM}
Y. Daldoul, T. Ahmed, D.E. Meddour,
\newblock{IEEE 802.11n Aggregation Performance Study for the Multicast},
\newblock{IFIP Wireless Days (WD), (2011), 1-6}

\bibitem{SOSH}
A. Saif, M. Othman, S. Subramaniam, N.A. Hamid,
\newblock{An Enhanced A-MSDU Frame Aggregation Scheme for
802.11n Wireless networks}, 
\newblock{
Wireless Personal Communications, 6(4),
(2012), 683-706 }

\bibitem{KKS}
J. Kolap, S. Krishnan, N. Shaha,
\newblock{Frame Aggregation Mechnism For High-Throughput
802.11n WLANs}, 
\newblock{International Journal of Wireless \& Mobile Networks (IJWMN),
4(3), (2012), 141-153}

\bibitem{KMLPC}
Y. Kim, E. Monroy, O. Lee, K.J. Park, S. Choi,
\newblock{Adaptive two-Level Frame Aggregation in IEEE 802.11n WLAN},
\newblock{18th Asia-Pacific Conference on Communication (APCC),
(2012), 658-663}

\bibitem{P}
N. Park,
\newblock{IEEE 802.11ac: Dynamic Bandwidth Channel Access}, 
\newblock{IEEE  International Conference on Communications (ICC), (2011),
1-5}

\bibitem{OKACHN}
E.H. Ong, J. Kneckt, O. Alanen, Z. Chang, T. Huovinen, T. Nihtila,
\newblock{IEEE 802.11ac: Enhancements for very High Throughput WLANs},
\newblock{IEEE 22nd International Symposium on Personal Indoor and
Mobile radio Communications (PIMRC), (2011), 849-853}

\bibitem{CAHNOKR}
Z. Chang, O. Alanen, T. Huovinen, T. Nihtila, 
E.H. Ong, J. Kneckt, T. Ristaniemi, 
\newblock{Performance Analysis of IEEE 802.11ac DCF with
Hidden Nodes}, 
\newblock{IEEE 75th Vehicular Technology Conference (VTC), 
(2012), 1-5}

\bibitem{BBSVO}
B. Bellalta, J. Barcelo, D. Staehle, A. Vinel, M. Oliver,
\newblock{On the Performance of Packet Aggregation
in IEEE 802.11ac MU-MIMO WLANs}, 
\newblock{IEEE Communication Letters, 16(10), (2012), 
1588-1591}

\bibitem{SA}
O. Sharon. Y. Alpert,
\newblock{MAC level Throughput comparison: 802.11ac vs. 802.11n,}
\newblock{Physical Communication Journal 12 (2014) 33-49 }


\bibitem{CS}
L. Changwen, A.P. Stephens,
\newblock{Delayed Channel Access for IEEE 802.11e Based WLAN,}
\newblock{IEEE International Conference on 
Communication 10 (2006) 4811-4817}

\bibitem{SNMB}
D. Skordoulis, Q. Ni, G. Min, K. Borg,
\newblock{Adaptive Delayed Channel Access for IEEE 802.11n WLANs,}
\newblock{IEEE International Conference on Circuits and
Systems for Commuincations (ICCSC) (2008) 167-171}

\bibitem{TQDDYT}
Li T, Ni Q, Malone D, Leith D, Xiao Y, Turletti T,
\newblock{Aggregation with fragment retransmission for
very high-speed WLANs},
\newblock{IEEE/ACM Trans. Networking, 17(2), (2009) 591-604}


\bibitem{SNZ}
D. Skordoulis, Q. Ni, C. Zarakovitis,
\newblock{A Selective Delayed Channel Access (SDCA) for
the high-throughput IEEE 802.11n},
\newblock{IEEE Wireless Communication and Networking
Conference (WCNC) (2009) 1-8}

\bibitem{MGCR}
D. Camps-Mur, M. D. Gomony, X. Perez-Costa, S. Sllent-Ribes,
\newblock{Leveraging 802.11n Frame Aggregation to enhance QoS and 
Power Consumption in Wi-Fi networks},
\newblock{Computer Networks, 56(12) (2012) 2896-2911}

\bibitem{SA1}
O. Sharon. Y. Alpert,
\newblock{The combination of Aggregation, ARQ, QoS Guarantee and
mapping of Application flows in Very High Throughput 802.11ac Networks,}
\newblock{Submitted, Physical Communication Journal }


\bibitem{WFA}
\newblock{ 
Wi-Fi CERTIFIED n  Interoperability Test Plan Version v2.0.38 published (2011)},

\bibitem{L1}
J. Lemmon,
\newblock{Wireless link statistical bit error rate model}, 
\newblock{Technical Report 02-934, U.S. Dept. of Commerce,
June, (2002)}


\bibitem{ZRM}
M. Zorzi, R. Rao, L.B. Milstein,
\newblock{On the accuracy of a first order Markov Model for data
transmission on  fadding channels}, 
\newblock{IEEE ICUPC'95, (1995)}




\end{thebibliography}
